\documentclass[showpacs,prx,superscriptaddress,twocolumn]{revtex4-1}
\usepackage[table]{xcolor}
\usepackage{pgf}
\usepackage[utf8]{inputenc}
\usepackage{amsmath}
\usepackage{braket}
\usepackage{amssymb}
\usepackage{amsmath}
\usepackage{bbold}
\usepackage{enumitem} 
\usepackage{appendix}
\usepackage{graphicx}
\usepackage{overpic}
\usepackage{epstopdf}
\usepackage[pdftex,bookmarks,colorlinks]{hyperref}
\usepackage{hyperref}

\usepackage{ulem}

\newcommand{\mean}[1]{\left\langle #1\right\rangle}

\newcommand{\nep}{\textrm{e}}

\newcommand{\bd}[1]{\hat b^{\dagger}_{#1}}
\newcommand{\Real}{\operatorname{\Re{\rm e}}}
\newcommand{\Aimag}{\operatorname{\Im{\rm m}}}
\newcommand{\Mod}[1]{\ (\mathrm{mod}\ #1)}

\newcommand{\green}[1]{{#1}}

\begin{document}
\title{Floquet time crystals in clock models}
\author{Federica Maria Surace}
\affiliation{SISSA, Via Bonomea 265, I-34136 Trieste, Italy}
\affiliation{Abdus Salam ICTP, Strada Costiera 11, I-34151 Trieste, Italy}

\author{Angelo Russomanno}
\affiliation{NEST, Scuola Normale Superiore \& Istituto Nanoscienze-CNR, I-56126 Pisa, Italy}
\affiliation{Abdus Salam ICTP, Strada Costiera 11, I-34151 Trieste, Italy}

\author{Marcello Dalmonte}
\affiliation{Abdus Salam ICTP, Strada Costiera 11, I-34151 Trieste, Italy}
\affiliation{SISSA, Via Bonomea 265, I-34136 Trieste, Italy}
\author{Alessandro Silva}
\affiliation{SISSA, Via Bonomea 265, I-34136 Trieste, Italy}
\author{Rosario Fazio}
\affiliation{Abdus Salam ICTP, Strada Costiera 11, I-34151 Trieste, Italy}
\affiliation{NEST, Scuola Normale Superiore \& Istituto Nanoscienze-CNR, I-56126 Pisa, Italy}

\author{Fernando Iemini}
\affiliation{Instituto de Física, Universidade Federal Fluminense, 24210-346 Niter\'oi, Brazil}
\affiliation{Abdus Salam ICTP, Strada Costiera 11, I-34151 Trieste, Italy}

\begin{abstract}
We construct a class of period-$n$-tupling discrete time crystals based on $\mathbb{Z}_n$ clock variables, for all the integers $n$. We consider two 
classes of systems where this phenomenology occurs, disordered models with short-range interactions and  fully connected models. In the case of  
short-range models we provide a complete classification of time-crystal phases for generic $n$. For the specific cases of $n=3$ and $n=4$ we 
study in details the dynamics by means of exact diagonalisation. In both  cases, through an extensive analysis of the Floquet spectrum, we are able to 
fully map the phase diagram.   In the case of infinite-range models, the mapping onto an effective bosonic Hamiltonian allows us to investigate the 
scaling to the thermodynamic limit. After a general discussion of the problem, we  focus on $n=3$ and $n=4$, representative 
examples of the generic behaviour. Remarkably, for $n=4$ we find clear evidence of a new  crystal-to-crystal transition between period $n$-tupling and period 
$n/2$-tupling.

\end{abstract}

\maketitle
\section{Introduction}
Classifying phases of matter in terms of symmetry breaking, one of the highlights of Landau's legacy, is a  fundamental pillar in our 
understanding of Nature~\cite{goldenfeld}. Its impact in modern physics spans over a multitude of fields, from condensed-matter to 
high-energy physics, embracing both equilibrium and non-equilibrium phenomena. Time-translation symmetry breaking has a special 
place in this saga. It has been considered for the first time only a few years ago, almost a century after Landau's work.

A time crystal is a state of matter where time-translation symmetry is spontaneously broken. Its possible existence has been 
proposed by  Wilczek~\cite{Wilczek2012,Shapere2012,Wilczek2013} generating immediately a  fervent debate~\cite{first-papers}.  
A no-go theorem~\cite{Watanabe2015} forbids time-translation symmetry breaking to take place in the ground or thermal state of a  
quantum system  (at least for not too long-ranged interacting systems).  A time crystal therefore emerges as a truly non-equilibrium 
phenomenon that cannot be understood as a simple analogue in time of an ordinary crystal. 
 
The intense theoretical effort to look for non-equilibrium time crystals has focused both on closed~\cite{Else2016a,Khemani_2016,
Khemani2016,Zhang2016,Choi2016,Russomanno2017,Yao2017,Ho2017,Berdanier2018,Lazarides2017,Huang2017,Else2017,Syrwid2017} 
and open many-body quantum systems~\cite{Fernando2017,Gong2018,Smale2018,Shammah2018}. So far, periodically driven systems 
have represented the most successful arena to study time crystals. Here, despite the quantum system being governed by a time-dependent 
Hamiltonian of period $T$, there are  observables that oscillate, in the thermodynamic limit, with a multiple period $qT$. Floquet time 
crystals~\cite{Else2016a} (also known as $\pi$-spin glasses~\cite{Khemani_2016}) were observed for the first time in 2017 with trapped 
ions~\cite{Zhang2016} and with Rydberg atoms~\cite{Choi2016} following earlier theoretical predictions~\cite{Else2016a,Khemani_2016}.  
New  experimental evidences appeared very recently in Refs.~\cite{PhysRevLett.120.180603,PhysRevB.97.184301,Pal2018}.

An essential requirement for the existence of Floquet time crystals is the presence of an ergodicity-breaking mechanism  which prevents the system 
from heating up to infinite temperature~\cite{Ponte2014, D_Alessio_2014, Huse2013}. Many-body localisation induced by disorder can hinder 
energy absorption in support of a discrete time-crystal phase~\cite{Else2016a}.  In the absence of disorder, solvable models with infinite-range 
interactions possess the necessary ingredients~\cite{Russomanno2017} as well. In specific cases, subharmonic oscillations can be exhibited by 
many-body systems with long-range interactions in a pre-thermal regime~\cite{Else2017} or with a slow critical dynamics~\cite{Ho2017, Choi2016}.

Until now, essentially all the theoretical activity on time crystals has focused on period doubling. In this case time-translation symmetry 
is spontaneously broken from a group $\mathbb{Z}$ to $2\mathbb{Z}$. This  is intimately connected to the fact that the system  
breaks also a discrete internal symmetry, the $\mathbb{Z}_2$ one~\cite{Else2016a} leading to the concept of  spatio-temporal 
ordering~\cite{Khemani2016,Khemani_2016,von_Keyserlingk_2016}.  It is natural to expect that a similar model with $\mathbb{Z}_n$ 
symmetry can produce oscillations with a multiple periodicity. Although mentioned in the literature~\cite{Keyserlingk2016,Else2016a,
von_Keyserlingk_2016, Sreejith_2016}, this possibility has not been analysed so far.  The early experimental observation of period 
tripling~\cite{Choi2016} adds further motivations to explore this issue.  

In this paper we tackle this problem by studying Floquet time crystals in driven $n-$states clock models. When  $n > 2$, the spontaneous 
breaking of  $\mathbb{Z}_n$ symmetry leads to a wealth of new phenomena.  The appearance of the time-crystal phases, as well 
as their properties, depends in a non-trivial way on the integer $n$ and on the symmetries of the periodic driving. Not all classes of clock
Hamiltonians allow for time-translation symmetry breaking. In this work we determine the conditions under which a time-crystal phase is 
possible and we provide a classification of the possible different phases for a generic $n$.   Furthermore  for $n \ge 4$, different phases can 
appear depending on the choices of the coupling constants of the underlying Hamiltonian.  We predict a new direct transition between time crystals 
of different periodicity. 

Some of the  recent impressive experimental advancements in the coherent evolution of interacting  models show that the building blocks to 
realise clock models are already available~\cite{Lukin_Nat}. These  new capabilities, together with the control in the unitary dynamics of 
periodically kicked many-body systems~\cite{Zhang2016, Choi2016},  make the experimental verification of our theoretical findings feasible.

The paper is organised as follows. In Section~\ref{witness:sec} we briefly review some properties of Floquet time crystals and introduce the 
observables employed to characterise the crystalline phase.
The clock Hamiltonian, studied throughout  the paper, is introduced in Section~\ref{model:sec}. We consider  two classes of models, 
a disordered short-range model where the time crystal is stabilised by  many-body localisation and the opposite limit of a fully-connected  model 
where this stabilisation comes from regular dynamics in an infinite-range interacting system. We first discuss the results for the  short-range case in 
Section~\ref{disordered:sec} and  give a complete classification of time crystals for generic $n$. In order to study the stability of the crystalline phase, 
we consider different types of perturbations. Furthermore, we provide arguments to support the persistence of the  period $n$-tupling oscillations 
for a time exponentially large with the system size.  
We support and complement our findings with numerical results based on exact diagonalisation for the cases $n=3$ and $n=4$.  In the case $n=3$ we 
are able to fully map the phase diagram using the spectral multiplet properties of the Floquet eigenvalues.  In the same Section we also discuss a model 
with $n=4$ clock variables which may lead to a transition between a time-translation symmetry breaking phase with $4$-tupling oscillations to a phase 
with period doubling oscillations.  We finally move to the study of the infinite-range clock models in Section~\ref{time_infty:sec}. 
In addition to exact diagonalisation, we also analyse the scaling to the thermodynamic limit of this model by employing a mapping onto 
a $n-$species bosonic model. This analysis is feasible because in the thermodynamic limit this model is described by a classical effective Hamiltonian
whose dynamics can be easily studied numerically. In this infinite-range case we are able to construct a model based on $\mathbb{Z}_n$ clock variables which undergoes a transition 
between a period-$n$-tupling phase and a period-$n/2$-tupling case.  We numerically verify the existence of this transition and study it in detail in the case $n=4$. 
To the best of our knowledge this is the first  example of a direct transition between two time-crystal phases. Finally, Section~\ref{concola:sec} is devoted 
to a summary and our concluding remarks. Various technical details are summarised in the Appendices.

\section{Properties of Floquet time crystals} 
\label{witness:sec}
 
Floquet time crystals have been introduced in~\cite{Else2016a}. In order to keep the presentation  self-contained it is useful to briefly recap those 
properties of Floquet time crystals that will be used in the rest of the paper.  The goal of this Section is also to introduce various indicators of discrete 
time-crystal phases,  skipping however the formal aspects of the definitions~\cite{Else2016a}. 

Given a periodic Hamiltonian $\hat{H}(t)=\hat{H}(t+T)$ a time-crystal is characterised by a local order parameter $\widehat{\mathcal{O}}_j$ whose 
time-evolved expectation value, in the thermodynamic limit $N\to\infty$, 
\begin{equation} \label{evolama:eqn}
  		\mathcal{O}_i(t)=\lim_{N\to\infty}  \bra{\psi(t)}\widehat{\mathcal{O}}_i\ket{\psi(t)} \;\;. 
\end{equation}
oscillates with a period $qT$ (for some integer $q>1$), for a generic class of initial states $\ket{\psi_0}$. In the previous definition $i$ labels a discrete space 
coordinate and $\ket{\psi(t)}=\hat{U}(t)\ket{\psi_0}$ (with $\hat{U}(t)$ the evolution operator).   It is important to stress  the importance of the thermodynamic 
limit. A time crystal is a collective phenomenon; like any other (standard) long-range order it can happen only  in this limit. 

A necessary ingredient to identify a Floquet time crystal is its robustness. The period $q$-tupling should not require, for its existence, any fine tuning 
of the parameters of the  Hamiltonian. This is important in order to distinguish a time crystal from periodic oscillations occurring at isolated  points in the 
parameter space that are however fragile, in the absence of interactions, against  arbitrarily tiny perturbations. 

In the time-crystal phase correlation functions have a peculiar temporal behaviour. The correlators will  show  persistent oscillations 
\begin{equation} \label{correlone:eqn}
  		\lim_{|i-j|\to\infty}\lim_{N\to\infty}\langle\widehat{\mathcal{O}}_i(t_1)\widehat{\mathcal{O}}_{j}(t_2)\rangle
  		= f(t_1-t_2)
\end{equation}
 when $|t_1-t_2|\to\infty$ and the separation between the sites $i,j$ grows [we define here $\widehat{\mathcal{O}}_i(t_{1,2})=\
\hat{U}^\dagger(t_{1,2})\widehat{\mathcal{O}}_i\hat{U}(t_{1,2})$]. 
 
In the rest of the paper we will restrict to stroboscopic times (multiples of the period $T$).  Moreover, we  will make extensive use of the Floquet 
states $\ket{\phi_\alpha}$, which are the eigenstates of the time-evolution operator over one period (the Floquet operator)
$$
		\hat{U}(T)\ket{\phi_\alpha}=\nep^{-i\mu_\alpha T}\ket{\phi_\alpha}  \; .
$$ 
There are two very important properties which characterise the Floquet spectrum of a time crystal and  that are intimately connected to its robustness. 
The first one concerns the eigenstates $\ket{\phi_\alpha}$; none of them can be short-range correlated  i.e. fulfilling the cluster property
\begin{equation} \label{correlalpha:eqn}
		\bra{\phi_\alpha}\widehat{\mathcal{O}}_i(t_1)\widehat{\mathcal{O}}_{j}(t_2)\ket{\phi_\alpha}
  		\sim  \bra{\phi_\alpha}\widehat{\mathcal{O}}_i(t_1)\ket{\phi_\alpha}
  		\bra{\phi_\alpha}\widehat{\mathcal{O}}_{j}(t_2)\ket{\phi_\alpha}
\end{equation} 
for $|i-j|$ larger than some correlation length. If this property is satisfied then the correlator in Eq.(\ref{correlalpha:eqn}) is time-independent since 
this is the case for each of the terms $\bra{\phi_\alpha}\widehat{\mathcal{O}}_{j}(t_i)\ket{\phi_\alpha}$. This implies that  the time-translation symmetry 
is not broken. In order to have time-translation symmetry breaking, {\it all} the Floquet states  must have quantum correlations extending 
macroscopically through the whole system and must therefore violate the cluster property~\cite{Else2016a}. For this sake, they have  to be superpositions of 
macroscopic classical configurations, the so-called cat states.  This requirement stands also behind the robustness of the time-crystal 
phase to changes of the system parameters. If the eigenstates of  the stroboscopic dynamics are non-local objects, then they do not constrain
the dynamics  of local observables, which can show in this way a behaviour  distinct from the time-periodic symmetry of the Hamiltonian. 
Particular attention must be paid to the case where the Floquet spectrum is degenerate. In this case the existence of a complete set of 
Floquet eigenstates violating cluster  property is not sufficient to identify a time crystal. In general, if the spectrum is degenerate the choice of 
a basis set is not unique: a linear combination of different Floquet states  with the same quasi-energy could in principle satisfy cluster property, 
even if the original Floquet states did not. A local perturbation can resolve this degeneracy selecting those  Floquet states in the manifold which 
have small entanglement and obey cluster property. Therefore, degeneracies break the robustness of the time crystal constraining the  time-translation 
symmetry breaking oscillations to a fine-tuned point. The undesired effect of degeneracies will clearly emerge in Sections~\ref{solvn} and~\ref{nodeg:sec}
where a complete classification of Floquet time crystals for $n$-state models will be discussed.  

Another important property concerns the Floquet  spectrum. If the periodicity of period $T$ is broken to a period $qT$ the Floquet 
spectrum will be structured in  multiplets $\mu_{\alpha}^\nu=\mu_\alpha+2\pi \nu/q$ (with $\nu=0,1,\dots, q-1$).  This property of the spectrum 
can be understood  as follows~\cite{Note4}. On expanding the time-evolving state in the Floquet basis  one gets  $\ket{\psi(t)}=\sum_{\alpha,\nu}  
R_\alpha^\nu\nep^{-i \mu_\alpha^\nu t/\tau}\ket{\phi_\alpha^\nu}$; then substituting in Eq.~\eqref{evolama:eqn}, one obtains
\begin{equation} \label{diagon_time:eqn}
  		\mathcal{O}(t)=\lim_{N\to\infty} \sum_{\alpha,\,\beta}\sum_{\nu,\,\nu'}(R_\alpha^\nu)^*
		R_\beta^{\nu'}\bra{\phi_\alpha^\nu}\hat{\mathcal{O}}\ket{\phi_\beta^{\nu'}}\nep^{i(\mu_\alpha^\nu-\mu_\beta^{\nu'})t}\,.
\end{equation}
It is convenient to analyse the various terms in the sum separately. The diagonal terms ($\alpha=\beta$, $\nu=\nu'$)  do not depend on the stroboscopic 
time and therefore are periodic with the same period of the driving. The off-diagonal ones ($\alpha\neq\beta$) will vanish in the long-time  
limit (possibly after a disorder average)~\cite{Note1} due to the destructive interference between the phase factors. Finally, the terms ($\alpha=\beta$, 
$\nu\neq\nu'$) are left; they have a  phase factor of the form $\nep^{it2\pi (\nu-\nu')/q }$. These terms are those that give rise to the period $q$-tupling 
oscillations and higher harmonics and hence to the time-crystal behaviour. 

For the purpose of analysing the numerical data, in order to see the persisting period $q$-tupling oscillations in the  order parameter of 
Eq.~\eqref{evolama:eqn},   two  quantities will be considered in the rest of the paper.  The time-correlator
\begin{equation} \label{super_zapp:eqn}
  		\mathcal{Z}_q^{[\mathcal{O}]}(t)=\nep^{-(2i\pi/q) t}\overline{\braket{\hat {\mathcal{O}}_i(t) \hat{ \mathcal{O}}_i^\dagger(0)}}
\end{equation}
is a constant  if there are  period $q$-tupling oscillations. In the previous definition the angle brackets indicate the expectation value over an initial 
state and the bar $\overline{\cdots}$  refers to the average  performed over disorder and a set of initial states (in some cases this average includes also 
a spatial average over the chain). 

Often it will be convenient also to  consider the discrete Fourier transform of Eq.(\ref{correlone:eqn}) of the oscillating quantities (followed over $N_T$ periods)
\begin{equation} \label{trasformazzi:eqn}
  		f^{[\mathcal{O}]}_\omega=T \sum_{k=0}^{N_T}\mean{\hat{\mathcal{O}}}_{kT}\nep^{i\omega kT} \;. 
\end{equation} 
where we denote $\mean{\hat{\mathcal{O}}}_t=\bra{\psi(t)}\hat{\mathcal{O}}\ket{\psi(t)}$.
Time-translation symmetry breaking appears if the position of the dominant peak in the Fourier transform tends
to the period $q$-tupling frequency 
\begin{equation} \label{eq:omega_q}
  		\omega(q)=\frac{2\pi}{qT}
\end{equation}
when the thermodynamic limit $N\to\infty$ is considered.

\section{Kicked clock models} \label{model:sec}

The dynamics  of the systems we are going to study in this paper is governed by a time-periodic Hamiltonian $\hat{{\cal H}}(t)$ of the form
\begin{equation}
	\hat{{\cal H}}(t) = \hat H_{\hspace{0.03cm}\left[\cdot \right]}^{\left[\cdot \right]} + 
	\sum_{k  \in \mathbb{Z}} \delta (t-kT) \hat K_{\hspace{0.03cm}\left[\cdot \right]}^{\left[\cdot \right]}\;. 
\label{total-ham}
\end{equation}
where both $ \hat H$ and $ \hat K$ are time-independent operators. The evolution in one period is defined by the Floquet operator
\begin{equation} \label{protocol:eqn}
	 \hat U(T) =  e^{-i T \hat H} e^{-i \hat K}.
\end{equation}
It is  characterised by a time-independent dynamics, dictated by $ \hat H$, spaced out  by kicks  (at intervals $T$) controlled by the 
operator  $\hat K$. Both $ \hat H$ and $ \hat K$ will depend on many different parameters (the various coupling constants, $n$, range of the 
couplings, $\dots$) and several different models will be analysed. The symbol $\left[\cdot \right]$ in the superscript and subscript of the 
Hamiltonian operators in Eq.(\ref{total-ham}) indicate the set of all these parameters needed to specify the evolution. The  form of $ \hat H$ 
and of $ \hat K$, together with their dependence on these various couplings will be specified in the forthcoming paragraphs. In order to simplify 
the notation, some of the indices may not always be indicated, whenever not necessary for  the understanding of the text.

\begin{figure}[b]
 \includegraphics[width=0.5\textwidth]{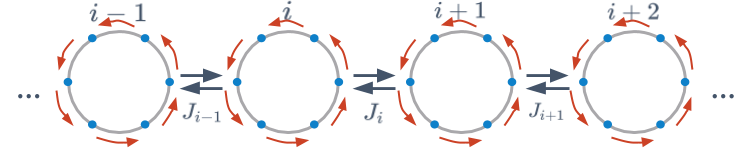}
 \caption{Pictorial representation of a clock model (with $n=6$) in a one-dimensional chain. Each site, labeled by the index $i$, has a 
 	$n$-dimensional local Hilbert space, the blue points on the circle indicate the possible states of the clocks. The red arrows 
	represent the action of the $\tau$ operators, and hence the action of the kicks, as discussed in the main text. In the case shown here, the 
	clocks interact through a nearest-neighbour coupling of amplitude $J_i$. The coupling between the sites is indicated by the double black arrow.}
 \label{clock}
\end{figure}

\paragraph*{Clock variables -} As sketched in  Fig.\ref{clock}, clock models~\cite{Baxter89} are defined on a lattice with   $L$  sites, each site having a local 
basis of $n$ states that can be  represented as $n$ positions on a circle  (the "hands" of the clock). This generalises the case $n=2$ where the canonical local 
basis is $\ket{\uparrow}$, $\ket{\downarrow}$.  The local Hilbert space is characterised by the operators $\hat \sigma$ and $\hat \tau$, satisfying the relations
\begin{equation}\label{eq:clockprop}
	\hat \sigma \hat \tau= \omega \hat \tau \hat \sigma,\qquad \hat \sigma^n=1,\qquad \hat \tau^n=1 
\end{equation}
with $\omega=e^{2\pi i/n}$. In the basis  where $\sigma$ is diagonal 
\begin{equation} \label{cappelli:eqn}
  	\hat \sigma\ket{\omega^k} = \omega^k \ket{\omega^k}, \qquad \hat \tau\ket{\omega^k}=\ket{\omega^{k+1}}
\end{equation}
for $k=0,1,...,n-1$, and 
\begin{equation} \label{tauvolata:eqn}
	\hat \sigma=\left( \begin{array}{cccc}
	1 & 0 & 0 & 0 \\
	0 & \omega & 0 & 0\\
	0 & 0 & \dots & 0\\
	0 & 0 & 0 & \omega^{n-1}\end{array} \right),\quad
	\hat \tau=\left( \begin{array}{cccc}
	0 & 0 & 0 & 1 \\
	1 & 0 & 0 & 0\\
	0 & \dots & 0 & 0\\
	0 & 0 & 1 & 0\end{array} \right).
\vspace{0.5cm}
\end{equation}
For later purposes, note that $(\hat{\sigma}^{\dagger})^m = \hat{\sigma}^{n-m}$ and $(\hat{\tau}^{\dagger})^m = \hat{\tau}^{n-m}$. Moreover, for $n=2$, $\hat \sigma$ 
and $\hat \tau$ become the Pauli matrices $\hat \sigma^z$ and $\hat \sigma^x$.  While in the Ising case the parity symmetry 
is related to the flipping of all the spins, in a clock model the $\mathbb{Z}_n$ symmetry operation is implemented by the operator  that moves all the 
hands of the clock one step forward.

The operators defined above will be used to construct the model Hamiltonians $\hat H$ and $\hat K$. In the rest of this Section we will first define  the 
time-independent Hamiltonian $ \hat H$ and afterwards  we will discuss the evolution due to the kicks.

\paragraph*{The model Hamiltonian $\hat H$ -} The evolution between two kicks is governed by the $n$-state clock 
Hamiltonian $\hat H_n$~\cite{Baxter89,Fendley2012}, see Fig.\ref{clock}, whose most general form is
\begin{eqnarray}
	 	\hat H_n &= & \sum_{i\neq j} J_{ij} \sum_{m=1}^{n-1} \alpha_m (\hat \sigma_i^\dagger \hat \sigma_{j})^m \nonumber \\
			      &+& \sum_i h_{z,i} \sum_{m=1}^{n-1} \gamma_m \,\hat \sigma_i^m 
    		+\sum_i h_{x,i} \sum_{m=1}^{n-1} \beta_m \,\hat \tau_i^m
\label{Hamiltonian-H}	
\end{eqnarray}
with real couplings $J_{ij}, h_{x,i},h_{z,i}$ and complex $\alpha_m, \beta_m, \gamma_m$~\cite{Note3}. The site-label $i$ runs from 1 to $L$. 
In the case of short-range interaction we will further assume periodic boundary condition.  In order for the Hamiltonian to be Hermitian, 
$\alpha_m^*=\alpha_{n-m}$,  $\beta_m^*=\beta_{n-m}$ and $\gamma_m^*=\gamma_{n-m}$. While  $J_{ij}$ accounts for the interaction between 
different sites, $h_{x,i}$ ($h_{z,i}$) represent a transverse (longitudinal) field. In the absence of longitudinal field ($h_{z,i}=0$, $\forall i$) the Hamiltonian 
has a $\mathbb{Z}_n$ symmetry generated by 
$$
\hat X=\prod_{i=1}^L \hat \tau_i \;.
$$
Together with the analysis for generic $n$, in the rest of the paper we will consider several different choices of the couplings, encompassing both a 
disordered short-range model as well as an infinite-range case.  In these specific cases we will perform explicit numerical/analytical calculations. 
For future reference these specific cases are summarised Table~\ref{table-1}.

\begin{center}
 {\def\arraystretch{1.5}
\begin{table*}
\begin{tabular}{|l|l|l|l|}
 		\hline
 		\; & \; &  \; \; $n=3$ &  \; \;  $n=4$\\
 		\hline
  		\; \;  Short-Range  interaction (SR) \; \; & \; \; $J_{ij}=J_i\delta_{i+1,j}$ \;\;  &  \; \;$\alpha_1=\alpha_2^*=e^{i\varphi}$  \; \; &  \; \; $\alpha_2=1$, 
		$\alpha_1=\alpha_3^*=(1-\eta)e^{i\varphi}/2$ \; \; \\
 		\; & \; &  \; \;$\beta_1=\beta_2^*=e^{i\varphi_x}$ &  \; \; $\beta_2=\eta$, $\beta_1=\beta_3^*=\delta$ \\
 		\;  & \; &  \;\;  $\gamma_1=\gamma_2^*=e^{i\varphi_z}$ &  \; \; $\gamma_2=1$, $\gamma_1=\gamma_3^*=\delta$\\
 		\hline
  		\; \; Long-Range  interaction  (LR)  & \;\;  $J_{ij}=-\frac{J}{L}$  &\; \;  $\alpha_1=\alpha_2^*=1/2$& \; \;  $\alpha_2=\eta J'/J$, $\alpha_1=\alpha_3^*=(1-\eta)/2$ \\
		 \; & \;\; $h_{x,i}=-h$  & \; \;  $\beta_1=\beta_2^*=1$ & \; \;  $\beta_2=2\eta h'/h$, $\beta_1=\beta_3^*=1-\eta$ \\
		\; &  \; & \; \;  $\gamma_1=\gamma_2^*=0$ & \; \;  $\gamma_1=\gamma_2=\gamma_3=0$\\
 		\hline
\end{tabular}
\caption{A summary of the various choices of couplings that we will analyse in the paper. We will consider both short- and long-range systems (SR and LR 
respectively). In the cases $n=3$ and $n=4$ the coefficients $\alpha_m, \beta_m, \gamma_m$ will parameterised as specified in the table. For $n=4$, $J,J',h,h'\ge 0$ 
and the parameter $0\le \eta\le 1$ will control the transition between time-crystals with different symmetries.}
\label{table-1}
\end{table*}}
\end{center}

More specifically the first model we will discuss is a short-range disordered $n$-state clock model. Both the nearest-neighbour coupling 
$J_i$  and $h_{x,i}$/ $h_{z,i}$ will be  real random numbers uniformly distributed  in the intervals $[J_z/2, 3J_z/2]$ and  $[0,h_z]$ respectively.  
Only the strength of the interactions and of the fields are allowed to vary over the chain,  $\alpha_m$, $\beta_m$ and 
$\gamma_m$ are site-independent.  For the long-range  case  we will  consider  a generalisation of the Lipkin-Meshkov-Glick~\cite{Lipkin} model. The Hamiltonian   
has  a $\mathbb{Z}_n$ symmetry generated by $\hat X$, as well as an invariance under sub-systems permutations. Despite its simplicity the model Hamiltonian 
contains, as we will show, the necessary ingredients  to realise a time-crystal, in particular an extensive number of symmetry breaking eigenstates. For 
$n=4$, the parameters of the Hamiltonian can be adjusted  to favour a phase either with $\mathbb{Z}_4$  spontaneously symmetry breaking states, or a phase 
with lower $\mathbb{Z}_{2}$ symmetry breaking states.   

\paragraph*{Time evolution during a kick -}  The kicks are local, acting on each site independently, i.e. $\hat K = \sum_i \tilde{K}_i$.  
It is convenient to discuss the evolution due to the kicks by introducing the operator $\hat X^{(n)}_{\epsilon}$ as 
\begin{equation}  
 		e^{-i \hat K^{(n)}_{\epsilon}} \equiv \left[\hat X^{(n)}_{\epsilon}\right]^p.
\end{equation}
(the superscript $n$ and the subscript $\epsilon$ are made explicit as they are essential in characterising the type of kick).
Indeed the generic kick will depend on the parameter $\epsilon$ that will be varied in order to probe the stability of the time-crystal phase.

In the ideal case,  the kicking is $p$-times the application of  the operator $\hat X^{(n)}_{\epsilon=0}=\prod_{i=1}^L \hat \tau_i$.  Assuming for 
simplicity $p=1$, if the operator $\hat{\tau}_i$ acts over an eigenstate of $\hat{\sigma}_i$  its effect is simply to exchange it with another 
eigenstate [see Eq.~\eqref{cappelli:eqn}]. The state returns back to itself after the action of $n$ times $\hat{\tau}$. A measure of  
the  expectation of $\hat{\sigma}_i$ witnesses  naturally the  period $n$-tupling. 

It is convenient to write the  perfect-swapping kicking operator $ \hat X^{(n)}_{\epsilon=0}$ as
\begin{equation} \label{kickolo:eqn}
  		\hat X^{(n)}_{\epsilon=0} \equiv \hat X = \prod_{i=1}^L \hat \tau_i = \prod_{i=1}^L \nep^{i(\pi/n) \hat{\Theta}_i^{(n)}}
\end{equation}
where $\hat{\Theta}^{(n)}_i$ is an Hermitian matrix acting in the $i$-th site. Specifically, for the cases $n=2,3$ or $4$, 
$ \hat{\Theta}$ has the form
 \begin{eqnarray} \label{perturbanti:eqn}
  		\hat \Theta^{(2)} &= &\left( \begin{array}{cc}
        	 	-1 & 1 \\ 1 & -1
         	\end{array}\right) ,\nonumber\\
  		\hat \Theta^{(3)} &=& \frac{2}{\sqrt{3}}
  		\left( \begin{array}{ccc}
		0 & i & -i\\
		-i & 0 & i\\
		i & -i & 0
  		\end{array}\right), \nonumber\\
 		\hat \Theta^{(4)} &=& \left( \begin{array}{cccc}
		3 & -1-i & -1 & -1+i\\
		1-i & 3 & -1-i & -1\\
		-1 & -1+i & 3 & -1-i\\
		-1-i & -1 & -1+i & 3
		\end{array} \right)\,.
 \end{eqnarray}
Using the previous parameterisation, the perturbed kicking operator is defined as
\begin{equation} \label{perturbato:eqn}
 	 	\hat X^{(n)}_\epsilon = \prod_{i=1}^L \nep^{ {i\left(\frac{\pi}{n}+\epsilon\right)\hat{\Theta}^{(n)}_i} }.
\end{equation}

In the next Sections we will discuss in details the phase diagram for the different versions of the clock Hamiltonian.  We first discuss the 
case of short-range interactions, the infinite-range interacting limit will be analysed in Sec.~\ref{time_infty:sec}.

\section{Disordered short-range model} \label{disordered:sec}

In this Section we are going to focus on the short-range disordered version of the Hamiltonian Eq.~\eqref{Hamiltonian-H} (see also 
Table~\ref{table-1}) and we denote it as $\hat{H}_n^{(SR)}$. Disorder is essential for the time-crystal physics in this context. It leads to many-body 
localisation thus preventing heating up to infinite temperature.  In this regime  all the eigenstates in the spectrum of $\hat{H}_n^{(SR)}$ posses a  
long-range glassy order in the thermodynamic limit~\cite{Pollmann14}. The absence of heating, starting from a state with long-range order and driving,
guarantees that such order persists in the dynamics. On passing, we also note that, to our knowledge, this is the first time the many-body localised state
has been analysed in a clock model.

Following in spirit the same approach used for the spin-1/2 case~\cite{Else2016a} we first consider a set of couplings in Eq.~\eqref{protocol:eqn} so 
that the Floquet eigenstates can be computed exactly. This is going to form the basis for the classification of possible time-crystal phases for generic 
$n$. We then move to the analysis of the robustness of such a phase under perturbations in the evolution.  In this case, as already mentioned, 
the presence of  many-body localisation is the key to stabilise the time-crystal. We will conclude this Section with a more detailed discussion of the 
specific cases $n=3$ and $n=4$.
 
\subsection{Classification of time-crystals: $h^x=0$ }\label{solvn}  
Let us start by considering the simplest possible situation:  zero transverse field ($h_{x,i} = 0$, $\forall i$) and an ideal-swapping kick operator 
as defined in Eq.~\eqref{kickolo:eqn}.  In this  case, the operator $\hat \sigma_i$ commutes with $\hat{H}_n^{(SR)}$. It evolves after one period 
$T$ according to the  Floquet operator $\hat U(T)$ 
as
\begin{equation}\label{eq:sigma}
		{\hat U}(T)^\dagger \hat \sigma_i {\hat U}(T)=\omega^p \hat \sigma_i
\end{equation}
and then goes back to itself after a time $qT$, where $q$ is the smallest positive integer such that $qp$ is a multiple of $n$. Before 
discussing whether these oscillations at subharmonic frequency are the manifestation of a period $q$ time-crystal, it is useful to analyse 
the properties of Floquet states and quasi-energies. In this case they can be written out explicitly and -- as we are going to show -- they obey 
the properties stated in Section~\ref{witness:sec} for time-translation symmetry breaking to occur. 

It is convenient to distinguish two cases: (i)  the integers $p$ and $n$ are coprime, and (ii)   the integers $p$ and $n$ have $\mbox{gcd}(p,n)=s>1$.\\
  
\begin{itemize}
 
 \item{The integers $p$ and $n$ are coprime - }  In this case it is not hard to see that $q=n$.
 As we discuss in Appendix~\ref{app1}, we note that $\hat U(T)^n= e^{-in\hat{\bar H}}$ 
 where
\begin{equation}\label{avgH1}
		\hat{\bar H}=\sum_i J_i \sum_{m=1}^{n-1} \alpha_m\, (\hat \sigma_i^\dagger \hat \sigma_{i+1})^m.
\end{equation}
The eigenstates of $\hat{\bar H}$ can be labeled by the sequence $\{s_i\}$, with $s_i=1,\omega,\dots, \omega^{n-1}$, such that 
$$
		\hat U^n\ket{\{s_i\}} =e^{-inT\mu^+(\{s_i\})}\ket{\{s_i\}}
$$ 
and 
$$
		\mu^+(\{s_i\})=\sum_i J_i \sum_{m=1}^{n-1} \alpha_m (s_i^* s_{i+1})^m \; .
$$ 
Given a configuration  $\{s_i\}$, the  states $\ket{\{s_i\}}$, $\hat U\ket{\{s_i\}}$, $\dots$, $\hat U(T)^{n-1}\ket{\{s_i\}}$ are degenerate (and inequivalent) 
eigenstates of $\hat U(T)^n$. They are not eigenstates of $\hat U(T)$. We denote as $\ket{\psi(\{s_i\},k)}$ (with $k=0,\dots, n-1$)
the linear combinations of these states that diagonalise $\hat U(T)$:
\begin{equation} \label{statuarium:eqn}
		\hspace{0.8cm} 
		\ket{\psi(\{s_i\}, k)}=\frac{1}{\sqrt{n}}\sum_{m=0}^{n-1} \omega^{km}e^{-imT\mu^+(\{s_i\})} \hat U^m\ket{\{s_i\}}
\end{equation}
which satisfy
\begin{equation} 
		\hspace{0.8cm}
		\hat U(T)\ket{\psi(\{s_i\},k)}=\omega^k e^{-iT\mu^+(\{s_i\})} \ket{\psi(\{s_i\},k)} . \nonumber 
\end{equation}
These eigenstates have quasi-energies $\mu^+(\{s_i\})-2\pi k/n$, forming multiplets of states with $2\pi/n$ splitting in quasi-energy.\\

 \item{ The integers $p$ and $n$ have $\mbox{gcd}(p,n)=s>1$ - }  In this case the period of the time-crystal is $q=n/s$. The 
Floquet operator satisfies $\hat U(T)^q= e^{-iq\hat{\bar H}}$ (see Appendix~\ref{app1}) where now
\begin{equation}\label{avgH2}
    		\hspace{0.8cm}
		\bar{H} = \sum_i  \hspace{-0.1cm}  J_i \hspace{-0.1cm} \sum_{m=1}^{n-1} \alpha_m \,(\hat \sigma_i^\dagger \hat \sigma_{i+1})^m
     		+ \sum_i \hspace{-0.04cm}h_{z,i} \hspace{-0.12cm} \sum_{m=1}^{n/q-1} \gamma_{m q}\, \hat \sigma_i^{mq}
\end{equation}
The states $\ket{\{s_i\}}, \hat U(T)\ket{\{s_i\}}, \dots,   \hat U(T)^{q-1}\ket{\{s_i\}}$ are all degenerate (and inequivalent) eigenstates  of $\hat U(T)^m$ 
but they are not eigenstates of $\hat U(T)$.  One can construct the $q$ linear combinations (labeled by $k=0,\dots,q-1$) that diagonalise $\hat U(T)$:
\begin{equation}
		\hspace{0.8cm}
		\ket{\psi(\{s_i\}, k)}=\frac{1}{\sqrt{q}}\sum_{m=0}^{q-1} \omega^{km}e^{-imT\mu^+(\{s_i\})} \hat U^m\ket{\{s_i\}}
\end{equation}
They satisfy
\begin{equation} 
 		\hspace{0.8cm}
		\hat U(T)\ket{\psi(\{s_i\},k)}=\omega^k e^{-iT\mu^+(\{s_i\})} \ket{\psi(\{s_i\},k)}  \nonumber
\end{equation}
forming  multiplets of states with $2\pi/q$ splitting in quasi-energy.

\end{itemize}
In both cases discussed above  Floquet states are cat states: the correlators of the local observable $\hat \sigma$ for two sites $i$ and $j$ is
\begin{equation}
\label{longrange3}
		\braket{\psi(\{s_i\},\omega^k)|\hat \sigma_i^\dagger \hat \sigma_j|\psi(\{s_i\},\omega^k)}=s_i^*s_j\neq 0, 
\end{equation}
while $\braket{\hat \sigma_i}=0$ for every site $i$. Correlations show a ``glassy'' long-range order, where $\braket{\hat \sigma_i^\dagger \hat \sigma_j}$ 
can assume the values $1, \omega, \dots, \omega^{n-1}$ depending on the sites. Therefore, $|\braket{\sigma_i^\dagger \sigma_j}|=1$ and correlations do 
not vanish  in the limit $|i-j|\rightarrow \infty$. Each state {$\ket{\psi(\{s_i\},k)}$} is a cat state consisting of a superposition of {$q$} product states. 
The condition of the Floquet states being long-range correlated in order to have the time-translation symmetry breaking is fulfilled. 

It is important  to check whether the Floquet spectrum is non-degenerate. Floquet states organise in multiplets, each one separated by $2\pi/q$ from 
the other (see Section~\ref{witness:sec}).    

The  cat states {$\ket{\psi(\{s_i\},k)}$} found above are eigenstates even for $J_i=0$, but  their long-range correlations 
cannot be the evidence of a truly many-body effect. In this case the correlations are a consequence  of an unusual choice of basis set. In the non-interacting case, 
the Floquet spectrum is extensively degenerate and many choices of Floquet states basis are possible. In particular, the Floquet operator can be diagonalised 
by tensor products of single-site states which are clearly not long-range correlated.  Even if at a particular point in the parameter space the systems shows 
a time-crystal dynamics, any tiny perturbation (for example by  slightly changing  the kicking and taking the one in Eq.~\eqref{perturbato:eqn} with $\epsilon\ll 1$) 
will destroy sub-harmonic oscillations. The perturbation splits the degeneracy  and selects a basis of Floquet states which are short-range correlated.  

Interactions are needed to remove {\it all} the degeneracies and stabilise the time-crystal phase. Furthermore, the interactions must be such that there are no 
degeneracies in the Floquet spectrum. If there are degeneracies in the spectrum, one could in principle construct a linear combination of different Floquet 
states with the same quasi-energy satisfying cluster property. As we are going to show in the next section, any local perturbation can resolve this degeneracy: 
it selects the Floquet states obeying the cluster property, therefore spoiling the time-translation symmetry breaking.  

In the presence of disordered couplings  degeneracies are quite unlikely. Nevertheless, as we are going to show, they can occur and one must choose 
certain parameters in order to avoid those cases. As before, we must distinguish two cases.

\begin{itemize}
 \item{The integers $p$ and $n$ are coprime - } In this case the quasi-energies are of the form
\begin{equation}  \label{cocoprime:eqn}
		\hspace{0.8cm}
		\mu(\{s_i\}, k)=\sum_i J_i \sum_{m=1}^{n-1} \alpha_m (s_i^* s_{i+1})^m-2\pi k/n.
\end{equation}
For each set of $\{s_i\}$, the quantity $s_i^* s_{i+1}$ assumes one of the $n$ possible values $1,\omega, \dots, \omega^{n-1}$, corresponding to the 
$n$ possible angles between the two hands of the clock. If two such values yield the same energy $\sum_{m=1}^{n-1} \alpha_m (s_i^* s_{i+1})^m$, then
the spectrum is degenerate. Therefore, we have degeneracies if there exist two integers $k_1$ and $k_2$ (with $0\le k_1<k_2 \le n-1 $) such that
\begin{equation} \label{cond0}
		\sum_{m=1}^{n-1} \alpha_m \omega^{mk_1}=\sum_{m=1}^{n-1} \alpha_m \omega^{mk_2}
\end{equation}
On the other hand, if no integers $k_1$ and $k_2$ satisfy this condition, the spectrum is not degenerate and a time crystal is possible. The same condition 
has been found in the context of parafermionic chains as a criterion for the existence of strong edge zero modes~\cite{Fendley2012}. Furthermore, in 
Ref.~\cite{Jermyn2014} the same condition for strong edge modes is discussed, especially for the case $n=3$, for which it coincides with the presence of 
chiral interactions (see section \ref{casen3}).

\item{The integers $p$ and $n$ have $\mbox{gcd}(p,n)=n/q>1$ - } In this case the quasi-energies are of the form
\begin{eqnarray}
 		\mu(\{s_i\}, k) &=&\sum_i J_i \sum_{m=1}^{n-1} \alpha_m (s_i^* s_{i+1})^m \nonumber \\
   		&+&\sum_i h_{z,i} \sum_{m=1}^{n/q-1} \gamma_{m q} \,s_i^{mq}-2\pi k/q\,.
\end{eqnarray}
With respect to the previous case, the condition that
\begin{equation}\label{cond1}
 		\sum_{m=1}^{n-1} \alpha_m \omega^{mk_1}\neq \sum_{m=1}^{n-1} \alpha_m \omega^{mk_2}
\end{equation}
for every pair of integers $0\le k_1<k_2 \le n-1 $ is sufficient but not necessary to have a time crystal. If, for any pair of integers $k_1$ and 
$k_2$ violating Eq.\eqref{cond1}, and for every $s_i=1,\dots, \omega^{n-1}$ the inequality
\begin{equation}\label{cond2}
 		\sum_{m=1}^{n/q-1} \gamma_{m q} (s_i)^{mq}\neq \sum_{m=1}^{n/q-1} \gamma_{m q}
 		\,s_i^{mq}\,\omega^{mq(k_2-k_1)}
\end{equation}
is satisfied, then no degeneracies occur. Note that if $k_2-k_1$ is a multiple of $n/q$, then Eq.\eqref{cond2} is an equality for every $s_i=1,\dots, \omega^{n-1}$
and the spectrum is still degenerate. Since the couplings and the local fields are taken from a random continuous distribution, no degeneracies occur in 
the spectrum  due to additional symmetries as e.g. translation invariance. Other degeneracies would require infinitely fine-tuned couplings.

\end{itemize}

\subsection{Robustness: $h_x \ne 0$, $\epsilon \ne 0$} 
\label{nodeg:sec}

In the case on which a transverse field is present, $h_x\neq 0$, and/or for a  general form of the kick (see Eq.~\eqref{perturbato:eqn})
it is not possible  to solve the model exactly. It is still possible to study the  system for small perturbations from the solvable case.

Let $\hat U_{\lambda}(T)$ be the perturbed Floquet operator ($\hat U_{0}(T)$ is the unperturbed case) where $\lambda$ generically parameterises the 
strength of the perturbation in the kicking and/or in $\hat H$. Following~\cite{Hastings2010},  the time crystal described above is robust for sufficiently 
small $\lambda$ if there is a non-zero local spectral gap.  A naive explanation of what local spectral gap means can be given using simple perturbation theory.  
Since the perturbation is local, it can have non-zero matrix elements only between pairs of states that differ locally. On the other hand, if two states differ globally 
they can only be connected at an order $O(L)$ in perturbation theory, where $L$ is the size of the system, so they do not mix at any perturbative order in the limit 
$L\rightarrow \infty$. We  define the local spectral gap as the gap between states which are connected at a finite order in perturbation theory, not scaling with $L$. 
This is an important point because, in the thermodynamic limit, the relevant parameter in  the perturbative expansion is not the ratio between $\lambda$ and 
the typical gap (which becomes exponentially small) but the ratio between $\lambda$ and the local spectral gap. If this ratio is sufficiently small, a unitary operator 
connecting unperturbed eigenstates with perturbed ones can be constructed order by order in perturbation theory.  Moreover, assuming that the Hamiltonian 
satisfies a Lieb-Robinson bound~\cite{Liebello}, it is possible to prove that the resulting transformation is local~\cite{Hastings2010,Else2016a,De_roeck_2015}.
For translationally invariant models, one does not expect to find local spectral gaps, and this unitary transformation is in general non local.
In the presence of disorder, on the other hand, the system can exhibit many-body localisation and local gaps can exist.

The presence of a non-zero local gap guarantees the existence  of a region of the parameter space where the eigenstates of the system are connected 
to  the unperturbed ones by a local unitary $\hat V_\lambda$:
\begin{equation}
		\hat V_\lambda \ket{\psi_0(\{s_i\}, k)}=\ket{\psi_\lambda(\{s_i\}, k)}
\end{equation}
where $\hat V_\lambda$ depends continuously on $\lambda$. The argument applies to a generic small perturbation of $\hat U(T)$, irrespective of its 
specific form~\cite{Keyserlingk2016}. As shown in Appendix \ref{app2}, in our model the existence of the local mapping  $\hat V_\lambda$ and its continuity 
with respect to $\lambda$ have the following relevant consequences:
\begin{enumerate}
[label=(\roman*)]
\item the dressed operators $\tilde{\sigma}_{i,\lambda}=\hat V_\lambda^\dagger \hat \sigma_i \hat V_\lambda$ 
are local operators exhibiting long range correlations on the eigenstates $\ket{\psi_\lambda(\{s_i\}, k)}$:
\begin{equation} 
		\braket{{\psi_\lambda}(\{s_i\}, k)
		|\tilde{\sigma}_{i,\lambda}^\dagger\tilde{\sigma}_{j,\lambda}
		|{\psi_\lambda}(\{s_i\}, k)}= s_i^*s_j.
\end{equation}
Hence, the perturbed system fulfills the definition of time crystal.
\item up to corrections that are exponentially small in the system size, the order parameter operator $\tilde{\sigma}_{i,\lambda}$ evolves by 
acquiring a phase $\omega$ at each period  
\begin{equation}
 		\hat U(T)^\dagger \tilde{\sigma}_{i,\lambda} \hat U(T)= \omega^p \tilde{\sigma}_{i,\lambda}+O(e^{-cL}).
 \end{equation}
After a time $mT$, corrections are of the order $mO(e^{-cL})$, meaning that for  sufficiently large $m$ they destroy the oscillations. Therefore,
the time scale at which we expect oscillations to decay grows exponentially with $L$. Due to locality, the undressed operator $\hat \sigma_i$ has some 
finite overlap with  $\tilde{\sigma}_{i,\lambda}$: it will also show persistent oscillations (just, with a smaller amplitude).

\item the spectrum is made of multiplets of states with exact $2\pi/q$ splitting in the thermodynamic limit. For finite size systems, this is only valid up to 
corrections of the order $O(e^{-cL})$.
\end{enumerate}
The arguments given above apply to  generic $n$, and are in agreement with what has been found numerically for the specific case of 
period doubling $n=2$ [see Ref.~\onlinecite{Else2016a}].

If the unperturbed spectrum has no local gap, the argument proving the  stability of the oscillations does not apply: states that differ only locally can 
have the same quasi-energy. A local perturbation mixes these states and splits the  degeneracy, such that the new eigenstates correspond to physical 
states with no long  range correlations. If the spectrum is (locally) degenerate, the oscillations  in Eq.~\ref{eq:sigma} can become unstable to some 
arbitrarily small perturbations, meaning that no time crystal can be observed in an experiment.
This point further clarifies the need for the absence of degeneracies in the Floquet spectrum and is in agreement with the fact that many-body localisation 
induced by disorder is needed in order to  have a non-zero local gap everywhere in the spectrum~\cite{Huse2013}. In the next subsections we are 
going to corroborate  the findings presented so far with a numerical analysis for the cases with $n=3$ and $n=4$.

\subsection{Phase diagram - $n=3$}
\label{casen3}
In this case the parameters $\alpha_m$, $\beta_m$, $\gamma_m$  can be expressed in terms of three angles $\varphi$, $\varphi_x$, $\varphi_z$ as 
indicated in the central column of Table~\ref{table-1}. The parameter $\varphi$ defines the chirality  of the model; when $\varphi = 0 \Mod {\pi/3}$, the model 
is non-chiral or Potts model otherwise it is termed  chiral-clock model~\cite{Zhuang2015, Samajdar2018}.

\begin{figure}[h]
    \includegraphics[width=0.47\textwidth]{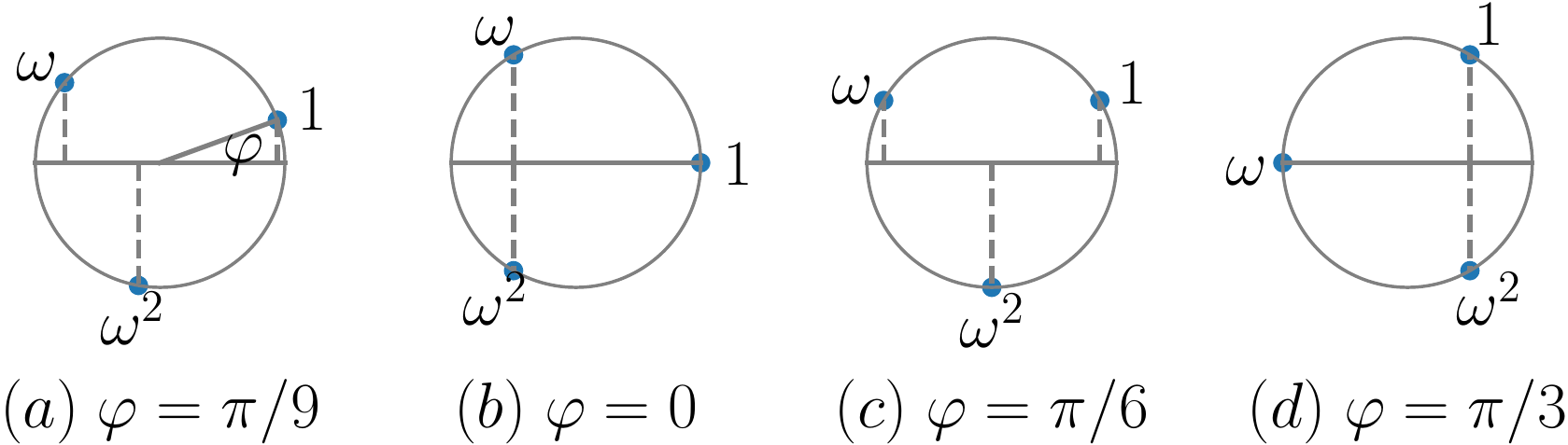}
    \caption{The dots on the circle indicate the possible values of $e^{i\varphi}s_i^* s_{i+1}$. In the non chiral 
    case (for example $\varphi=0$ and $\varphi=\pi/3$) different values of $e^{i\varphi}s_i^* s_{i+1}$ have 
    the same real part: the spectrum is degenerate. In the chiral case (for
    example $\varphi=\pi/9$ and $\varphi=\pi/6$) there is no degeneracy.}
    \label{chirality}
\end{figure}

It is useful to recap how the general analysis of Section~\ref{solvn} applies to this specific model when the solvable point ($h_x=0$, 
$\epsilon=0$) is considered. The Floquet states appear in triplets given by Eq.~\eqref{statuarium:eqn} with $n=3$ whose quasi-energies  
are respectively $\mu^+(\{s_i\})$, $\mu^+(\{s_i\})-2\pi/3$ and $\mu^+(\{s_i\})+2\pi/3$ with $\mu^+(\{s_i\})=\sum_i J_i (e^{i\varphi}s_i^* s_{i+1}
+{\rm H.\,c.})$ [see Eq.~\eqref{cocoprime:eqn}]. For each pair, $s_i^* s_{i+1}$ can assume three possible values $1, \omega, \omega^2$ and 
the corresponding interaction energies of the pair are $2J_i \Real (e^{i\varphi})$, $2J_i \Real (\omega e^{i\varphi})$ and $2J_i \Real (\omega^2 
e^{i\varphi})$. Because of the disorder in $J_i$ (which makes other degeneracies unlikely), a degeneracy in the Floquet spectrum is possible only 
if the model is non-chiral and $\varphi=0\Mod {\pi/3}$ [Fig.~\eqref{chirality}]. 

We numerically simulate the  dynamics of this model with the kicks defined in  Eq.~\eqref{perturbato:eqn} using exact diagonalisation of finite-size 
systems and then  we extrapolate to the thermodynamic limit utilizing finite size scaling.  We do that by probing the order parameter 
$\mathcal{Z}_3^{[\hat{\sigma}]}(t)$ defined in Eq.~\eqref{super_zapp:eqn} (we remind that $t$ is discrete and is a multiple of the driving period $T$). 
In the solvable case ($h_x=0$, $\epsilon=0$) it is easy to use the analysis of Sec.~\ref{solvn} and see that $\mathcal{Z}_3^{[\hat{\sigma}]}(t)=1$ for every 
$t$ and therefore period 3 oscillations last forever. The data shown in the following are typically averaged over 100 disorder configurations, the 
variance is small on the scale of the figures.
We start considering the effect of a transverse field $h_x$ for different values of the chirality parameter 
$\varphi$.

\begin{figure*}
\centering
\includegraphics[width=\textwidth]{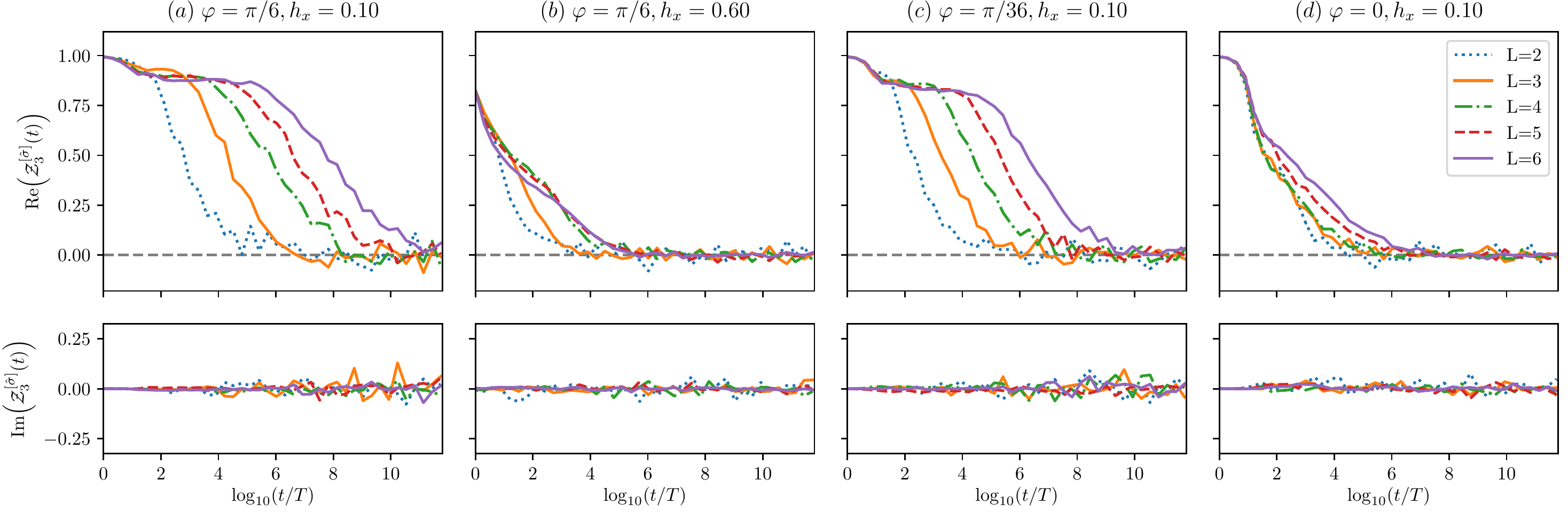}
 		\caption{Time evolution of the order parameter $\mathcal{Z}_3^{[\hat{\sigma}]}(t)$ for $n=3$ with distinct 
 		transverse fields $h_x$ and distinct chiralities $\varphi$. We notice the absence of time crystal in the non-chiral case $\varphi=0$. Numerical parameters: 
 		$\epsilon = 0$, $T=1$, $J_z=1$, $h_z=0.9$, $\varphi_z=0$, $\varphi_x=0$ and the results are 
		averaged over 100 disorder realizations. Error bars (not shown in the plots) are of the order $10^{-2}$.}
\label{fig.n3tc.Zt}
\end{figure*}

For a sufficiently small $h_x $, $\mathcal{Z}_3^{[\hat{\sigma}]}(t)$ reaches a plateau after a small time, with $\Real{\mathcal{Z}_3^{[\hat{\sigma}]}}<1$ 
and $\Aimag{\mathcal{Z}_3^{[\hat{\sigma}]}} \sim 0$ (see of Fig.~\eqref{fig.n3tc.Zt}-(a)). Oscillations with respect to the value of the plateau are observed for a 
single configuration of disorder. They tend to disappear when we take the disorder-averaged values~\cite{Note3}. The order parameter
$\Real{\mathcal{Z}_3^{[\hat{\sigma}]}(t)}$ decays from the constant value of the plateau to $0$ after a time $t^*$ which increases with the system size. 
For increasing values of $h_x$, the time-crystal behavior is destroyed and the plateau disappears: we can see an instance of that in Fig. \eqref{fig.n3tc.Zt}-(b). 

\begin{figure*}
\includegraphics[width=\textwidth]{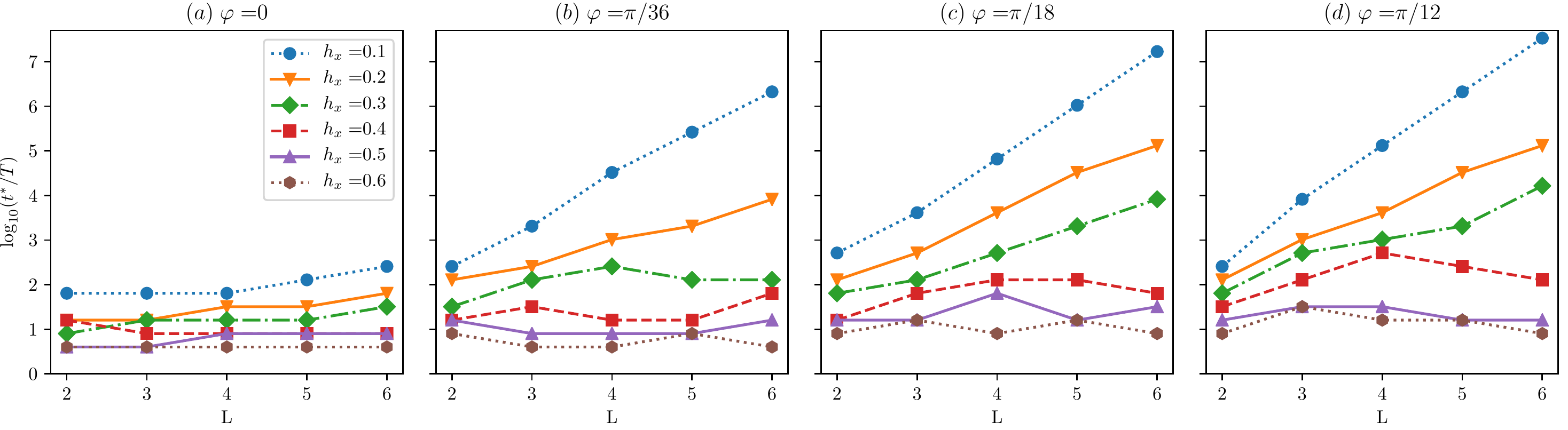}
    		\caption{Dependence of $t^*$ as defined in the text on the size of the system for different values of $h_x$ and $\varphi$. In the chiral case 
    		$t^*$ grows exponentially for sufficiently small $h_x$, and becomes independent on the size for large $h_x$. As we get closer to the non-chiral case, the 
    		dependence gets flatter and flatter until at $\varphi=0$ we do not see the exponential growth for any value of $h_x$.} 
\label{tdecay}
\end{figure*}

In  Fig.~\ref{fig.n3tc.Zt}-(c) and Fig.~\ref{fig.n3tc.Zt}-(d) we consider the effect of the chirality parameter $\varphi$. We show the time dependence of 
$\mathcal{Z}_3^{[\hat{\sigma}]}(t)$ for different values of $\varphi$. When $\varphi$ is close to the non-chiral case oscillations are less stable. We compare the 
case $\varphi=\pi/36$ [Fig.~\ref{fig.n3tc.Zt}-(c)] and $\varphi=\pi/6$ [Fig.~\ref{fig.n3tc.Zt}-(a)] for the same value of $h_x$: we see that the exponential increase 
of $t^*$ with the size $L$ is slower. As predicted, when $\varphi=0$ and the solvable Hamiltonian is degenerate, no time-crystal was observed, even for small 
values of $h_x$, [see Fig.~\ref{fig.n3tc.Zt}-(d)]. Here we have a numerical confirmation of the role of degeneracies is in making time-translation 
symmetry oscillations extremely fragile to perturbations.

A more accurate analysis, where we estimate $t^*$ as the time at which $\Real{\mathcal{Z}_3^{[\hat{\sigma}]}(t)}$ reaches 0.5, indicates that $t^*$ 
exponentially increases with the system size  when $\varphi \neq 0$ (see panels (b), (c) and (d) of Fig.~\ref{tdecay}). In the thermodynamic limit $t^*\to\infty$ and 
the period-tripling oscillations are persistent: the system is a time crystal as we predicted in Section \ref{nodeg:sec}. As we can see in panel (a) of Fig.~\ref{tdecay}, 
no exponential growth is found in the non-chiral case: $t^*$ is essentially independent on the size of the system, thus no time crystal in the thermodynamic 
limit. Based on these results, we can infer that the critical value of $h_x$ that represents the transition to a normal phase gets smaller and tends to 0 as $\varphi$
approaches the non-chiral value $\varphi=0$. We will  confirm this picture by studying the spectral-triplet properties and mapping a full phase diagram in 
$h_x$-$\varphi$ plane.

As we discussed in Section~\ref{witness:sec}, the presence of triplets in the spectrum with $2\pi/3$ quasi-energy splitting is necessary in order to have 
a period-tripling behavior. We expect to  see finite-size corrections to the splitting of the order $O(e^{-cL})$, as we have discussed in Sec.~\ref{nodeg:sec}.
In order to probe spectral triplets we study the quantities
\begin{equation}
 		\Delta^\alpha_0 = \mu_{\alpha+1}-\mu_\alpha
\end{equation} 
\begin{equation}
 		\Delta^\alpha = |\mu_{\alpha+\mathcal{N}}-\mu_\alpha-2\pi/3|
\end{equation} 
where the quasi-energies $\mu_\alpha$ are sorted from the  lowest to the greatest value in the first  Floquet Brillouin zone $[0,2\pi/T]$ and $\mathcal{N}=3^{L-1}$. 
Since the total number of  states is $3^L$, the quasi-energies $\mu_{\alpha+\mathcal{N}}$ and $\mu_\alpha$ are separated by one third of the levels of the spectrum. 
If the system is a time crystal, for a finite (but large) $L$ we expect  to find values of $\Delta^\alpha$ much smaller than the level 
spacing between two subsequent quasi-energies $\Delta_0^\alpha \sim 2\pi/3^L$.

\begin{figure}
\includegraphics[width=0.48\textwidth]{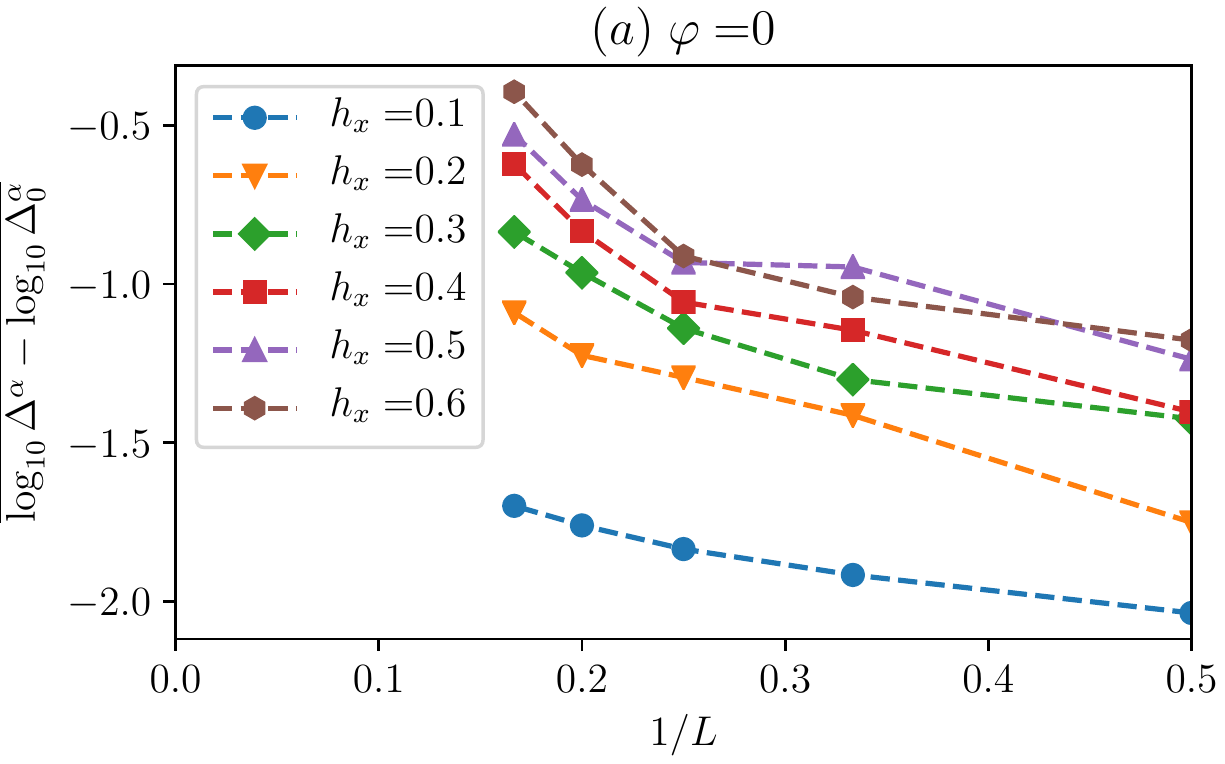}
\includegraphics[width=0.48\textwidth]{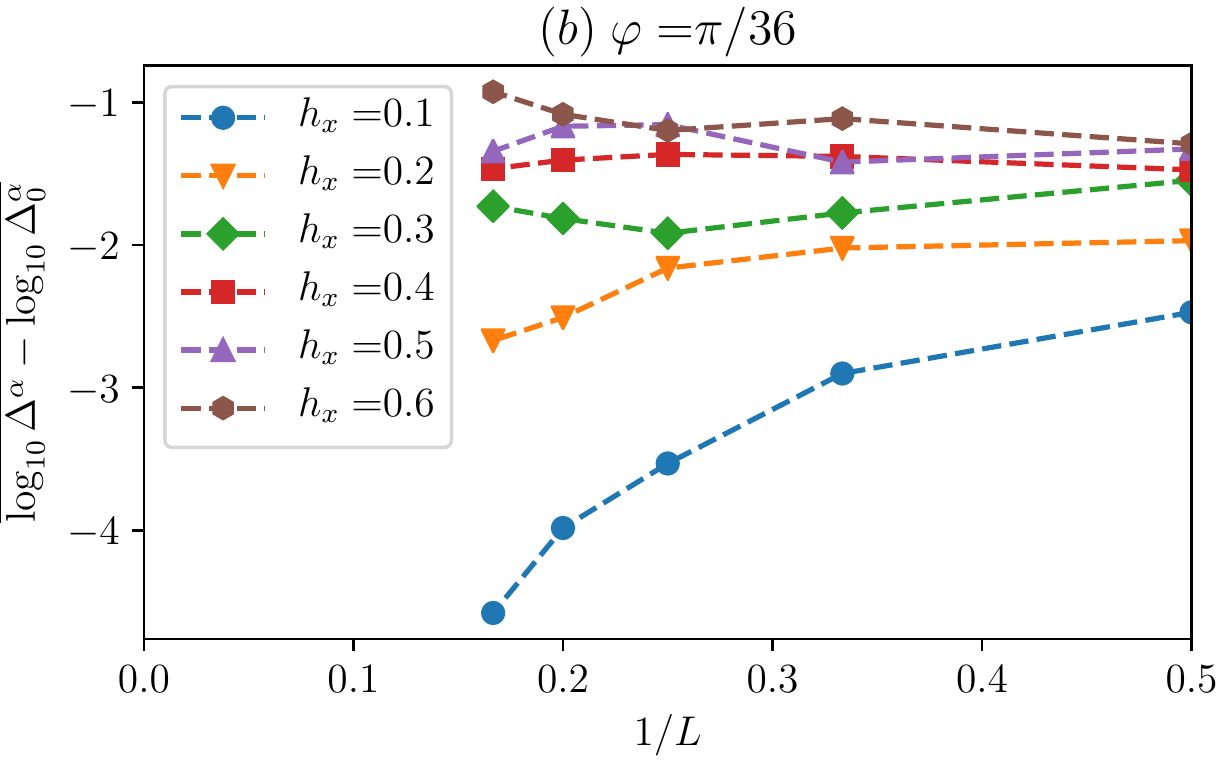}
   		\caption{Scaling of the spectral ratio $\overline{\log_{10}\Delta^\alpha-\log_{10}\Delta_0^\alpha}$ with the system size,
   		for different values of the chirality parameter $\varphi$.}
\label{DN}
\end{figure}

In Fig.~\ref{DN} we plot the dependence  of $\log_{10}\Delta^\alpha-\log_{10}\Delta_0^\alpha$ as a function of $1/L$. The quantity is averaged over all 
the Floquet quasi-energies $1 \leq \alpha \leq 3 \mathcal N$  and over different disorder  configurations. When the  parameters $\varphi$ and $h_x$ are chosen 
such that the system is a  time crystal, we expect to find by extrapolation  that $\overline{\log_{10}\Delta^\alpha}-\overline{\log_{10}\Delta_0^\alpha} \rightarrow -\infty$ 
in the thermodynamic limit. On the contrary, for a generic spectrum with Poisson statistics (but no $2\pi/3$ triplets) this quantity should diverge with increasing $L$.
Fig.~\ref{DN}-(a) refers to the non-chiral model. The plot shows that, for  every value of $h_x$ in the range selected, 
$\overline{\log_{10}\Delta^\alpha-\log_{10}\Delta_0^\alpha}$ does not  converge to $0$ as we increase the system size. On the contrary, this quantity 
increases with $L$. This confirms the absence of a time-crystal for the non-chiral clock model.

For the chiral clock model with $\varphi=\pi/36$, the results shown  in Fig.~\ref{DN}-(b) are consistent with the presence of a time crystal 
phase for $h_x$ sufficiently small ($h_x\simeq 0.1\div 0.2$). A transition from the time crystal phase to a normal phase 
is suggested for larger values of $h_x$: $\overline{\log_{10}\Delta^\alpha-\log_{10}\Delta_0^\alpha}$ is expected 
to increase as $1/L$ goes to $0$ for $h_x\gtrsim 0.3$, and decrease  for $h_x \lesssim 0.1$. However, the small size of the systems that
can be analysed  is a serious constraint to the possibility to make precise predictions.

In order to systematically analyse the dependence of the spectral gaps on the strength of the perturbation $h_x$, we 
study the quantities  $\overline{\log_{10}\Delta^\alpha}$ and $\overline{\log_{10}\Delta_0^\alpha}$   as functions of $\log_{10} h_x$.

\begin{figure}[h]
\hspace{-0.3cm}
\includegraphics[width=85mm]{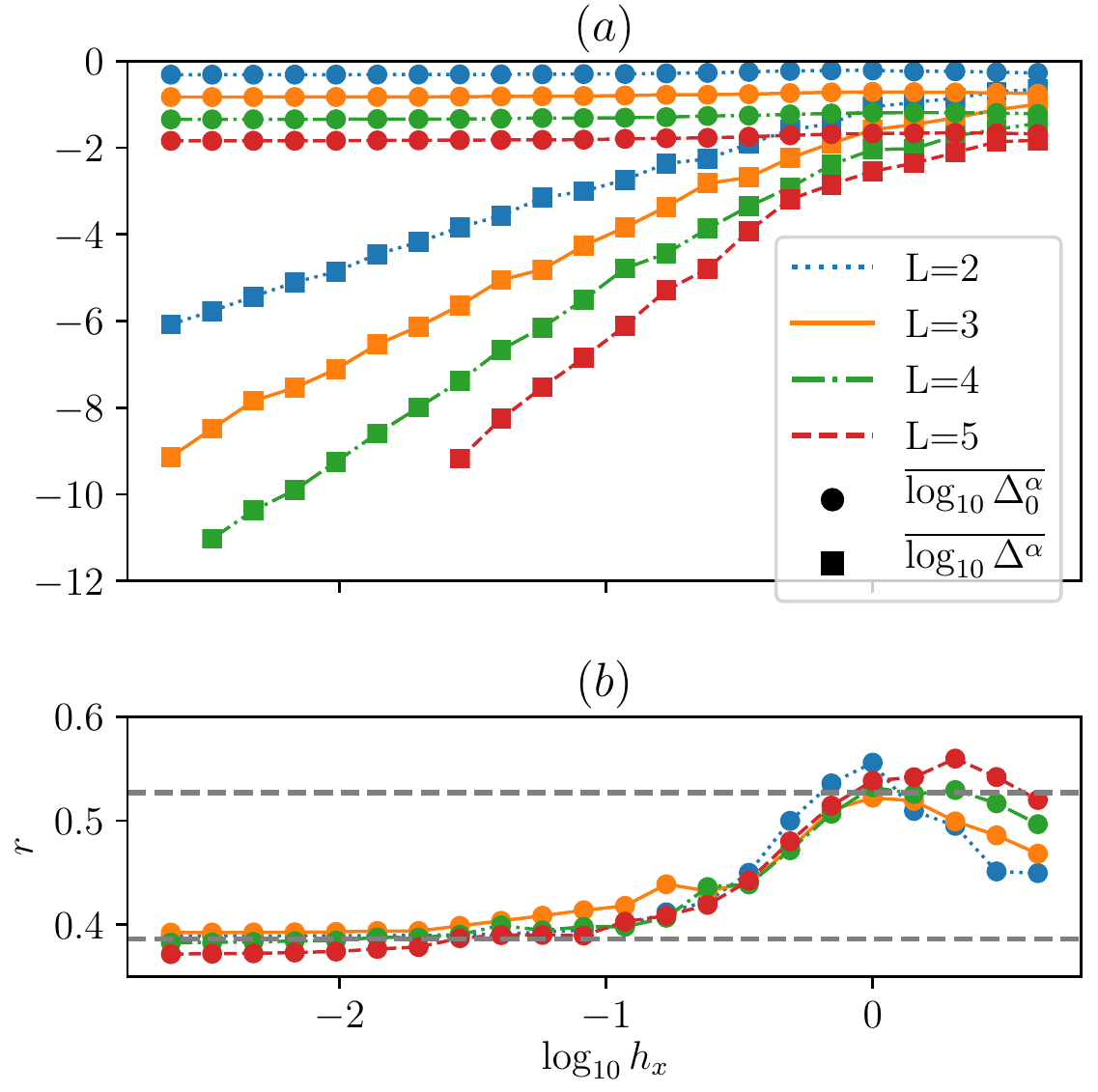}
 		\caption{(a) Averaged value of the logarithms of the spectral gaps $\overline{ \log_{10} \Delta_0^\alpha }$,  
 		$\overline{\log_{10} \Delta^\alpha}$ as a function of $\log_{10} h_x$. (b) quasi-energy level statistic 
 		ratio $r$ as a function of $\log_{10} h_x$. Dashed lines represent the value for Poisson statistics (0.386) and for Wigner-Dyson 
		(0.527). Data were obtained for the chiral case  $\varphi=\pi/18$.}
 \label{delta_r}
\end{figure}

\begin{figure}
 \includegraphics[width=85mm]{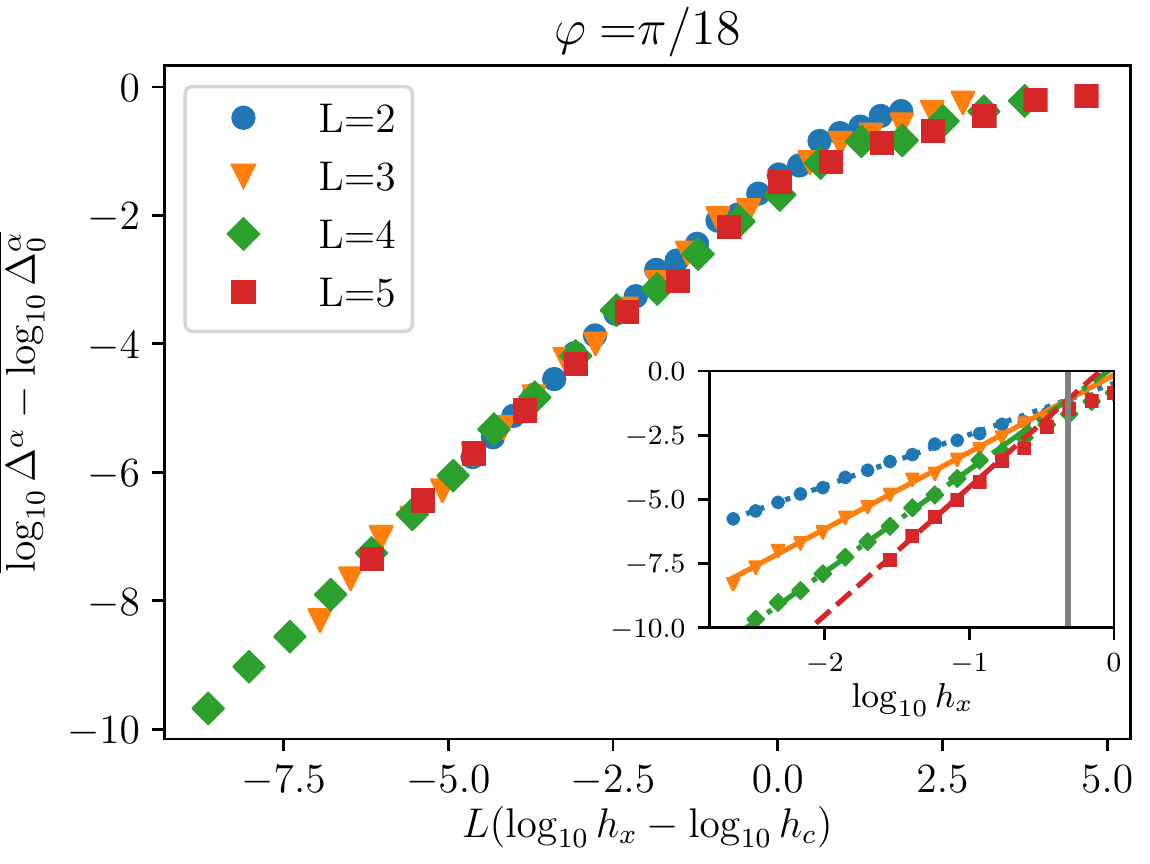}
 		\caption{Averaged values of $\overline{\log_{10} \Delta^\alpha} - \overline{\log_{10} \Delta_0^\alpha }$ as a function of $L(\log_{10} h_x-\log_{10} h_c)$, 
 		for different system sizes and $\varphi=\pi/18$. In the inset, the same quantity is plotted versus $\log_{10} h_x$: dashed lines are the result of the 
		fitting procedure in the region of small $h_x$. The vertical grey line corresponds to the critical value $h_c$.}
 \label{log_delta}
\end{figure}

In Fig.~\ref{delta_r}-(a) we consider a chiral case ($\varphi=\pi/18$). We first notice that $\overline{ \log_{10} \Delta_0 }$ does 
not depend on $h_x$, consistently with the fact  that $\Delta_0\sim 2\pi/3^L$ for every value of $h_x$.  On the opposite, $\overline{ \log_{10} \Delta }$ linearly 
increases with $\log h_x$ with an angular coefficient linear in $L$  up to a critical value $h_c$  (a clearer evidence of this fact will be given in Fig.~\ref{log_delta}).
These results are consistent with a dependence of the form $\Delta \propto (h_x)^L$ for $h_x$ much smaller than a critical value $h_c$. For large $h_x$ the 
triplets disappear and $\Delta$ will tend to a constant value.

The transition is also revealed by the quasi-energy average spectral ratio defined as
\green{
\begin{equation}  
		r=\overline{\left(\frac{\min(\delta_\alpha,\delta_{\alpha+1})}{\max(\delta_\alpha,\delta_{\alpha+1})}\right)}
\end{equation}
}%
with $\mu_\alpha$ in increasing order and $\delta_\alpha=\mu_{\alpha+1}-\mu_\alpha$. The average is performed over the whole spectrum and over disorder.
This quantity is a useful signature of the level statistics and can be used to discriminate ergodic  from many-body localised phases ~\cite{Oganesyan2006,
D_Alessio_2014,Ponte2014}.  For small $h_x$, $r$ is close to the value of $0.386$ expected for a Poisson statistics (Fig.~\ref{delta_r}-(b)). This is an 
evidence for many-body  localization, because it shows the absence of level repulsion. When $h_x$ approaches the critical value, significant deviations 
from the  Poisson limit can be observed, signaling a transition in the level statistics. Therefore the melting of the time crystal is accompanied by a transition 
of the dynamics towards an ergodic behaviour.

The non-chiral model, where there is no time-crystal, has substantially different spectral properties from the chiral model.
\begin{figure}[h]
 \includegraphics[width=85mm]{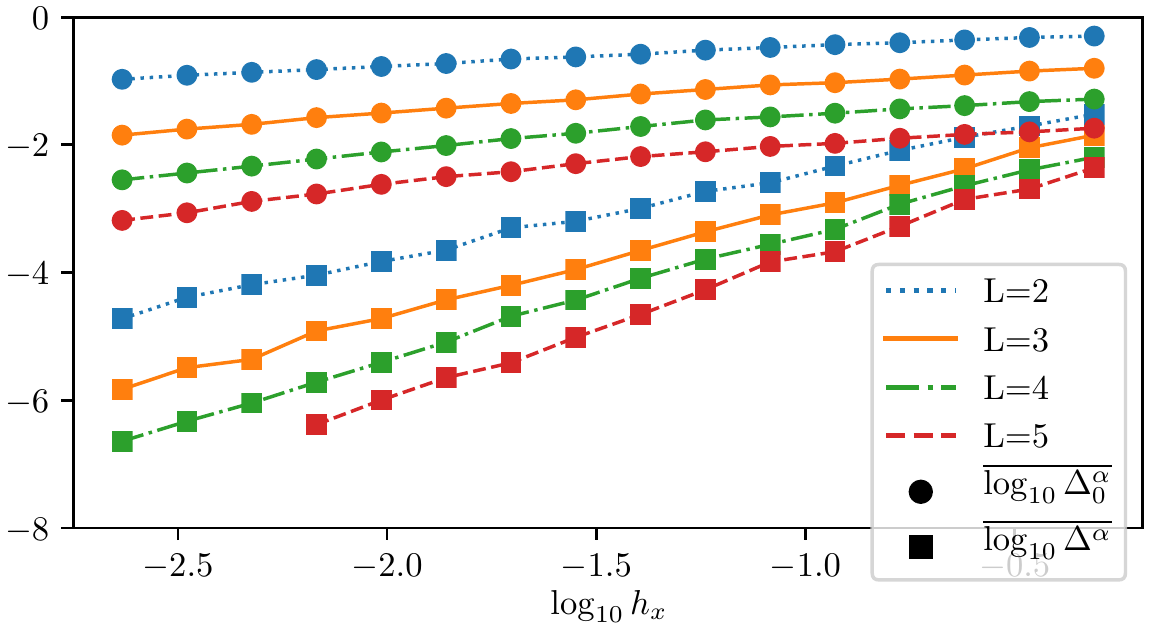}
 		\caption{Averaged value of the logarithms of the spectral gaps $\overline{ \log_{10} \Delta_0^\alpha}$,  $\overline{ \log_{10} \Delta^\alpha }$ as a 
 		function of $\log_{10} h_x$. Data were obtained for the non-chiral case $\varphi=0$.}
 \label{delta_r0}
\end{figure}
In Fig.~\ref{delta_r0} we show the dependence of $\overline{\log_{10} \Delta_0^\alpha }$ and  $\overline{ \log_{10} \Delta^\alpha }$ on $\log h_x$. A comparison with 
Fig.~\ref{delta_r} highlights some significant differences. The  gaps $\log \Delta^\alpha$ have a weaker dependence on $L$ than in the chiral 
case and the quantity $\overline{ \log_{10} \Delta_0^\alpha }$ is not constant with respect  to $\log h_x$. The dependence of the gap $\Delta_0$ between two 
consecutive  levels on $h_x$ is due to the fact that some eigenstates are degenerate in the absence of the perturbation: when $h_x\neq 0$ a gap that depends 
on the perturbation strength is opened between them.

A rough estimate of the critical value of $h_x$ can be obtained in the chiral case from the scaling $\Delta/\Delta_0 \propto (h_x/h_c)^L$. From an analysis of the 
plots, we can assume that this relation is valid when $h_x$ is much smaller than $h_c$. The data in the linear region (for small $h_x$) of Fig.~\ref{delta_r} are 
fitted with the expression \[\overline{\log \Delta} - \overline{ \log \Delta_0}= \log c+L\log h_x-L\log h_c\] with $\log c$ and $\log h_c$ as fitting parameters. 
In the inset of Fig.~\ref{log_delta} the dashed lines represent the linear relation derived  from the fit. From the fitting parameter $\log h_c$ (the grey vertical line
in Fig.~\ref{log_delta}) we obtain $h_c \simeq 0.48$. In Fig.~\ref{log_delta} we show the collapse of the curves in the inset when we  rescale the quantities 
with the system size: this confirms the validity of the scaling we assumed for $\Delta/\Delta_0$.

In order to further prove that the time-crystal phase disappears  in the non-chiral model, it is possible to use the same
fitting procedure to extrapolate an estimate of the critical value $h_c$ for different values of $\varphi$. We expect
that stability is lost in the proximity of the non-chiral case,  so $h_c\rightarrow 0$ as $\varphi$ approaches the value $\varphi=0$.
\begin{figure}
\includegraphics[width=0.5\textwidth]{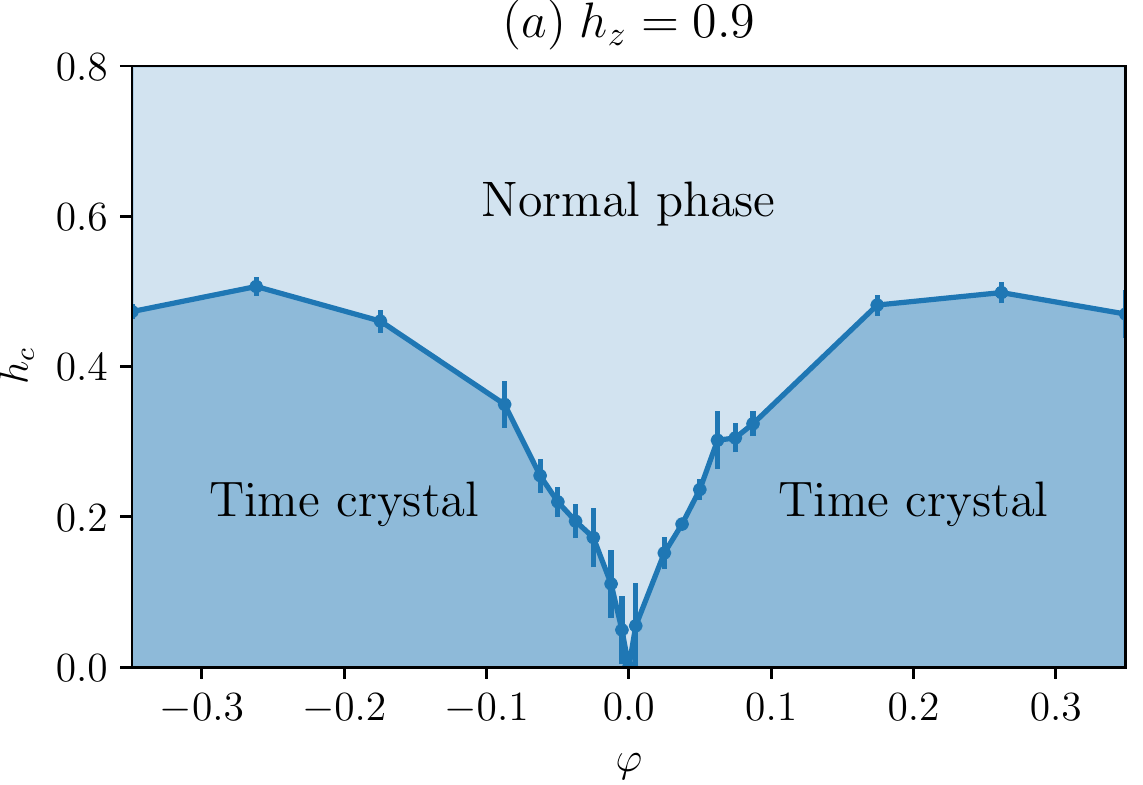}
\includegraphics[width=0.5\textwidth]{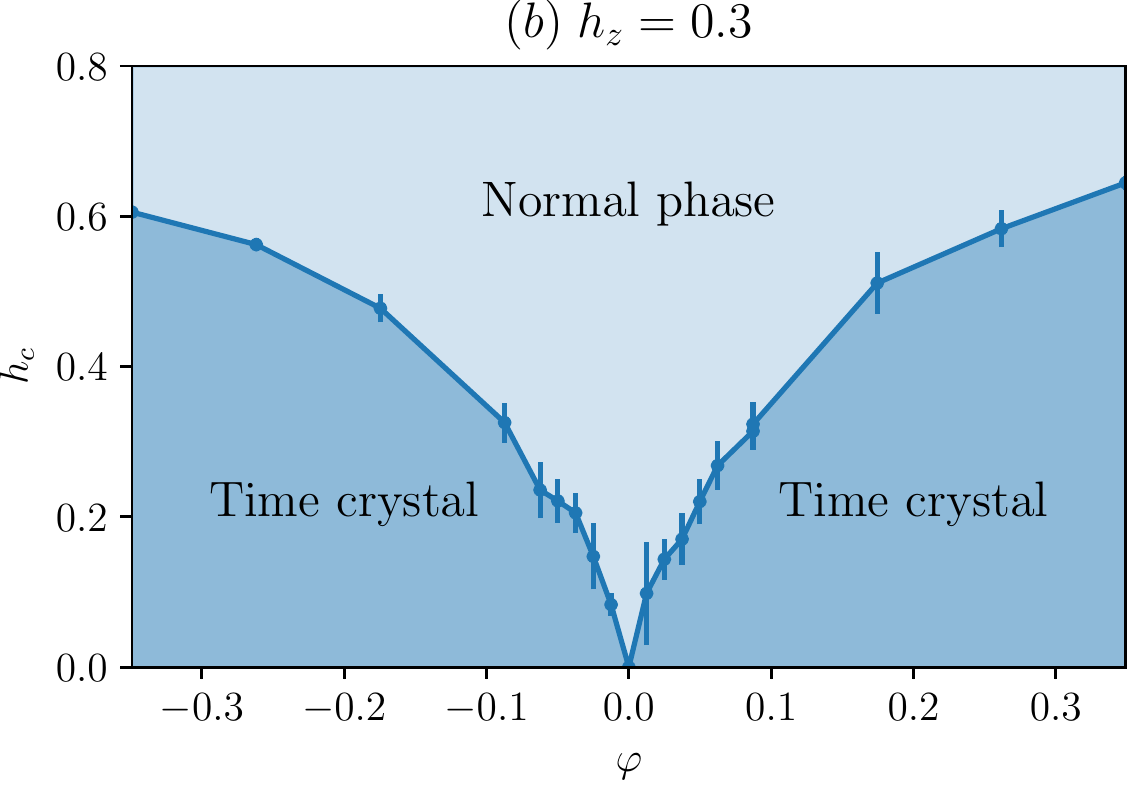}        
		\caption{The curve represents the critical value $h_c$ as a function of the chirality parameter $\varphi$. It corresponds 
		to the transition from the time crystal phase to a normal phase.}
\label{pd}
\end{figure}
An estimate of the critical value $h_c$ is derived as we vary $\varphi$ and it is shown in Fig. (\ref{pd}) for two different values of $h_z$. The curve 
that we get with this procedure represents the transition from the time crystal phase to a normal phase. Both plots confirm that the time crystal is 
less and less stable as $\varphi$ tends to 0. 

\begin{figure}
\includegraphics[width=.35\textwidth]{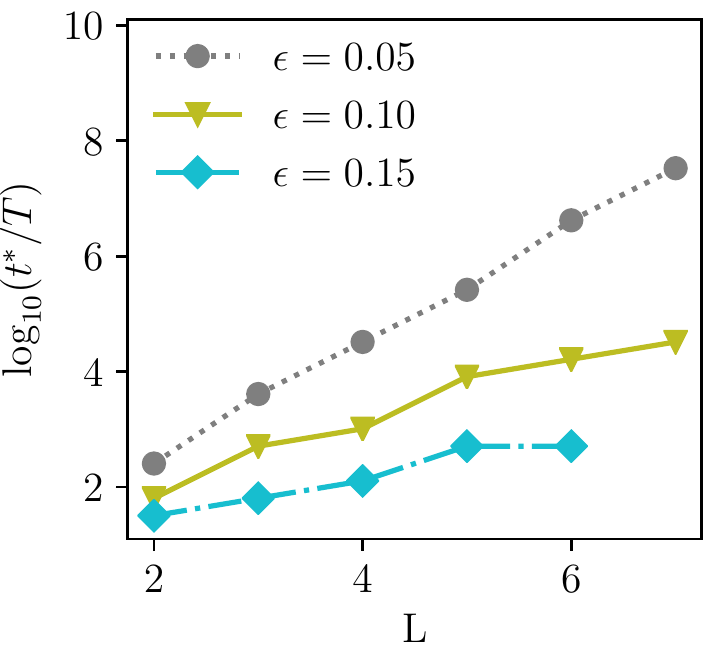}
		\caption{Dependence of $t^*$ (time before period-tripling oscillations decay) on the system size for distinct kicking 
		perturbations $\epsilon$. We consider here the chiral case $\varphi=\pi/6$. }
\label{pk}
\end{figure}

In the non-chiral case we  also checked the stability of the time crystal  to perturbations in the kicking (the case $\epsilon\neq 0$ in 
Eq.~\eqref{perturbato:eqn}) and there is no transverse field. Similarly to the case $h_x \neq 0$, numerical simulations show that oscillations of 
the order parameter decay after a time that grows exponentially in the system  size if the perturbation amplitude $\epsilon$ is sufficiently small 
(Fig.~\ref{pk}).  For larger values of $\epsilon$ oscillations decay much faster until time crystal behaviour is lost. 

\subsection{Phase diagram - $n=4$}\label{sec:disorder4}
The $n=4$ case is the minimal model where it is possible to investigate transitions between time-crystals of different periodicity. To this end we need to 
consider also terms with $m=2$ in Eq.\eqref{Hamiltonian-H} (see the corresponding entry in the Table~\ref{table-1}). The  Hamiltonian $\hat H$
is composed of different competing terms:  a term favouring states breaking spontaneously a $\mathbb{Z}_4$ symmetry, and another term favouring 
states breaking a lower $\mathbb{Z}_{2}$ symmetry, we write here explicitly for convenience the case $\delta=0$ (see Table~\ref{table-1}):
\begin{align}
\label{sr-eta}
 		\hat  H_{4,\eta}^{(SR)} &=\sum_i J_i \left[\hat  \sigma_i^{2} \hat \sigma_{i+1}^2+ \frac{(1-\eta)}{2}(e^{i\varphi}
		\hat \sigma_i^\dagger \hat \sigma_{i+1}+{\rm H.\,c.})\right]\nonumber\\
  		&+ \sum_i h_{z,i} \hat \sigma_i^2+\eta \sum_i h_{x,i} \hat \tau_i^2.
\end{align}
where $\eta$ parameterises the competing symmetry broken phases. The Floquet operator is of the type $\hat U_\eta(T) =  e^{-i T\hat H_\eta^{(SR)}} \hat X$, 
with $n=4$ clock variables and kick operator $\hat X$ as given by Eq.~\eqref{kickolo:eqn}. 
 
In the limit $\eta=0$, the Hamiltonian in Eq.\eqref{sr-eta} is of the type discussed in Section \ref{solvn} and is expected to support a time crystal with period $4$. 
On the other hand, for $\eta=1$ the operators $\hat \sigma_i^2$  and $\hat \tau_i^2$ commute among themselves and with $\hat H_\eta^{(SR)}$. Given the common 
eigenstates of these operators   $\ket{\{s_i, t_i\}}$, they satisfy $\hat \sigma_i^2\ket{\{s_i, t_i\}}= s_i\ket{\{s_i, t_i\}}$, $\hat \tau_i^2\ket{\{s_i, t_i\}}=t_i\ket{\{s_i, t_i\}}$.
 These states are eigenstates of ${\hat U_{\eta=1}(T)}^2$   with eigenvalue~\cite{Note5} $\left(\prod_i t_i \right)\exp\left[-2iTE(\{s_i, t_i\})\right]$
where we have defined $E(\{s_i, t_i\})=\sum_i J_i s_i s_{i+1} + \sum_i h_{x,i} t_i$. The Floquet states are
\begin{align}
  		&\ket{\psi(\{s_i, t_i\},\pm)}=\nonumber\\
  		&=\frac{1}{\sqrt{2}}\left(1\pm \left(\prod_i t_i \right)^{1/2} \nep^{iT\,E(\{s_i, t_i\})}\hat U_{\eta=1}\right) \ket{\{s_i, t_i\}}
\end{align}
and the corresponding quasi-energies $E(\{s_i, t_i\})+\frac{\pi}{2}\left(\prod_i t_i \right)\mp \frac{\pi}{2}$. Floquet states are indeed long-range correlated and 
there is $\pi$-spectral pairing. Moreover, due to disorder and the presence of the $\hat \tau^2$ term, the spectrum is not degenerate. Therefore we expect  to 
have a time crystal with period doubling.
 
 Let us consider now  the behaviour of the system for intermediate values of $\eta$. The Hamiltonian $\hat H_\eta^{(SR)}$ has the property
 that $\hat U_\eta^\dagger \hat \sigma^2_j \hat U_\eta=-\hat \sigma^2_j$ for every value of $\eta$. This suggests to take
$\mathcal{Z}_{2}^{[\sigma^2]}(t)$   [see Eq.~\eqref{super_zapp:eqn}] as the appropriate measure to study the robustness of the
 period-doubling oscillations since  $\hat U_\eta^\dagger \hat \sigma_j \hat U_\eta=i\hat \sigma_j$ only holds 
 for $\eta=0$.
 
In order to study a generic situation, we include a small perturbation $\hat V$ in the Floquet  operator  $ e^{-i(\hat H_\eta+\delta \hat V)} \hat X$ and study 
numerically  the robustness of oscillations for  different values of $\eta$.  We considered as perturbation
 $\hat V=\sum_i h_{z,i} (\hat \sigma_i +\hat \sigma_i^\dagger)+\sum_i h_{x,i} (\hat \tau_i+\hat \tau_i^\dagger)$. The reason for this choice is due to  
 $[ \hat \sigma^2_j, \hat V]\neq 0$,  so that the perturbation will affect the dynamics of $\hat \sigma^2_j$ in a non-trivial way. 
 
\begin{figure*}
\includegraphics[width=.33\textwidth]{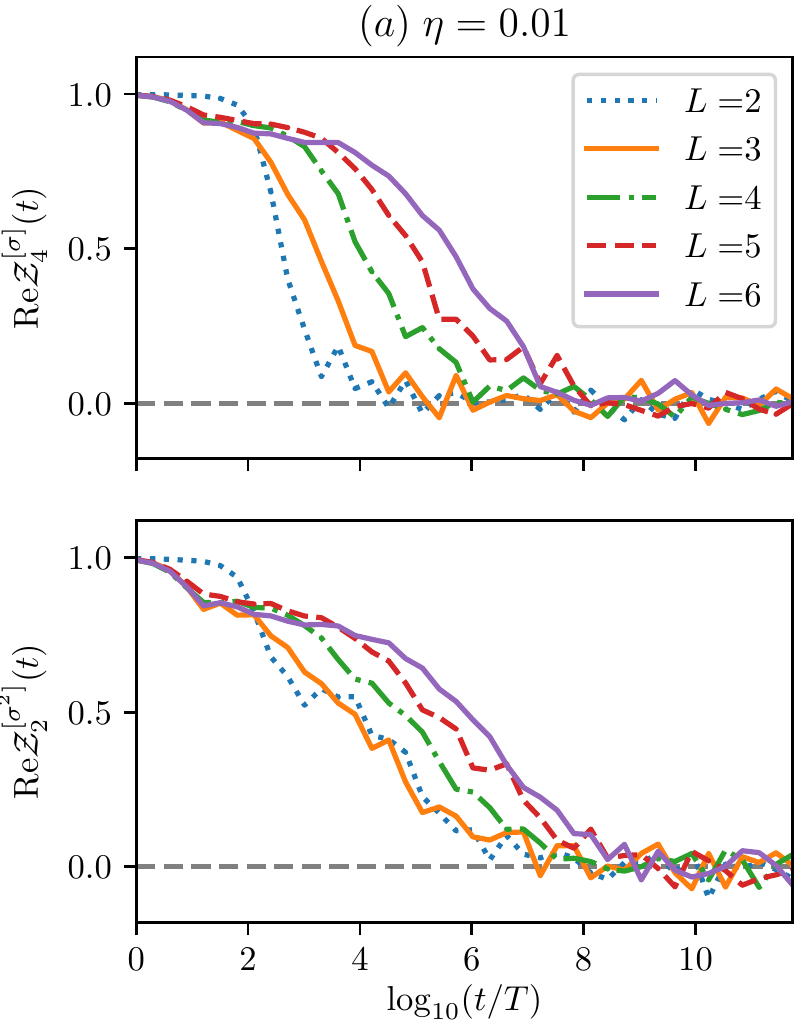}
\includegraphics[width=.33\textwidth]{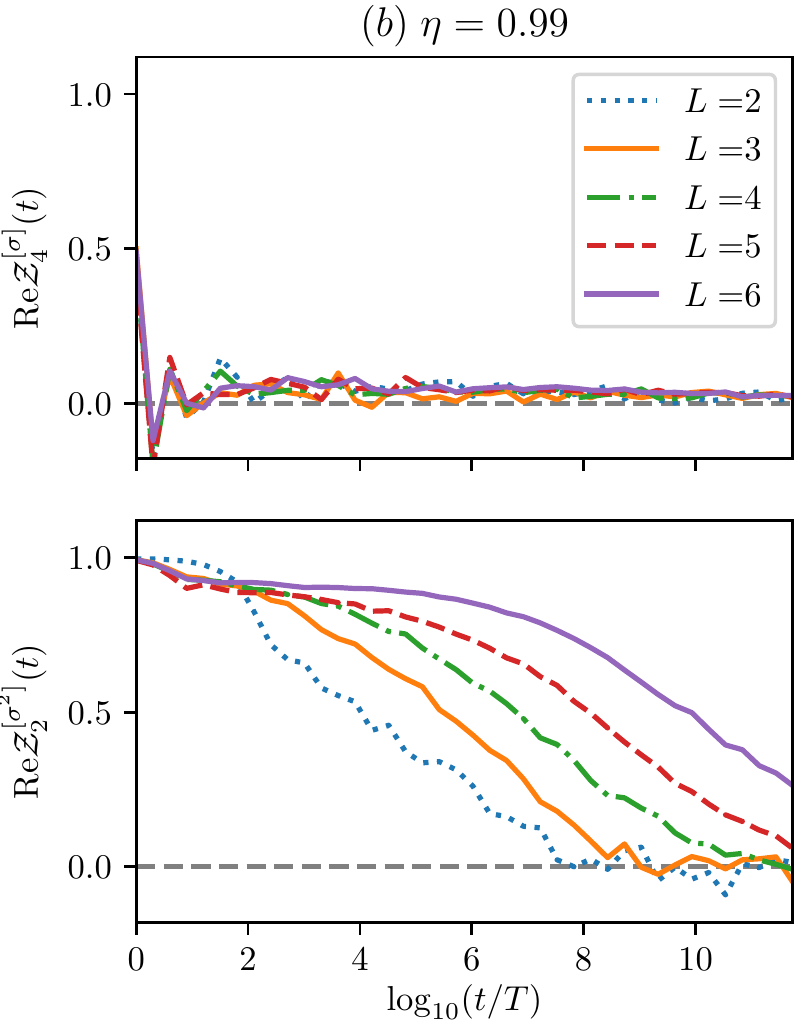}
\includegraphics[width=.31\textwidth]{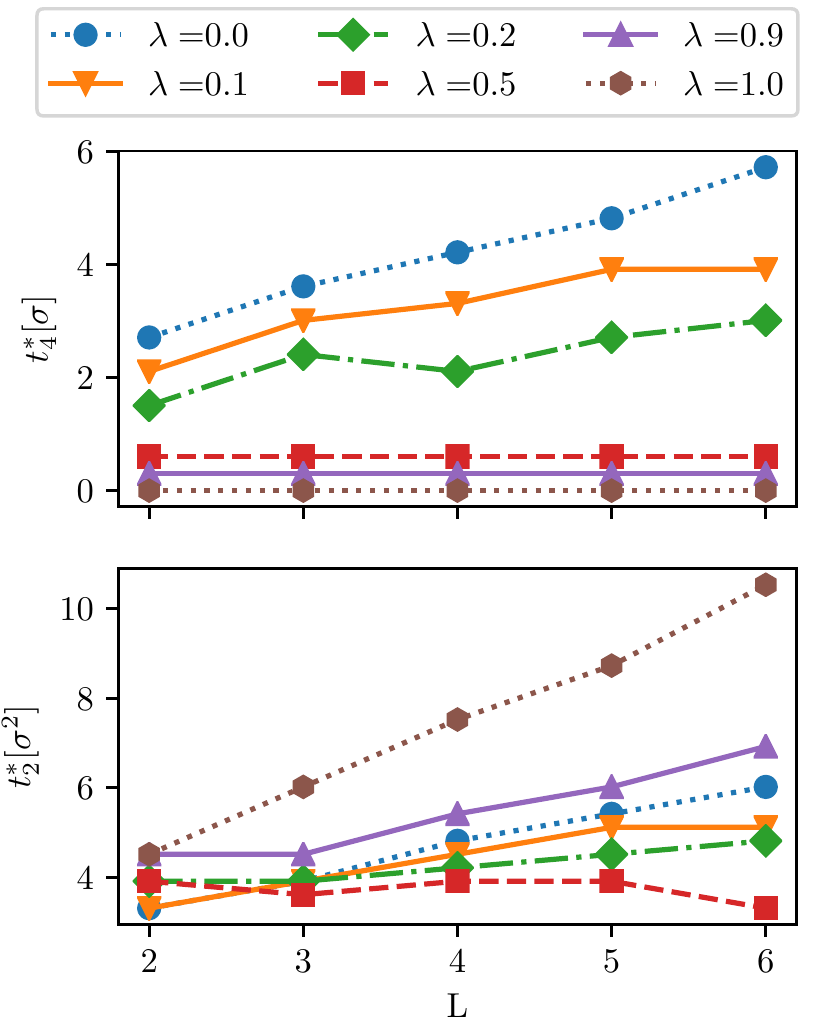}
    		\caption{(a), (b) Time evolution of the order parameters $\mathcal{Z}_4^{[\sigma]}(t)$ (period $4$ time crystal) and
     		$\mathcal{Z}_2^{[\sigma^2]}(t)$ (period doubling time crystal), for varying $\eta$ parameters. (c) The upper (lower) plot shows the time 
     		of the decay of period 4 (period 2) oscillations. Results are obtained with the following choice of parameters: $J_i$ from the uniform 
     		distribution $[1/2,3/2]$, $h_{z,i}$ from $[0,1]$, $h_{x,i}$ from $[0,1]$, $\varphi=\pi/3$, $\delta=0.1$}
\label{fig.z4toz2}
\end{figure*}

We show some of the results in Fig.~\ref{fig.z4toz2}. Additional data are discussed  in Appendix \ref{app:plot} [Fig.~\ref{fig.z4toz2complete}]. 
As expected, $\hat \sigma_j$ has  oscillations (with period $4$) only in a region close to $\eta=0$, while for $\hat \sigma_j^2$ we find stable 
oscillations (with period $2$) both  close to $\eta=0$ and $\eta=1$. A period 4-tupling time crystal is found in a finite region 
 of parameter space around {$\eta=0$} [Fig.~\ref{fig.z4toz2}-(a)], while a period doubling time crystal is found close to $\eta=1$ [Fig.~\ref{fig.z4toz2}-(b)]. 
 Our numerical analysis does not allow to draw reliable conclusions at intermediate values of $\eta$, because of the small system sizes. 
Although the model could in principle support a direct transition between period-doubling and period 4-tupling, it seems that in the short-range case, defined 
by Eq.(\ref{sr-eta}), the two phases appear to be probably separated by an intermediate normal region. In the next Section we will show that the situation 
is dramatically different in the long-range case where a direct transition between the two time-crystal phases is indeed found.

\section{Infinite-range model} 
\label{time_infty:sec}

We now turn  to the analysis of the Floquet dynamics with the infinite-range version of the Hamiltonian in Eq.~\eqref{Hamiltonian-H} and denote it as
\green{$\hat H_{n}^{(LR)}$}. Here, the physical origin of the time crystal \green{with period $qT$} lies in the existence of a phase of $\hat H_{n}^{(LR)}$ where a \green{$\mathbb{Z}_n$ 
symmetry of the Hamiltonian is broken to a lower symmetry $\mathbb{Z}_{n/q}$ (if $n/q>1$ is an integer) or fully broken (if $q=n$)} by an extensive amount of energy eigenstates. On initialising the 
system in one of the symmetry breaking manifolds, the state is brought cyclically between those manifolds even if the kick is not perfectly swapping. 
Consequently the order parameter of the symmetry breaking cycles among $q$ values. This mechanism was behind the 
time crystal with $q=n=2$ considered in Ref.~\cite{Russomanno2017} and applies also to the more general cases we discuss here.

The analysis of the infinite-range case will proceed as follows. In Section~\ref{boson_you:sec} we discuss how to use the permutation symmetry 
of the Hamiltonian to restrict to the even symmetry sector and -- in that sector -- map the Hamiltonian to a $n$-site bosonic model. 
The $\mathbb{Z}_n$ symmetry is mapped to a discrete translation symmetry of the boson model. Details of this mapping will be presented  in 
Appendix~\ref{app_bosonazzi:sec}. A detailed analysis of spontaneous symmetry breaking occurring in \green{$\hat H_{n}^{(LR)}$} is reported in  
Appendix~\ref{subsec.SSB.LR}.  Here we focus on the time-crystal behaviour. In Section~\ref{nettre:sec}  we analyse specifically  the cases 
with $n=3$ and $4$. As in the previous cases  the time crystal is detected by analysing the peak in the Fourier spectrum of the order 
parameter  at the characteristic q-tupling frequency (see Eq.~\eqref{trasformazzi:eqn} and the related discussion). Because we restrict to the even 
symmetry sector, we can study quite large system sizes and perform a finite-size scaling of the height of the peak and of its position showing that there 
is a time crystal in cases where the interaction Hamiltonian  shows symmetry breaking. In the same section we report on a direct transition between different 
time-crystal phases, by varying the $\eta$ parameter in the Hamiltonian (see Table~\ref{table-1}). More specifically, we study the transition from a 
period-doubling to a period 4-tupling time crystal. 
In Section~\ref{semiclass:sec} we study the dynamics of the local observables of 
these models in the semiclassical limit. In this way, 
we can study the existence of 
the period $n$-tupling directly in the thermodynamic limit.
 
\subsection{Mapping to a bosonic Hamiltonian and the semiclassical limit} 
\label{boson_you:sec}
Due to the infinite-range nature of the interactions in the model Hamiltonian, and the form of the kicking term, the Floquet operator has a symmetry 
generated by the invariance under permutation of its subsystems.  We focus our analysis on the symmetric subspace, here  the Hamiltonian  can  be 
represented in terms of boson operators,  providing in this way a description of the system which is simpler and more manageable for numerical 
implementation.  The main idea is to associate to each position of the clock-variable a bosonic mode. 
More precisely, given a set of bosonic operators $\{\hat b_j\}$, satisfying the usual commutation relations,
\begin{equation}
\label{eq:bosons.commutation}
  		\left[ \hat b_\ell, \hat b^\dagger _{\ell '}\right] = \delta_{\ell,\ell '},\quad \left[ \hat b_{\ell}, \hat b_{\ell '} \right] = 0,
 \end{equation}
for $\ell=1,...,n$, the Hamiltonian operators are described in this bosonic representation as follows {(see Appendix~\ref{app_bosonazzi:sec} for details)}
\begin{eqnarray} 
\label{STERRO:eqn}
   	 	\sum_{i=1}^L\hat\sigma_i & = &  \sum_{\ell=1}^n \hat n_\ell \omega^{\ell-1} \;\;\;\;\;\;\;\;
  	 	\sum_{i=1}^L\hat\sigma_i^2  =    \sum_{\ell=1}^n \hat n_\ell \omega^{2(\ell-1)}  \\
  	 	\sum_{i=1}^L\hat\tau_i & = &  \sum_{\ell=1}^n \hat b_{\ell} \bd{\ell+1}  \;\;\;\;\;\;\;\;\;\;
  	 	\sum_{j=1}^L \hat \tau_j^2  =  \sum_{\ell=1}^n \hat b_{\ell} \bd{\ell+2}
\end{eqnarray}
where $\hat n_\ell = \hat b_\ell^\dagger \hat b_\ell$. In the bosonic variables the Hamiltonian in Eq.~\eqref{Hamiltonian-H} is represented 
as a closed chain of $n$ bosonic sites, with fixed number of $L$ bosonic particles.
\green{
Its explicit expression is
\begin{eqnarray}
 \label{bosonic_ham:eqn}
 \hat H_n^{(LR)}=-\frac{J}{L} \sum_{\ell, \ell '=1}^n \sum_{m=1}^{n-1} \alpha_m \omega^{m(\ell-\ell')} \hat n_\ell \hat n_{\ell '}+\nonumber\\
 -h \sum_{\ell=1}^n \sum_{m=1}^{n-1}\beta_m \hat b_\ell \bd{\ell+m}
\end{eqnarray}
(see also Table~\ref{table-1}).
}

It is important to emphasise that the $\mathbb{Z}_n$ symmetry breaking in the clock representation is mapped to the breaking of the invariance 
under translation of the sites in the bosonic representation. As an illustrative example, the states breaking the rotational symmetry $\hat X$  with fully 
aligned clock operators $|s\rangle^{\otimes L}$, for  $s = \omega^\ell$,  are represented in the bosonic language by states in which all $L$ bosons occupy 
a single site $\ell$ (``$|0_1... L_\ell ... 0_n \rangle$'' in number representation). From now on we will consider only the bosonic representation of 
this Hamiltonian.
  
In this representation the kicking operator corresponds to  a global translation in the sites of the chain. Indicating with $b'$ the bosonic operators after 
the kick,  the unperturbed kicking, Eq.~\eqref{kickolo:eqn}, reads
\begin{equation} \label{eq:kicko31}
 		\left(\begin{array}{c}\hat{b}_1'\\\hat{b}_2'\\\vdots\\\hat{b}_n'\end{array}\right)=
 		\hat \tau \left(\begin{array}{c}\hat{b}_1\\\hat{b}_2\\\vdots\\\hat{b}_n\end{array}\right),\,
		\vspace*{0.3cm}
\end{equation}
where $\hat \tau$ is the $n\times n$ matrix defined in Eq.~\eqref{tauvolata:eqn}. In other words, the kicking corresponds to a global translation by a 
single site ($\ell\rightarrow \ell+1$) in  the bosonic chain.
In the  general case, the kicking acquires a more intricate form, 
\begin{equation} \label{eq:kicko3e}
 		\left(\begin{array}{c}\hat{b}_1'\\\hat{b}_2'\\\vdots\\\hat{b}_n'\end{array}\right)= \exp\left[{i\left(\frac{\pi}n+\epsilon\right)\hat{\Theta}^{(n)}}\right]\left(
 		\begin{array}{c}\hat{b}_1\\\hat{b}_2\\\vdots\\\hat{b}_n\end{array}\right)\,.
\end{equation}
where $\hat{\Theta}^{(n)}$ is the $n\times n$ matrix  defined in Eq.~\eqref{perturbanti:eqn}.

The limit of $L\to\infty$ is equivalent to the limit where the bosonic  modes are macroscopically occupied 
and the dynamics is described by a semiclassical equation like the Gross-Pitaevski one. In this limit  we can show that the dynamics of the 
bosonic model is governed by a classical effective  Hamiltonian, generalising the analysis done for the Bose-Hubbard dimer reported in~\cite{smerza}.
To this aim we use the transformation $\hat{b}_\ell=\sqrt{L\,\hat{p}_{\,\ell}/n}\,\nep^{i\hat{\phi}_\ell}$ where, in order to preserve the bosonic 
commutation relations, we have to assume $[\hat{\phi}_\ell,\,\hat{p}_{\,\ell'}]=in\delta_{\ell\,\ell'}/L$. In the limit $L\to\infty$ the commutators are 
vanishing and the dynamics is classical. It is induced by the effective Hamiltonian~\cite{Note7} ${\cal H}_{n}^{(LR)}$
\begin{eqnarray}
 \label{accadi:eqn}
 {\cal H}_n^{(LR)}=-\frac{J}{n} \sum_{\ell, \ell '=1}^n \sum_{m=1}^{n-1} \alpha_m \omega^{m(\ell-\ell')} p_\ell p_{\ell '}+\nonumber\\
 -h \sum_{\ell=1}^n \sum_{m=1}^{n-1}\beta_m \sqrt{p_\ell p_{\ell+m}}e^{i(\phi_\ell-\phi_{\ell+m})}
\end{eqnarray}
where the Poisson brackets between the canonical coordinates and momenta are $\{\phi_\ell,\,p_{\,\ell'}\}=\delta_{\ell\,\ell'}$, $\{\phi_\ell,\,\phi_{\ell'}\}=0$, 
$\{p_{\,\ell},\,p_{\,\ell'}\}=0$ .
 The Hamiltonian~\eqref{bosonic_ham:eqn} conserves the total number of bosons to the value $L$, this reflects in the classical 
Hamiltonian conserving the sum of the momenta to the value 1. This fact allows to restrict the dynamics to $n-1$ pairs of canonical coordinates and momenta. 

The kicking operator is described in the bosonic language by Eq.~\eqref{eq:kicko3e}. Using the relation $\hat{b}_\ell=\sqrt{L\hat{p}_{\,\ell}/n}\,\nep^{i\hat{\phi}_\ell}$ 
 this peaceful linear transformation becomes a strongly  non-linear object when expressed in terms of the variables $p_{\,\ell}$ and $\phi_{\ell}$.
In conclusion we can study if the model shows time-translation symmetry breaking in the thermodynamic limit  looking at the classical dynamics of an Hamiltonian 
system with $n-1$ degrees of freedom; we are going to perform this analysis first in the case  $n=3$ and $n=4$ with $\eta=0$
in the next subsection and then in the  case $n=4$ with $\eta \neq 0$, studying a transition between distinct time-crystal phases.

\begin{figure*}
\centering
\includegraphics[width=0.32\textwidth]{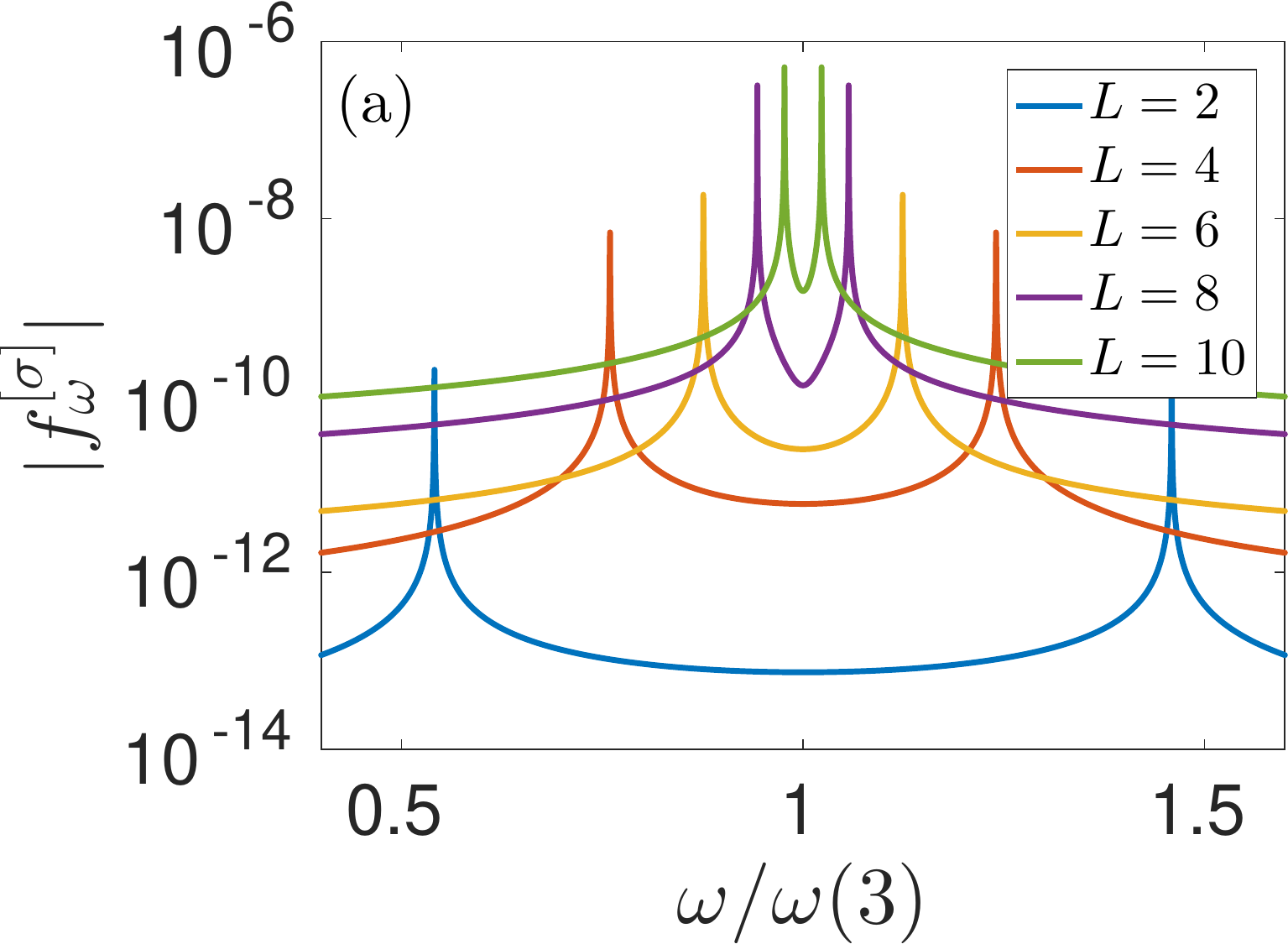}
\includegraphics[width=0.32\textwidth]{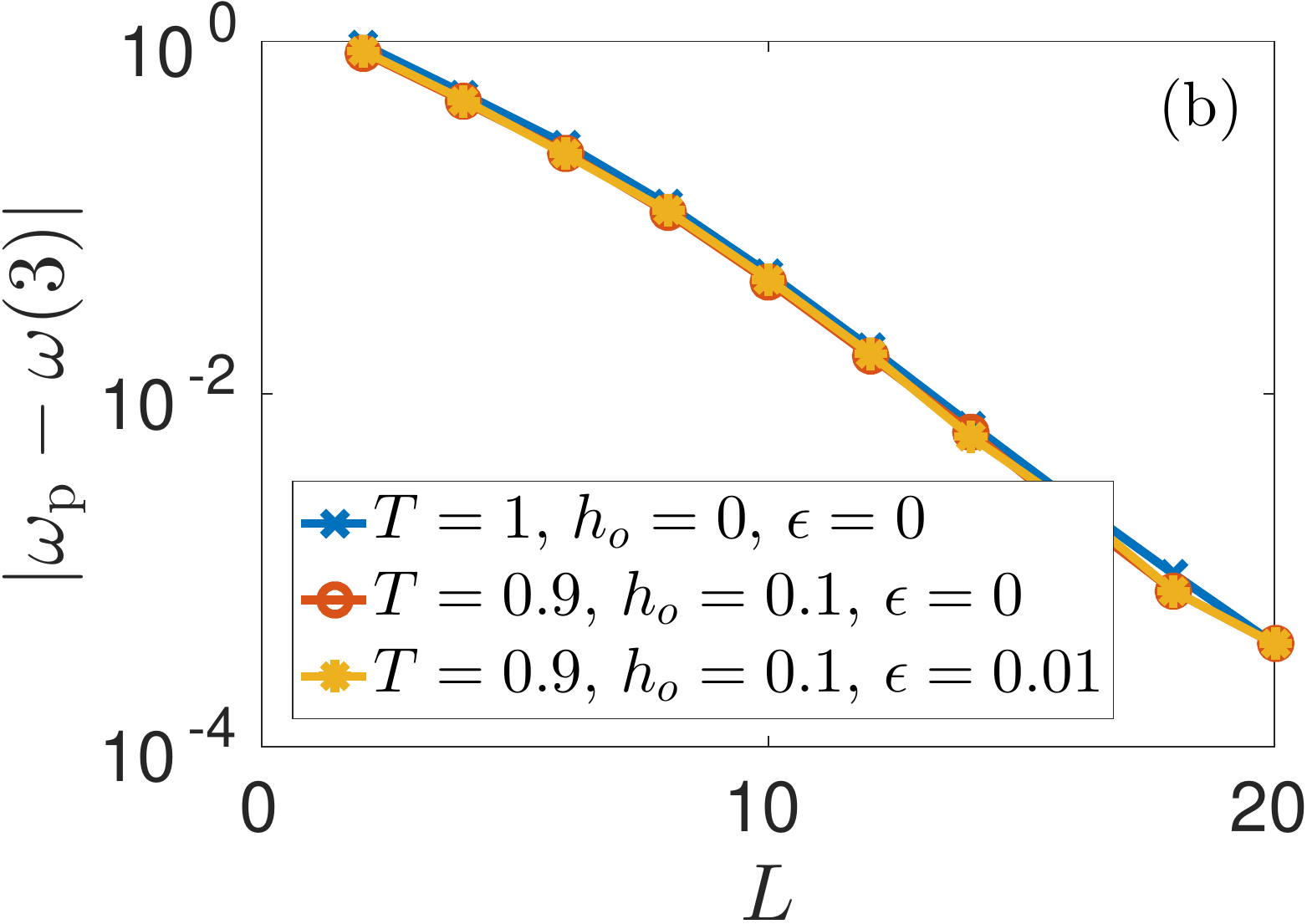}
\includegraphics[width=0.32\textwidth]{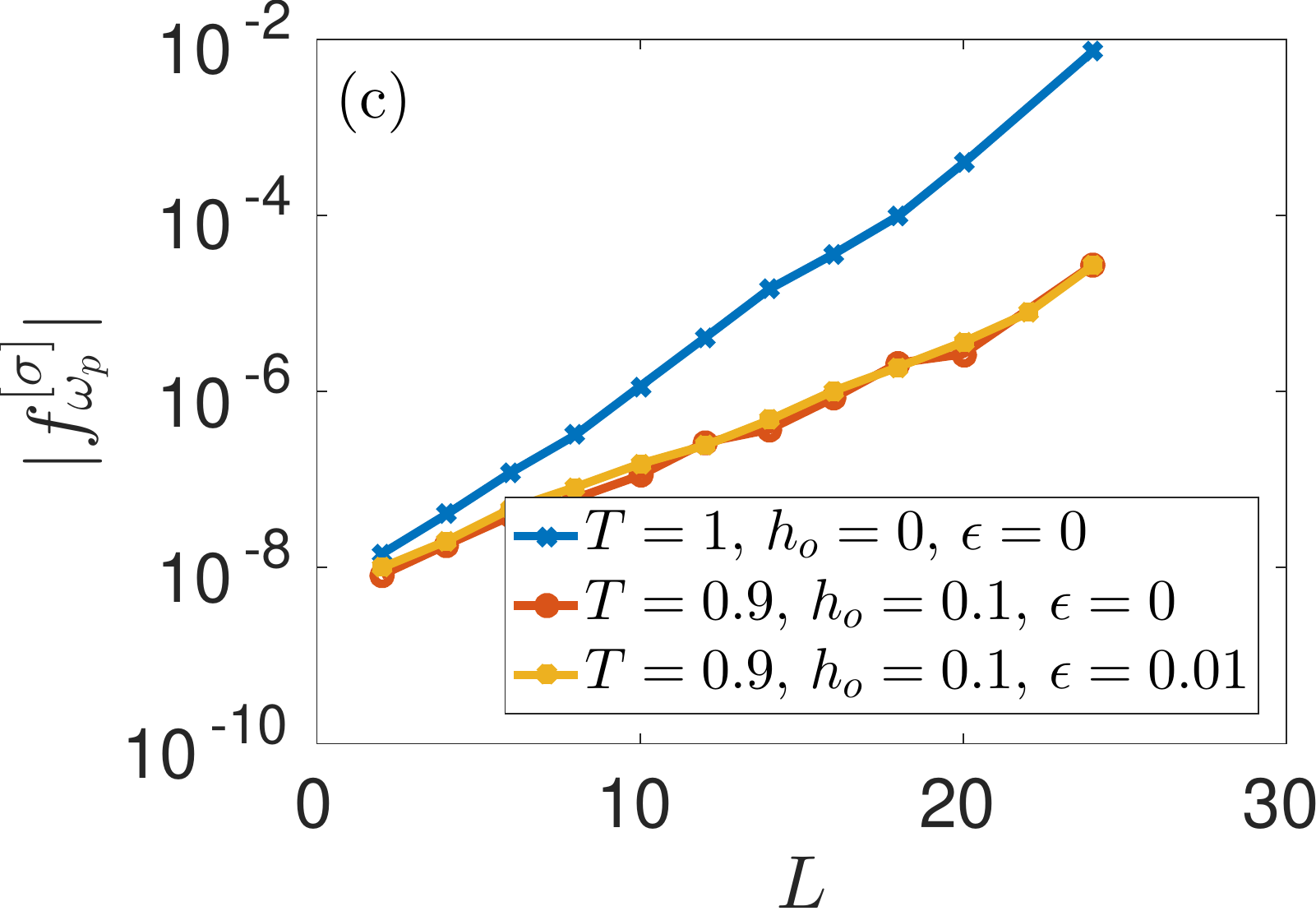}
 		\caption{ Time-crystal behavior for a system with $n=3$. 
 		(Numerical parameters: $J=1$, $h=0.5$ and $N_T=2^{15}$). (a) Fourier transform 
 		of the order parameter $\mean{\hat{\sigma}}_t$. The initial state here is
  		given by a symmetry-breaking ground state of the static Hamiltonian \green{$\hat H_{n=3}^{(LR)}$}, the period $T=1$ and 
  		there is no perturbation in the kicking operator ($\epsilon=0$).
 		(b-c) Finite-size scaling for the position of the dominant frequencies 
 		($\omega_p$) and the height of the corresponding peak ($f^{[\sigma]}_{\omega_p}$); different initial
		states and  different perturbations to the dynamics are considered.
		}
 \label{fig.tc3}
 \end{figure*}

\subsection{Time-crystal phases} \label{nettre:sec}

 \begin{figure*}
\centering
\includegraphics[width=0.32\textwidth]{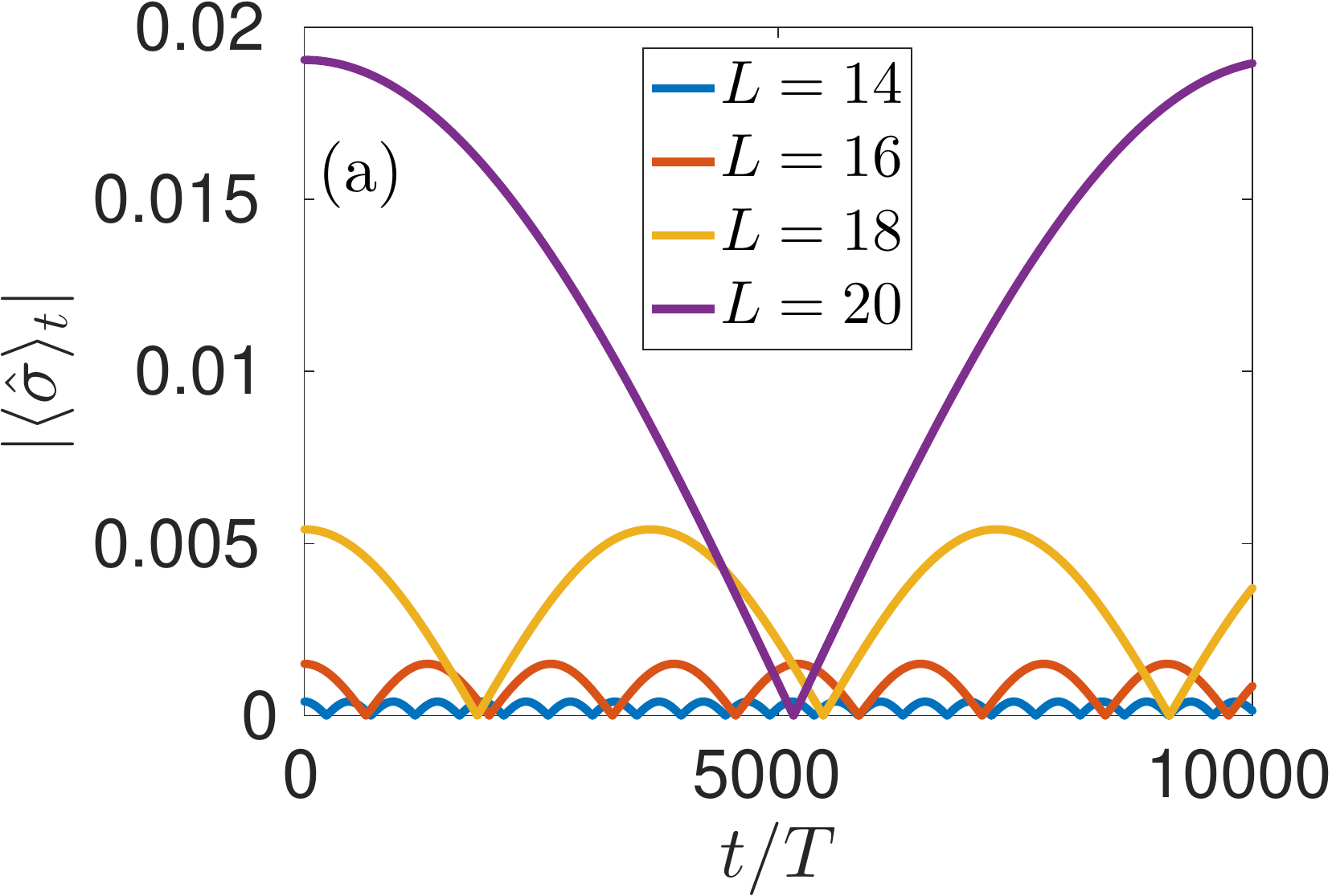}
\includegraphics[width=0.32\textwidth]{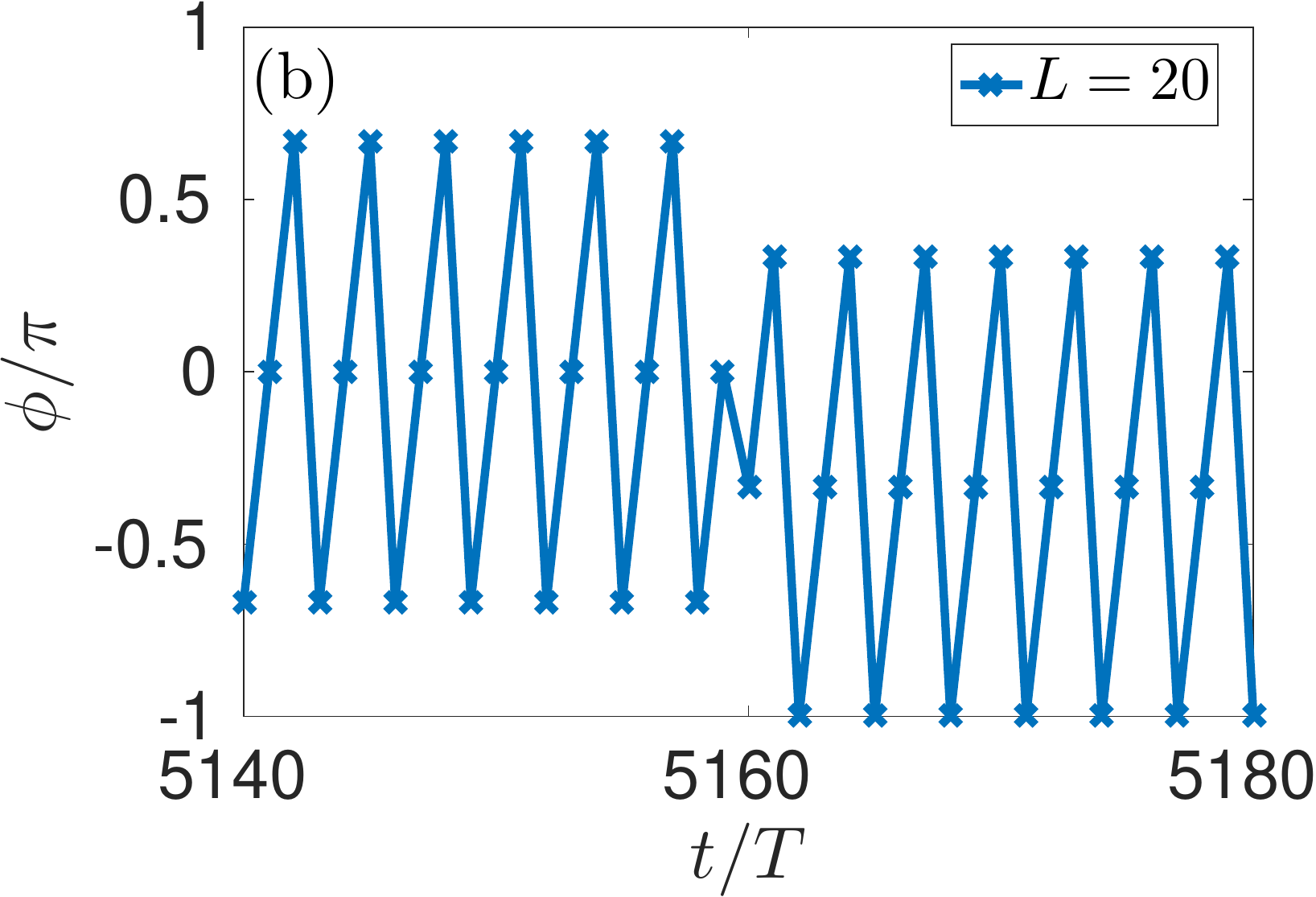}
\includegraphics[width=0.32\textwidth]{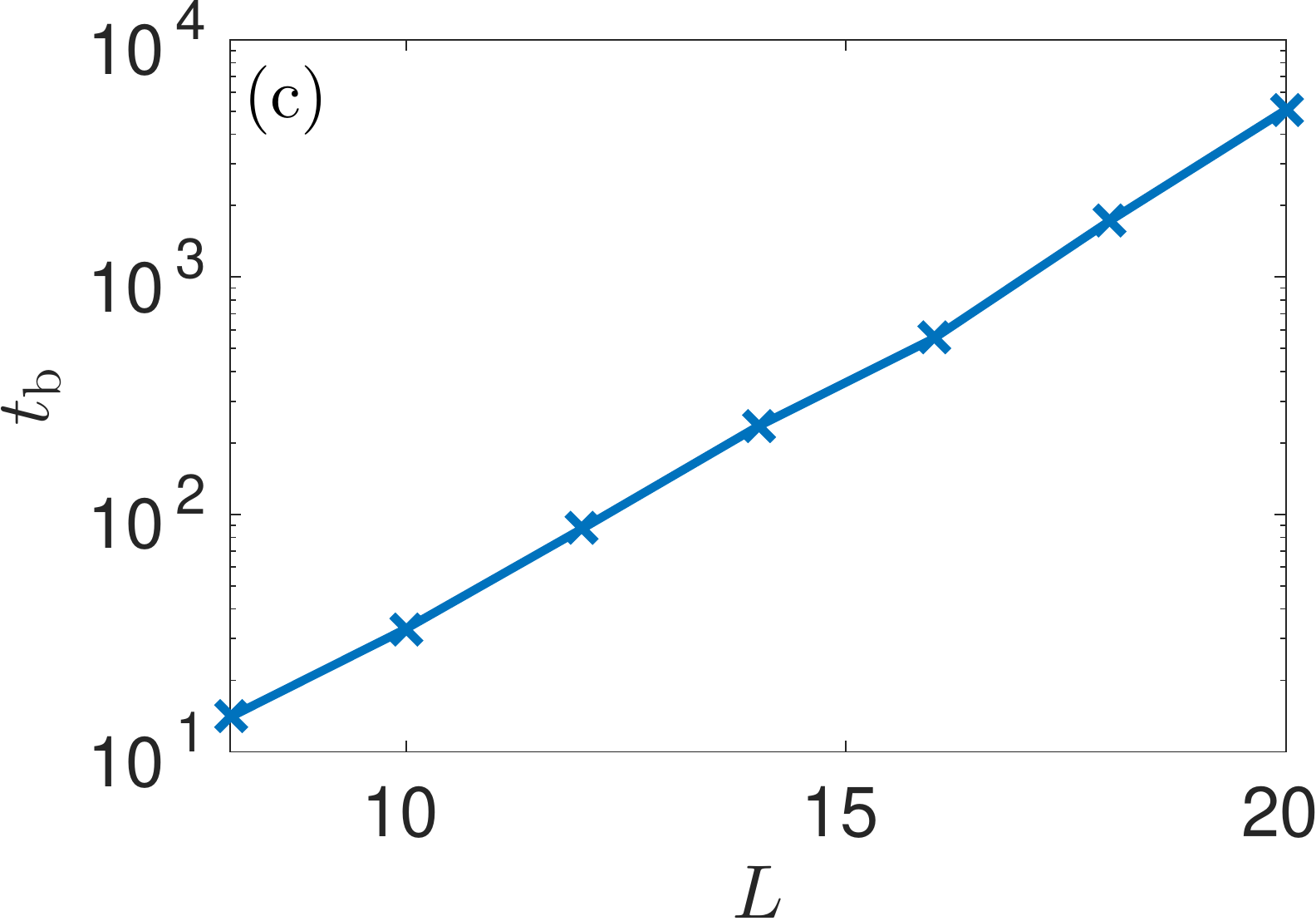}
 		\caption{ Time-evolution for the order parameter Eq.~\eqref{fernandeum:eqn} for a system with $n=3$, $J=1$, $h=0.5$ and $T=1$.
 		(a) Time-evolution for the absolute value $|\mean{\hat \sigma}_t|$ for different system sizes.
 		(b) Time-evolution for the phase $\phi(t)$ across the Rabi oscillation period $t_b$.
 		(c) Scaling of the Rabi-oscillation period $t_b$ with the system size.
 		}
\label{fig.tc3.Rabi}
\end{figure*}

We first focus on the analysis \green{of the cases $n=3$ and $n=4$ with $\eta=0$}, and study the existence of a discrete time crystal 
fully breaking the $\mathbb{Z}_n$ symmetry. Later on we will consider the case \green{$n=4$ with $\eta \neq 0$} which can show a transition between 
distinct time crystal orders. We consider the Floquet operator Eq.~\eqref{protocol:eqn} with $p=1$ and infinite-range interactions, expressed in 
the bosonic representation. In the rest of this Section we will study the dynamics of the sets of operators $\hat{\sigma}= L^{-1}\sum_i \hat{\sigma}_i$, 
and $\hat{\sigma}^2= L^{-1}\sum_i  \hat{\sigma}_i^2$. The expectation values of  $\hat{\sigma}_i$, $\hat{\sigma}^2_i$ are 
independent of the site-index $i$. They are therefore equivalent to the site-averages  which have a simple expression in terms of the bosonic 
operators, see Eq.~\eqref{STERRO:eqn}. 

 \subsubsection{$n=3$, $n=4$ with $\eta = 0$}

We considering the dynamics of $\hat\sigma$ we expect the system to pass cyclically between different symmetry-breaking subspaces, where the expectation of this 
operator is markedly different. As we have explained in Section~\ref{witness:sec}, we consider the expectation value at stroboscopic times 
$\mean{\hat{\sigma}}_t$ and perform its discrete Fourier transform over $N_T$ periods [see Eq.~\eqref{trasformazzi:eqn}].  

We start with a detailed numerical analysis in the case $n=3$. We first initialise the system in a symmetry-breaking ground state of the static 
Hamiltonian~\cite{Note4}. We first consider the perfect-swapping kick given in Eq.~\eqref{eq:kicko31}. In Fig.~\ref{fig.tc3}-(a) we plot the power spectrum 
$|f_\omega^{[\sigma]}|$ for a finite-size case  and see that, for a coupling $h/J$ smaller than the critical field value, there are two peaks that tend to  
$\omega(3)$ (Eq.\eqref{eq:omega_q}) as the system size is increased, Fig.~\ref{fig.tc3}-(b). The height of the corresponding peaks increases with the system size, 
see Fig.~\ref{fig.tc3}-(c). 
The system  breaks the discrete time-translation symmetry $\mathbb{Z}$ to $3 \mathbb{Z}$.
The height of the peaks is related to the initial state of the evolution and
 its expectation value for the order 
 parameter $\langle \hat \sigma \rangle_{t=0}$.
  For small system sizes the order parameter shows exponential corrections 
  with $L$  due to finite-size effects, while for larger system sizes it 
  scales polynomially to a finite value. We expect the peaks of the Fourier spectrum
   to behave in a similar way.
The separation between the two peaks is exponentially small in 
the system size [Fig.~\ref{fig.tc3}-(b)]. This gives rise to oscillations of period exponentially long in $L$, which appear in the Fourier 
spectrum as a splitting in two of the period-tripling peak. This behavior can be seen in Fig.~\ref{fig.tc3.Rabi}, where we show the time evolution of the order parameter 
\begin{equation} 
\label{fernandeum:eqn}
  		\langle \hat  \sigma \rangle_t = |\mean{\hat \sigma}_t|e^{i\phi(t)}\,.
\end{equation}
In Fig.~\ref{fig.tc3.Rabi}-(a) we show its absolute value, where we see a periodic behavior with period $t_b(L)$ related to the  oscillations. The phase $\phi(kT)$ of 
the order parameter shows period-tripling oscillations, as seen in Fig.\ref{fig.tc3.Rabi}-(b),  suffering a shift after every period $t_b(L)$.
In Fig.\ref{fig.tc3.Rabi}-(c) it is evident that the corresponding 
periods are exponentially large with the system size, and thus 
are effectively absent in the thermodynamic limit.
  
In order to verify that these period-tripling 
oscillations are not a fine-tuned behavior, we apply different perturbations to the dynamics, varying the period $T$, 
considering the perturbed kicking operator [Eq.~\eqref{eq:kicko3e} with $\epsilon\neq 0$], and considering different initial states (taken as symmetry-breaking 
ground states of \green{$\hat H_{n=3}^{(LR)} - h_o \sum_i ( \hat \tau_i + \hat \tau_i^\dagger)$}). In Figs.\eqref{fig.tc3}-(b,c) we see that the time-crystal behaviour is indeed robust 
to such perturbations, whenever $h/J$ is smaller than the critical field. We see therefore that there is a time-crystal behaviour and is intimately connected to the 
symmetry breaking of the interaction Hamiltonian.
 
We have also studied the case $n=4$, which \green{for $\eta=0$} shows essentially the same  behaviour as in the previous case $n=3$, but with period 4-tupling. 
The situation changes drastically when $\eta \ne 0$, in this case at $n=4$ a new dynamics phase transition  between period doubling and 
period 4-tupling appears. This will be the topic of the next subsection.

 \subsubsection{$n=4$, $\eta \neq 0$  - Transition between different time-crystal phases}  \label{TC_trans:sec}
 
\begin{figure*}
\includegraphics[width=0.38\textwidth]{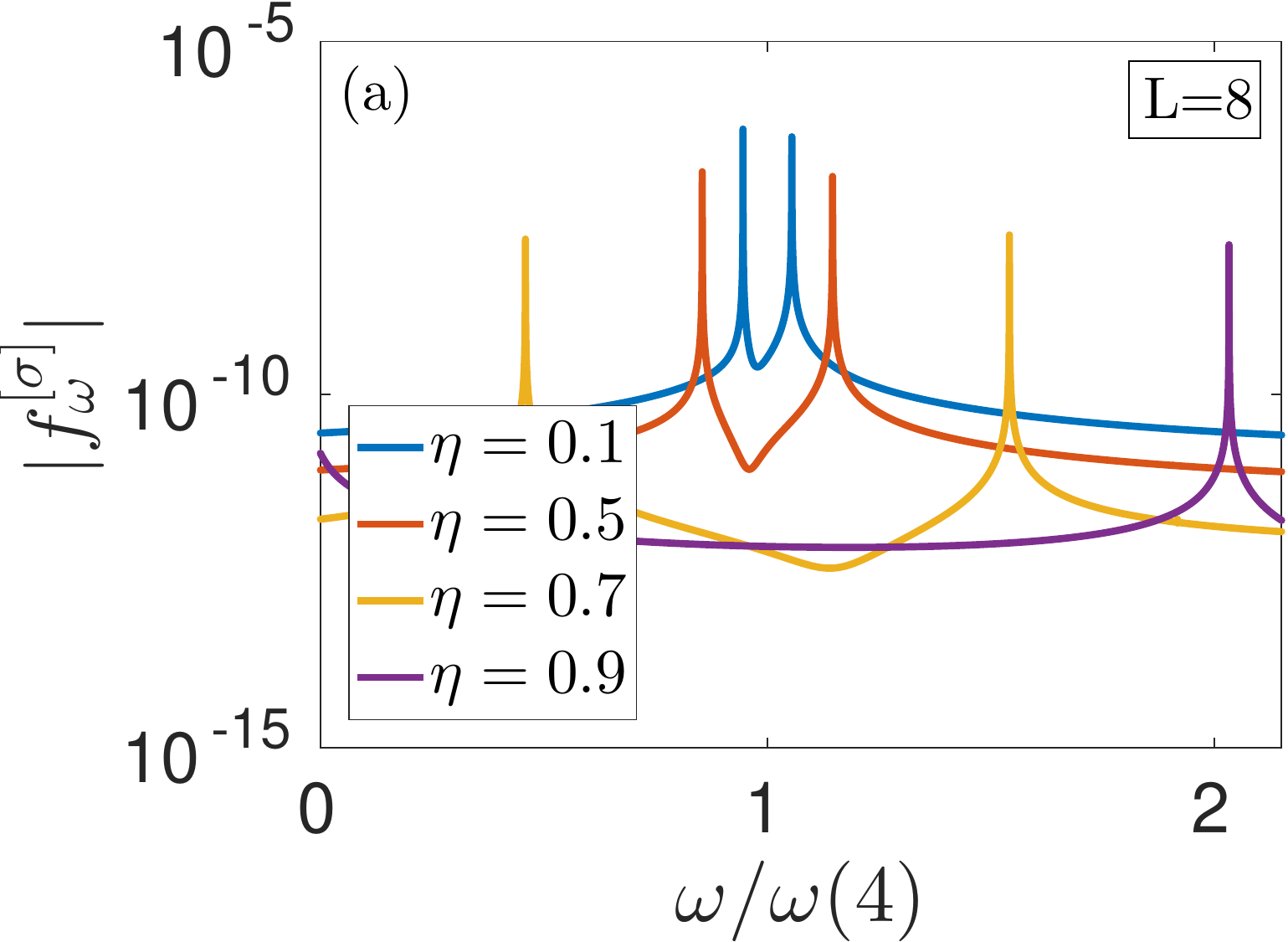}\quad 
\includegraphics[width=0.38\textwidth]{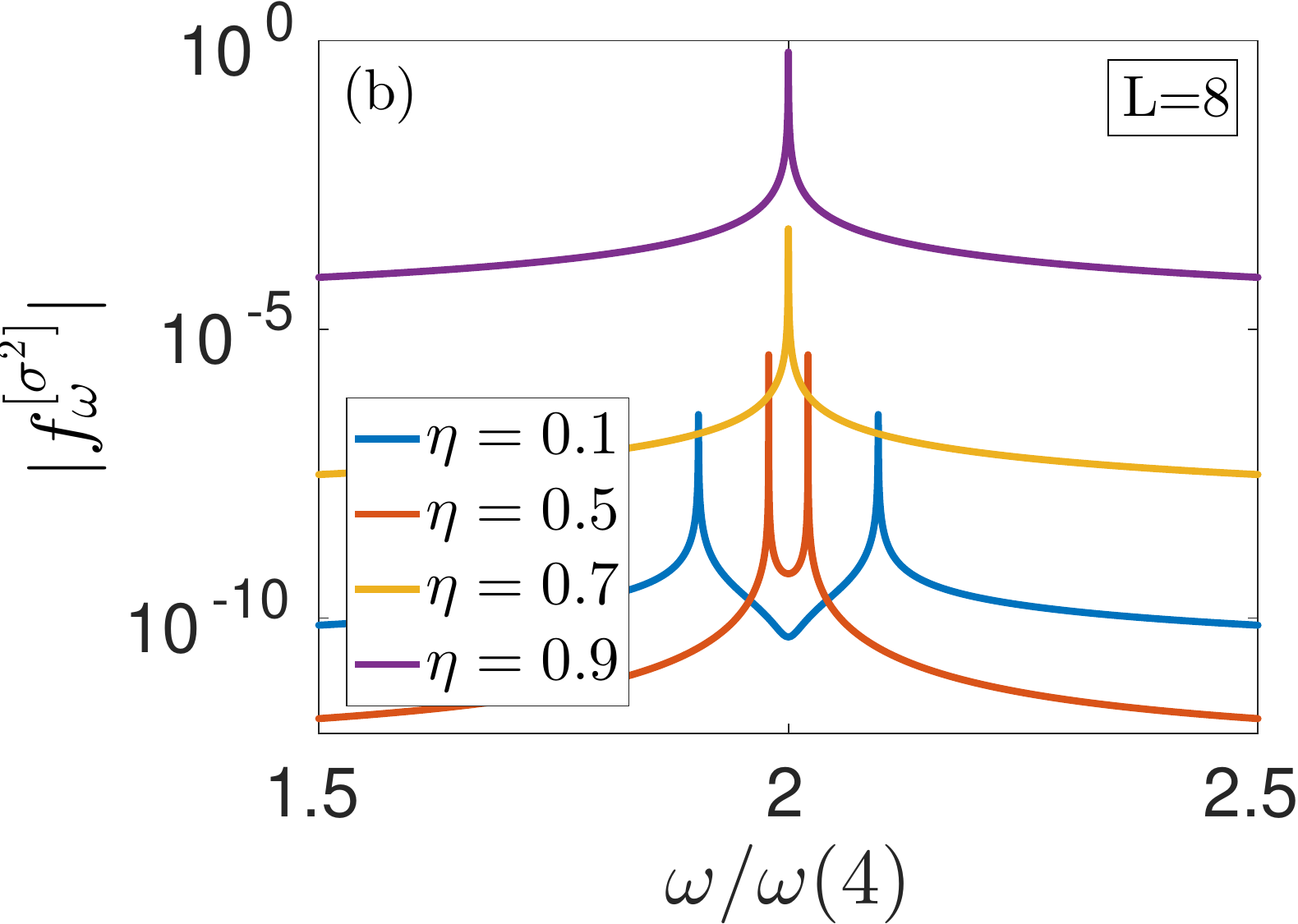}\\
\quad \qquad \includegraphics[width=0.35\textwidth]{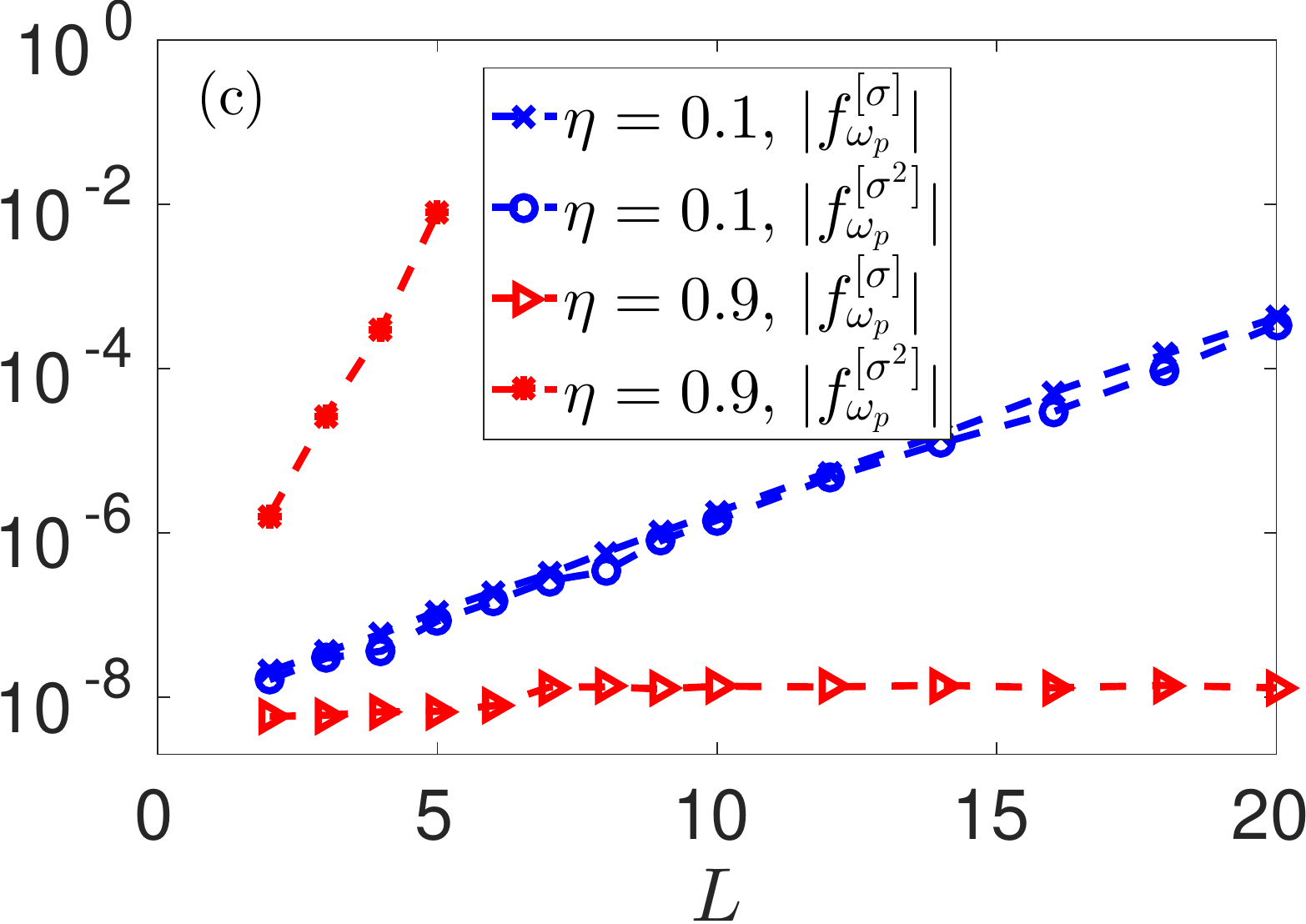}\qquad 
\includegraphics[width=0.35\textwidth]{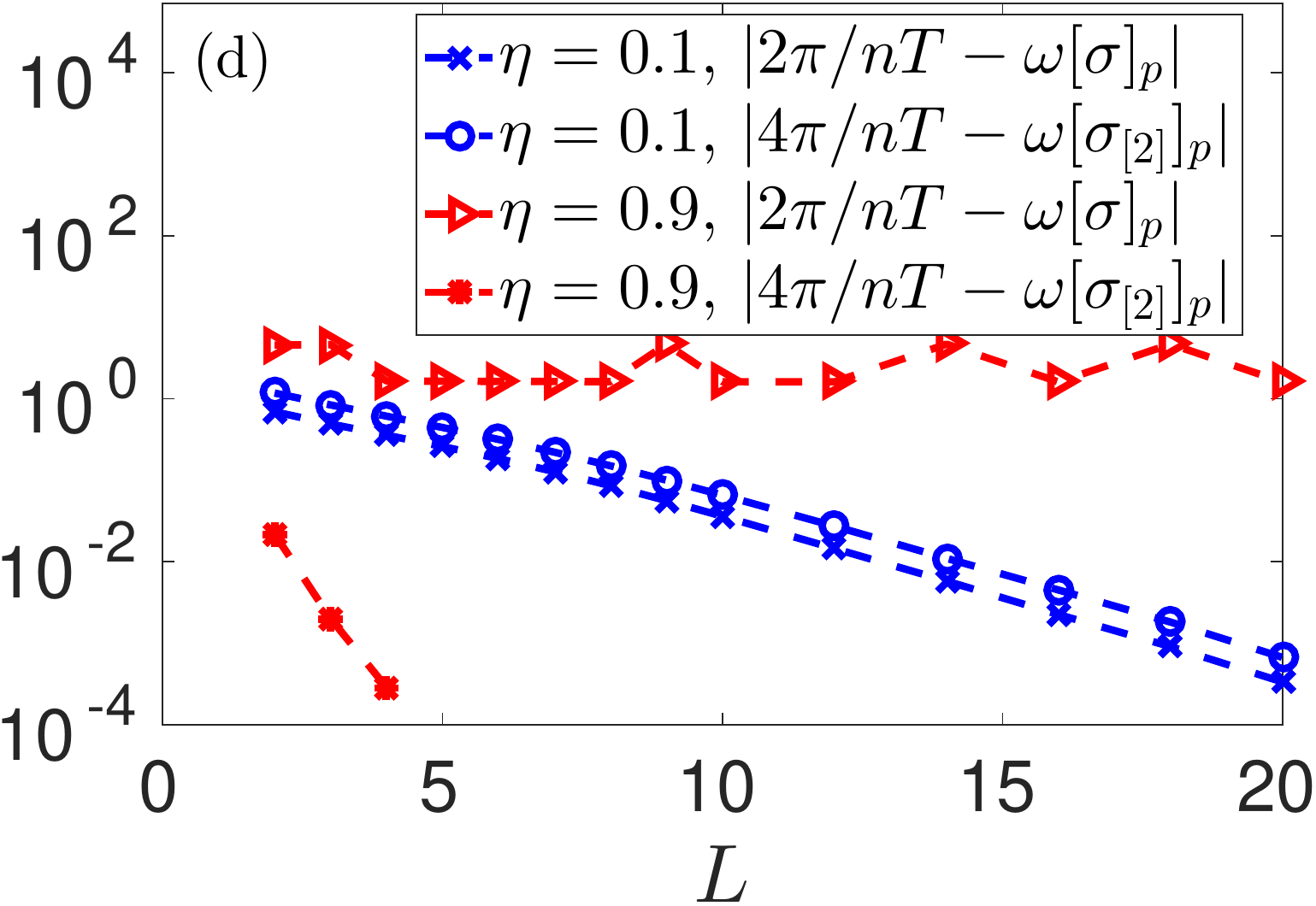}
 		\caption{ Distinct time-crystal phases in the model Hamiltonian $\hat H_{n=4,\eta}^{(LR)}$.
 		(a,b) Fourier transforms for the 
 		order parameters $\hat \sigma$ and $\hat \sigma_{[2]}$, considering a fixed system size $L=8$.
 		(c-d) Scaling with the system size for (c) the height of the dominant peak in the Fourier transforms and 
 		(d) its corresponding frequencies, for $\eta = 0.1$ and $0.9$. (Numerical parameters: $n=4, J=J'=1, h=h'=0.5$ and 
		$N_T = 2^{15}$).
		}
\label{fig.HptoHp2}
\end{figure*}

In this case the dynamics can generate distinct time-crystal phases. For $\eta $ not too large the system can break the \green{$\mathbb{Z}_4$} time translation 
symmetry, while for larger $\eta$ it breaks a lower \green{$\mathbb{Z}_{2}$} symmetry. We set $J=J'=1$ and $h=h'=1/2$, initialise the system in a 
symmetry-breaking ground state of \green{$\hat H_{n=4,\eta}^{(LR)}$}~\cite{Note6}  and perform a time evolution with $2^{15}$ periods.  
 
In Fig.~\ref{fig.HptoHp2}-(a,b) we show the Fourier power spectrum for the  order parameters $\mean{\hat {\sigma}}_t$ and $\mean{\hat {\sigma}^2}_t$ 
at a fixed system size $L=8$. For $\eta$ small we see two dominant peaks in $|f^{[\sigma]}_\omega|$ around the period 4-tupling frequency. As we 
increase $\eta$, the two dominant peaks of  $|f^{[\sigma]}_\omega|$ decrease their magnitude and become farther apart from each other.
On the other hand, the dominant peaks of $|f^{[\sigma^2]}_\omega|$ increase their magnitude and get closer to each other, around the period doubling frequency.    
This analysis for finite size suggests that there is at some point a transition from a period 4-tupling  at small $\eta$ witnessed by $\hat{\sigma}$ and a period doubling 
at large $\eta$ witnessed by $\hat{\sigma}^2$.
    
In Fig.~\ref{fig.HptoHp2} we show the finite-size scaling analysis 
for the frequency of the dominant peak (Panel c) and its magnitude (Panel d) for 
the cases $\eta = 0.1$ and $\eta=0.9$.  For 
$\eta=0.1$ we see the behaviour of a period $4$-tupling 
time crystal, in which the magnitude of the dominant
peak $|f^{[\sigma]}_{\omega_{\rm p}}|$ increases 
 with the system size, with the corresponding frequency $\omega[\sigma]_{\rm p}$ 
approaching the period $4$-tupling frequency. The order parameter $\hat \sigma^{2}$ displays a period-doubling response with a similar scaling 
behaviour of the dominant peak $|f^{[\sigma^{2}]}_{\omega_{\rm p}}|$ and of its frequency $\omega[\sigma^2]_{\rm p}$. In this case, therefore, the system is 
a period 4-tupling time crystal.

On the opposite limit of $\eta \sim 1$ the behaviour is different. We consider the case $\eta=0.9$. Here the magnitude of the dominant peak of $f^{[\sigma]}_\omega$ 
is  rather small and independent of the system size, marking the absence of a period 4-tupling time crystal phase. Furthermore, its frequency does not approach 
the period $4$-tupling frequency.  The Fourier transform $f^{[\sigma^2]}_\omega$, however, shows the expected behaviour for a time crystal, with the dominant frequencies 
approaching the period doubling frequency in the thermodynamic limit and the magnitude of the corresponding peak  increasing with $L$. In 
this case the system shows a period doubling.
 
The system supports distinct nontrivial time-crystal phases, breaking for $\eta \sim 0$ a discrete time-translation symmetry $\mathbb{Z}$ to $4 \mathbb{Z}$, 
while for $\eta \sim 1$ it breaks $\mathbb{Z}$ to $2 \mathbb{Z}$. The exact position of the transition between these two phases is difficult to 
locate using exact diagonalisation due to the limitations in the system sizes. For this goal we will use a semiclassical approach, which allows us to study 
the thermodynamic limit in a easier way. It is important to note that although the semiclassical approach allows us to obtain the exact behaviour in the 
thermodynamic limit, the finite-size scaling we have done until now was a crucial point in order to show that these symmetries are spontaneously broken \textit{only}
in the thermodynamic limit, as appropriate for a time crystal.

\subsection{Results for the semiclassical limit} 
\label{semiclass:sec}
\subsubsection{Case $n=3$}
\begin{figure}
\includegraphics[width=0.45\textwidth]{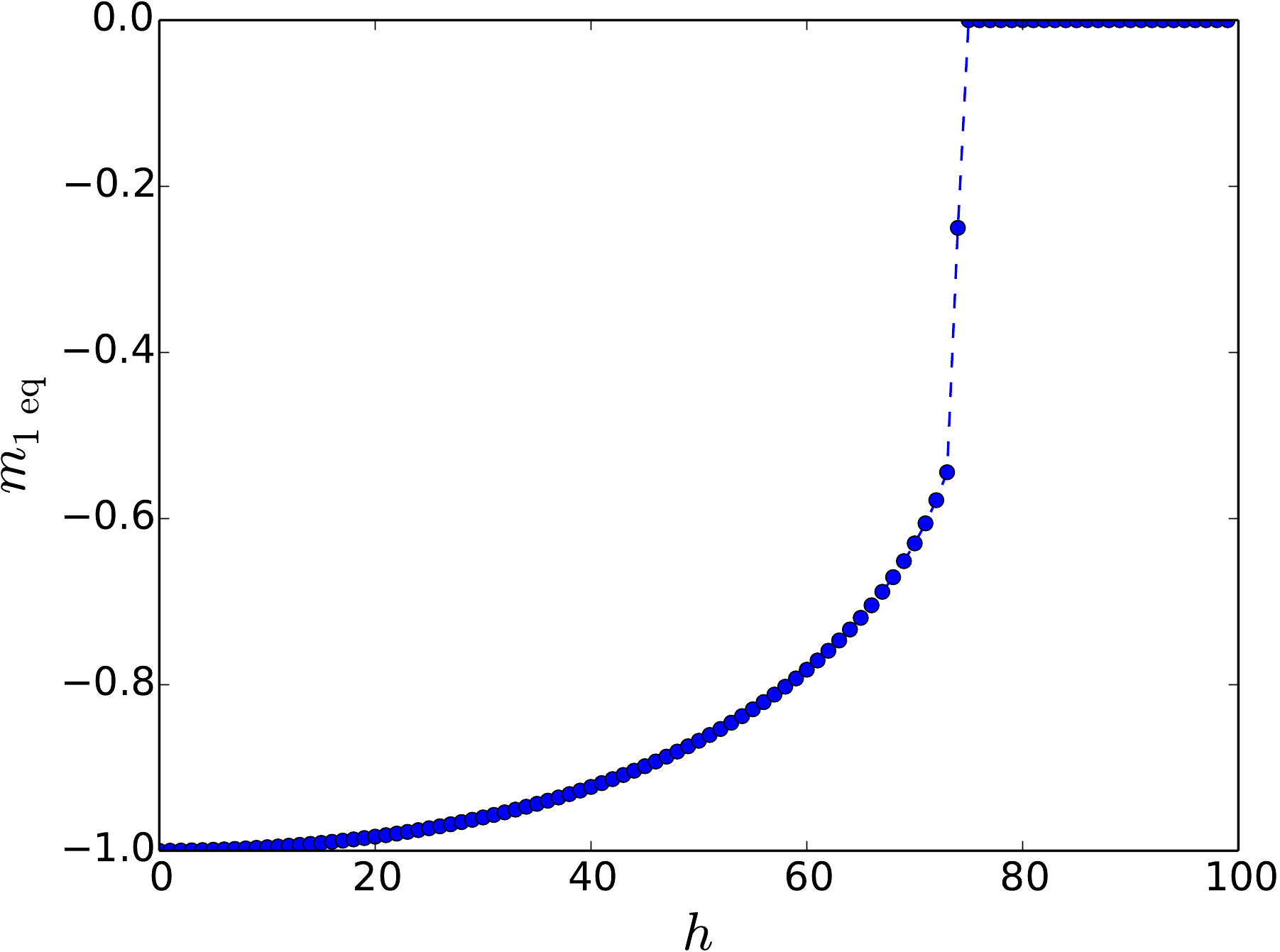}
		\caption{ The value of $m_{1\,\rm eq}=m_{2\,\rm eq}$ at the minimum point of 
 		the Hamiltonian Eq.~\eqref{accaccona:eqn} versus $h$. For $h<0.77$ it is non-vanishing, marking 
 		a $\mathbb{Z}_3$ symmetry-breaking phase (Numerical parameters $J=1$).} 
\label{minimalia:fig}
\end{figure}

In this case, exploiting the conservation of $p_1+p_2+p_3=1$, it is convenient to apply a linear canonical transformation in the following way
\begin{align} 
 		p_1&={\mathcal{N}}+{m_1} \label{eq:cambio1}\\
 		p_2&={\mathcal{N}}+{m_2} \label{eq:cambio2}\\
 		p_3&={\mathcal{N}}-{m_1}-{m_2} \label{eq:cambio3}\,,
\end{align}
\begin{align} 
  		{\phi}_1&=\frac{1}{3}\left(2{\theta_2}-{\theta_1}+{\Gamma}\right) \label{eq:cambio_phi1}\\
  		{\phi}_2&=\frac{1}{3}\left(-{\theta_2}+2{\theta_1}+{\Gamma}\right) \label{eq:cambio_phi2}\\
  		{\phi}_3&=\frac{1}{3}\left(-{\theta_2}-{\theta_1}+{\Gamma}\right) \label{eq:cambio_phi3}\,,
\end{align}
where $\{\theta_\ell,\,m_{\ell'}\}=\delta_{\ell\,\ell'}$, $\{\Gamma,\,\mathcal{N}\}=1$ and all the other Poisson brackets are vanishing. It is easy to see that the 
Hamiltonian written in the new variables does not depend on $\Gamma$ and therefore $\mathcal{N}$ is conserved to $1/3$. The Hamiltonian in 
the new variables acquires the form
\begin{align} 
\label{accaccona:eqn}
  		\green{{\cal H}_{3}^{(LR)}}&=-{J}\left({m_2}^{2}+{m_1}^{2}+{m_1}{m_2}\right)\nonumber\\
     		&-{2}h\bigg[\sqrt{(1+{m_2})(1+m_1)}\cos({\theta_1}-{\theta_2})\nonumber\\
         	&+ \sqrt{(1+{m_1})(1-m_1-m_2)}\cos\theta_1\nonumber\\
         	&+\sqrt{(1-m_1-m_2)(1+m_2)}\cos\theta_2\,\bigg]\,.
\end{align}
The order parameter for the static and the time-translation symmetry breaking can be written in terms of $m_1(t)$ and $m_2(t)$ using 
Eq.~\eqref{STERRO:eqn} and has the form
\begin{equation} \label{sigma:eqn}
		\sigma = \lim_{L\to\infty}\mean{\hat\sigma} = \frac{1}{6}\left(1+3m_2\right)+i\frac{\sqrt{3}}{6}\left(m_2+2m_1\right)\,.
\end{equation}
It is possible to find the state of minimum energy imposing $\theta_1=\theta_2=0$ and minimising  the energy along the line $m_1=m_2$. 
There is an interval of parameters where this state has $m_{1\,{\rm eq}}=m_{2\,{\rm eq}}\neq 0$ (see Fig.~\ref{minimalia:fig}) and therefore is triple 
degenerate (this can be easily seen repeating the same argument on the Hamiltonians which are obtained permuting cyclically the indices $1,2,3$ 
on the left side of the transformations Eqs.~\eqref{eq:cambio1}-\eqref{eq:cambio_phi3}). 
This fact marks the existence of a phase where there is a 
spontaneous breaking of the $\mathbb{Z}_3$ symmetry of the Hamiltonian Eq.~\eqref{accadi:eqn} for $n=3$; indeed in this phase the order
parameter Eq.~\eqref{sigma:eqn} is different from zero.  The critical field here is $h_c = 0.77$ and lies within the estimate
predicted using a finite-size scaling analysis  (see Appendix~\ref{subsec.SSB.LR}).

\begin{figure}
\begin{overpic}[width=0.45\textwidth]{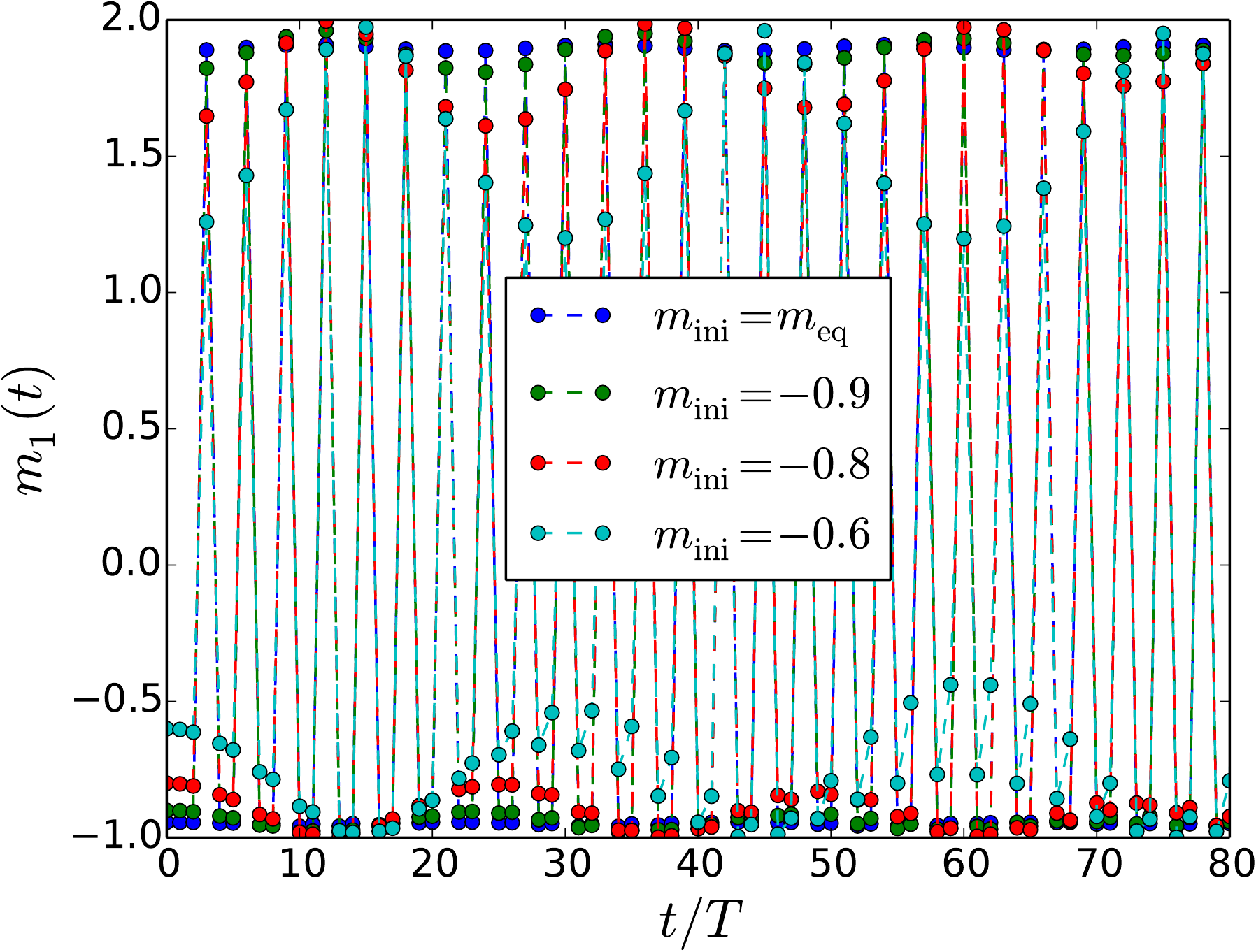}\put(-1,71){(a)}\end{overpic}\\
\begin{overpic}[width=0.45\textwidth]{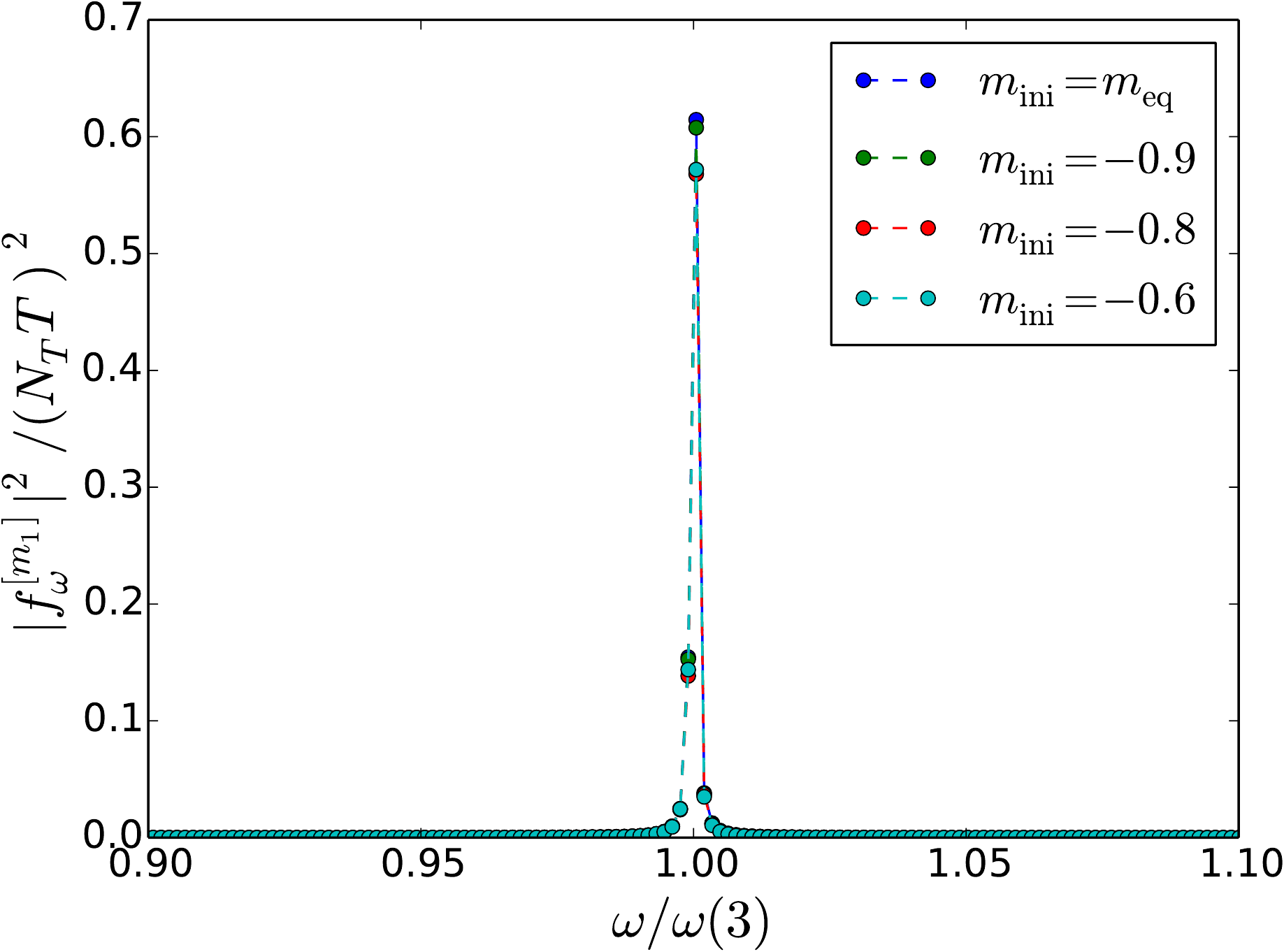}\put(-1,71){(b)}\end{overpic}
 		\caption{ (a) Evolution of $m_1(t)$ with  the Hamiltonian~\eqref{accaccona:eqn} and 
 		the kicking~\eqref{eq:kicko31}, with $n=3$. (b) Corresponding Fourier transform [see Eq.~\eqref{trasformazzi:eqn}]: we see a marked peak 
 		at the period-tripling frequency $\omega(3)=2\pi/(3T)$ (Numerical 
	 	parameters: $h=0.36$, $J=1.0$, $T=0.1$, $N_T=2048$). } 
\label{oscill:fig}
\end{figure}

\begin{figure}
\includegraphics[width=0.45\textwidth]{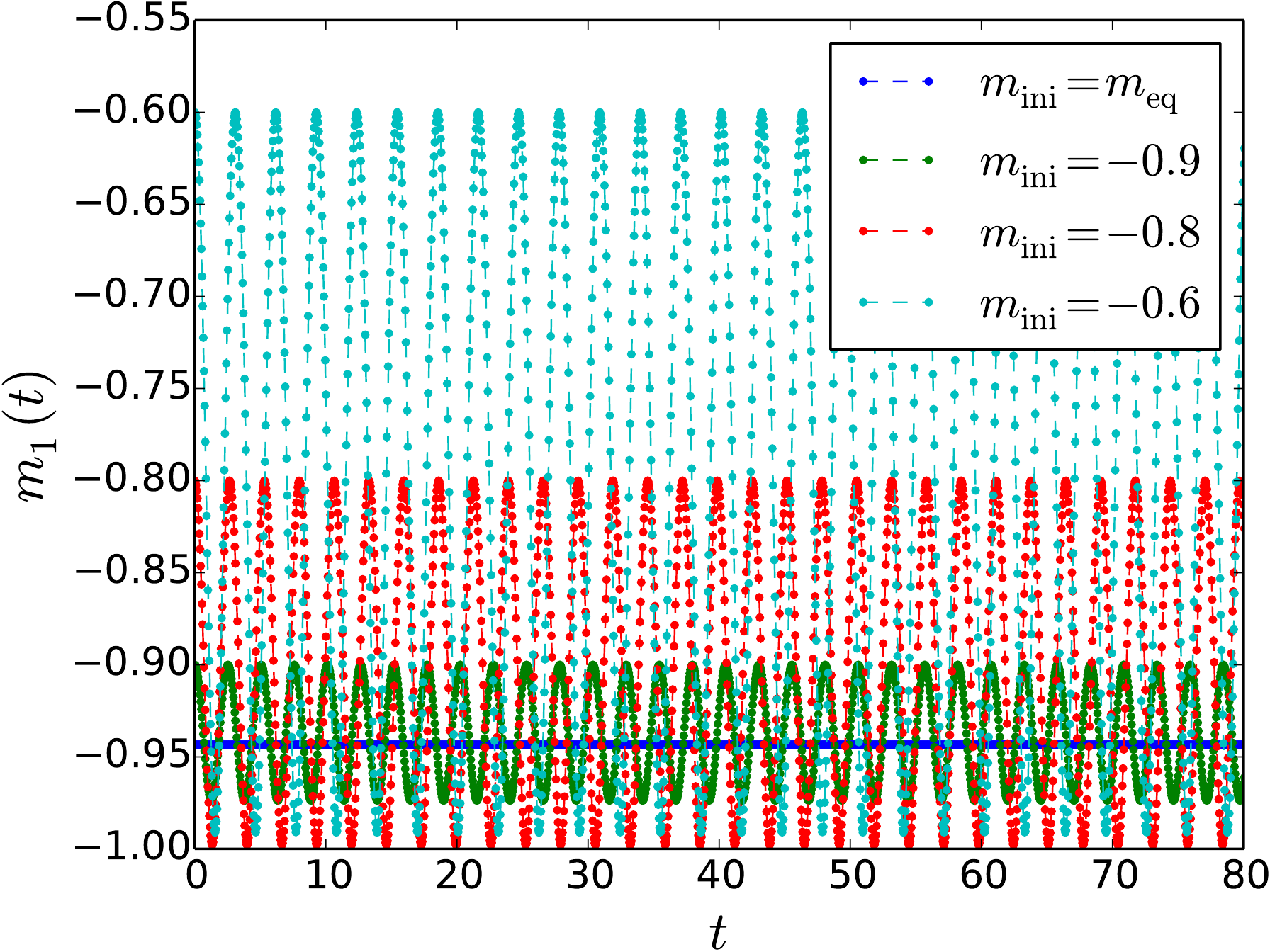}
 		\caption{ Evolution of $m_1(t)$ with the  Hamiltonian~\eqref{accaccona:eqn} without any kicking, for different initial conditions. 
		Notice the oscillations around an average  value different from 0, proving the existence of an interval of energies above the 
 		minimum where the corresponding trajectories are $\mathbb{Z}_3$ symmetry breaking (Numerical parameters: $h=0.36$, $J=1.0$).} 
\label{oscill_temp:fig}
\end{figure}

After the necessary introduction to the properties of the Hamiltonian, we now focus on the kicked dynamics and the period-tripling oscillations.
We apply the kicking Eq.~\eqref{eq:kicko31} to this Hamiltonian, solve the Hamilton differential equations and see if there are period-tripling oscillations. 
Because $m_1(t)$ and $m_2(t)$ have a very similar behavior, we will discuss in detail the behaviour of $m_1(t)$ (our conclusions hold for $m_2(t)$ 
and then for the order parameter $\sigma(t)$ exactly in the same way). Let us focus on a case where the $\mathbb{Z}_3$ symmetry is broken in the static 
part of the Hamiltonian ($h=0.36J$) and let us look for the period-tripling oscillations. If present, these oscillations appear as a marked peak at the period-tripling 
frequency  in the power spectrum of the Fourier transform of $m_1(t)$ [see Eq.~\eqref{trasformazzi:eqn}]. Remarkably we see those  oscillations 
both in time domain (upper panel of Fig.~\ref{oscill:fig}) and in frequency (lower panel of Fig.~\ref{oscill:fig}) if we initialise the system in one of the 
symmetry-breaking  ground states ($\theta_1(0)=\theta_2(0)=0$ and $m_1(0)=m_2(0)=m_{\rm eq}$) or if we initialise it with $\theta_1(0)=\theta_2(0)=0$ 
and a value of $m_1(0)=m_2(0)=m_{\rm ini}$ near to $m_{\rm eq}$. This robustness with respect to the initial state is due to the existence of an interval of 
energies where all the trajectories break the $\mathbb{Z}_3$ symmetry, as it occurs  in the period doubling case (see Ref.~\onlinecite{Russomanno2017}). 
We have checked this fact studying the dynamics of Hamiltonian~\eqref{accaccona:eqn}  \textit{without} a kicking: for the values of $m_1(0)=m_2(0)=m_{\rm ini}$ 
considered in Fig.~\ref{oscill:fig} we can see oscillations of  $m_1(t)$ around a non-vanishing value (see Fig.~\ref{oscill_temp:fig}). This interval of energies
where the trajectories break the $\mathbb{Z}_3$ symmetry directly corresponds to the extensive amount of eigenstates below an energy threshold which break 
the symmetry in the finite-size case (see Appendix~\ref{subsec.SSB.LR}).

\begin{figure*}
\begin{center}
\begin{tabular}{cc}
\includegraphics[width=8cm]{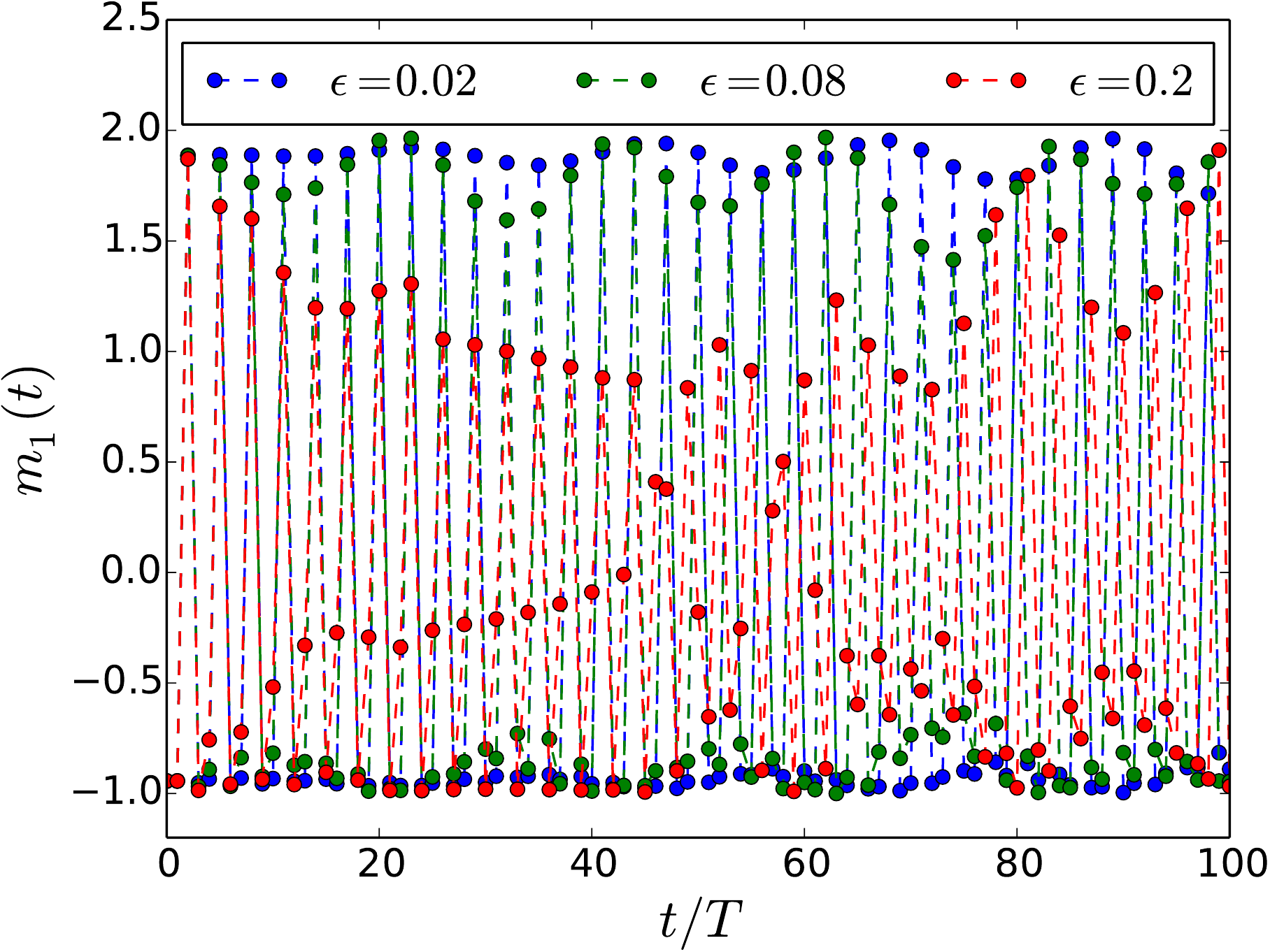}
&
\includegraphics[width=8cm]{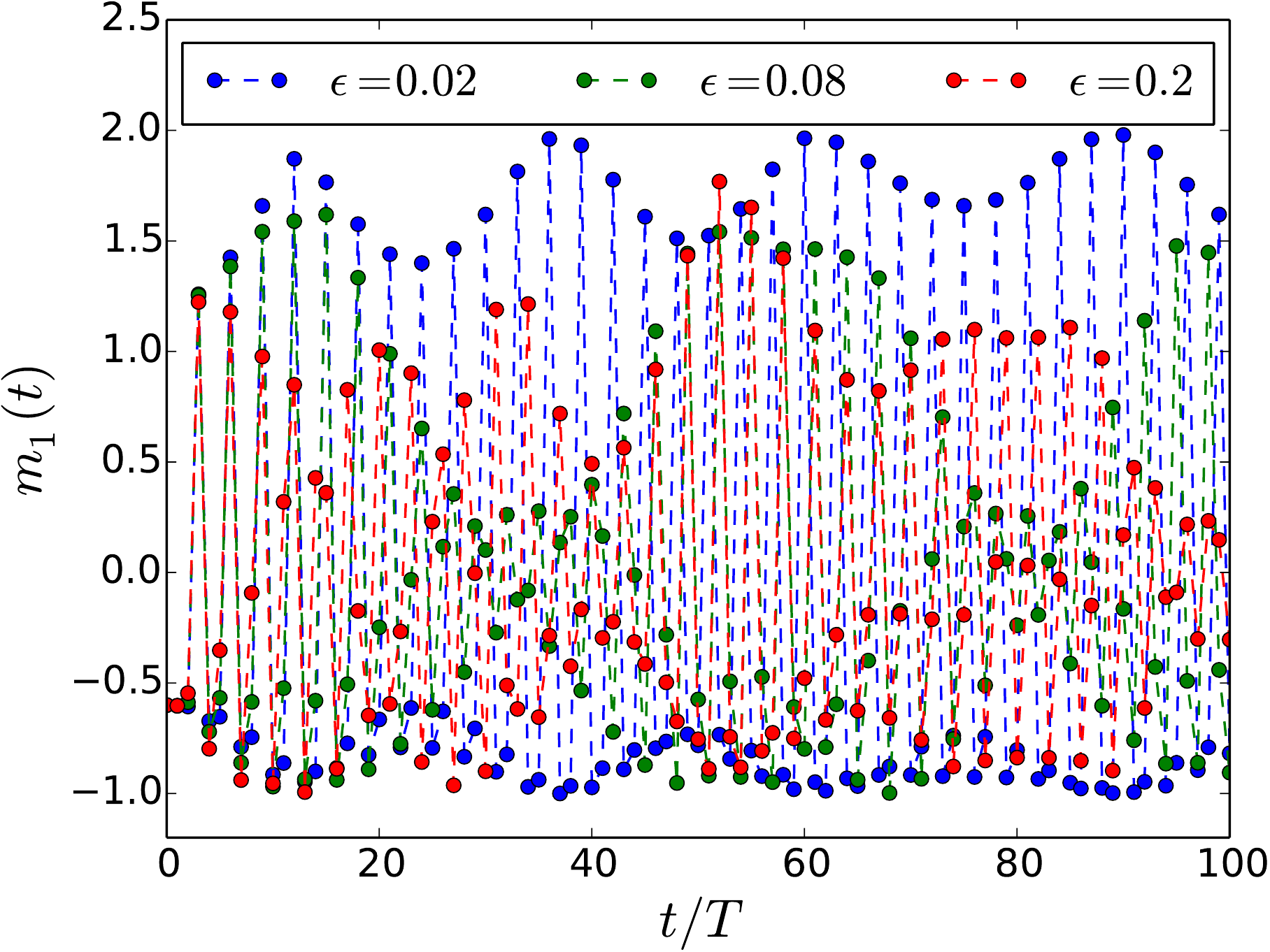}
\\
\hspace{0cm}
\includegraphics[width=8cm]{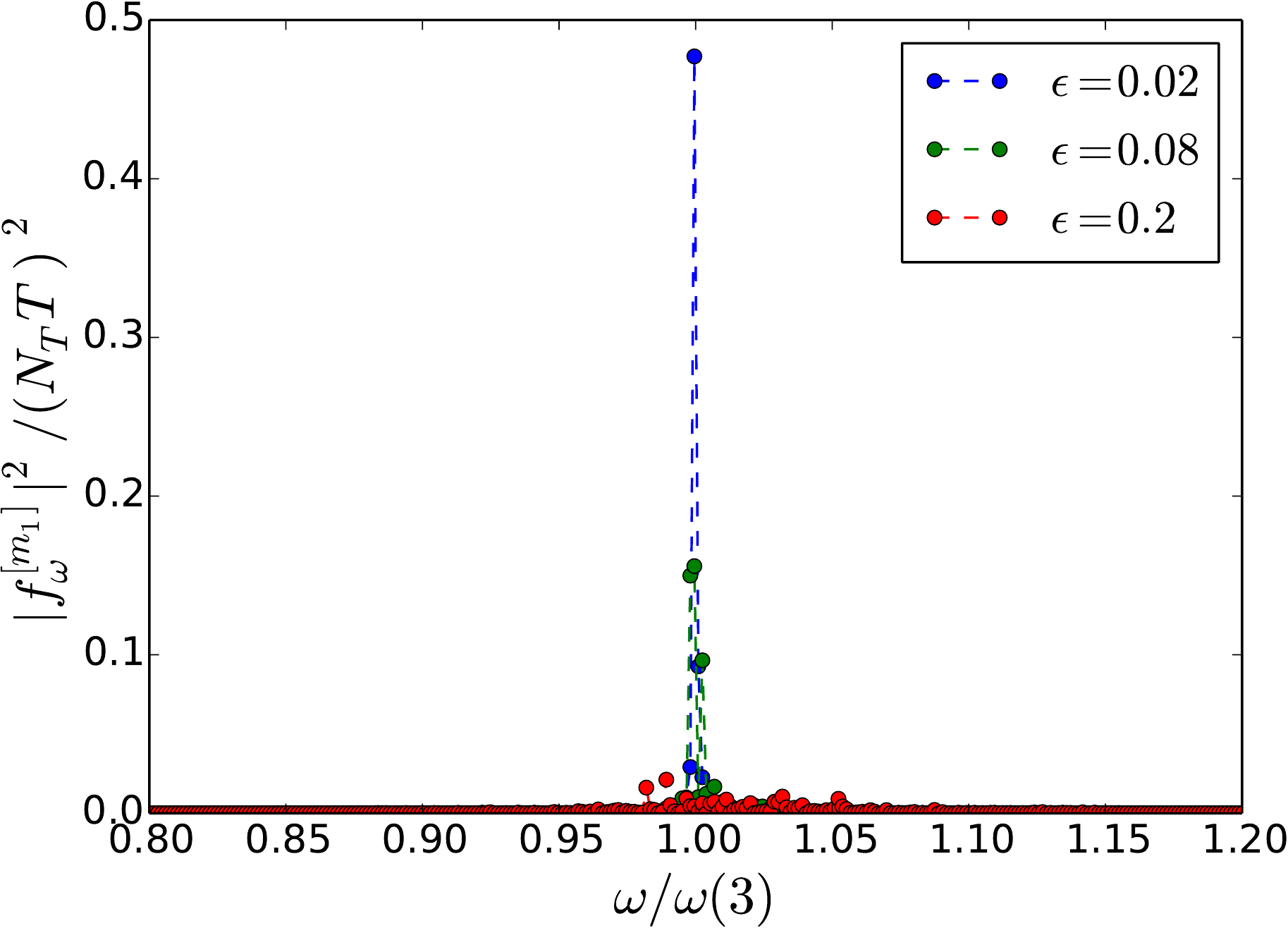}&
\includegraphics[width=8cm]{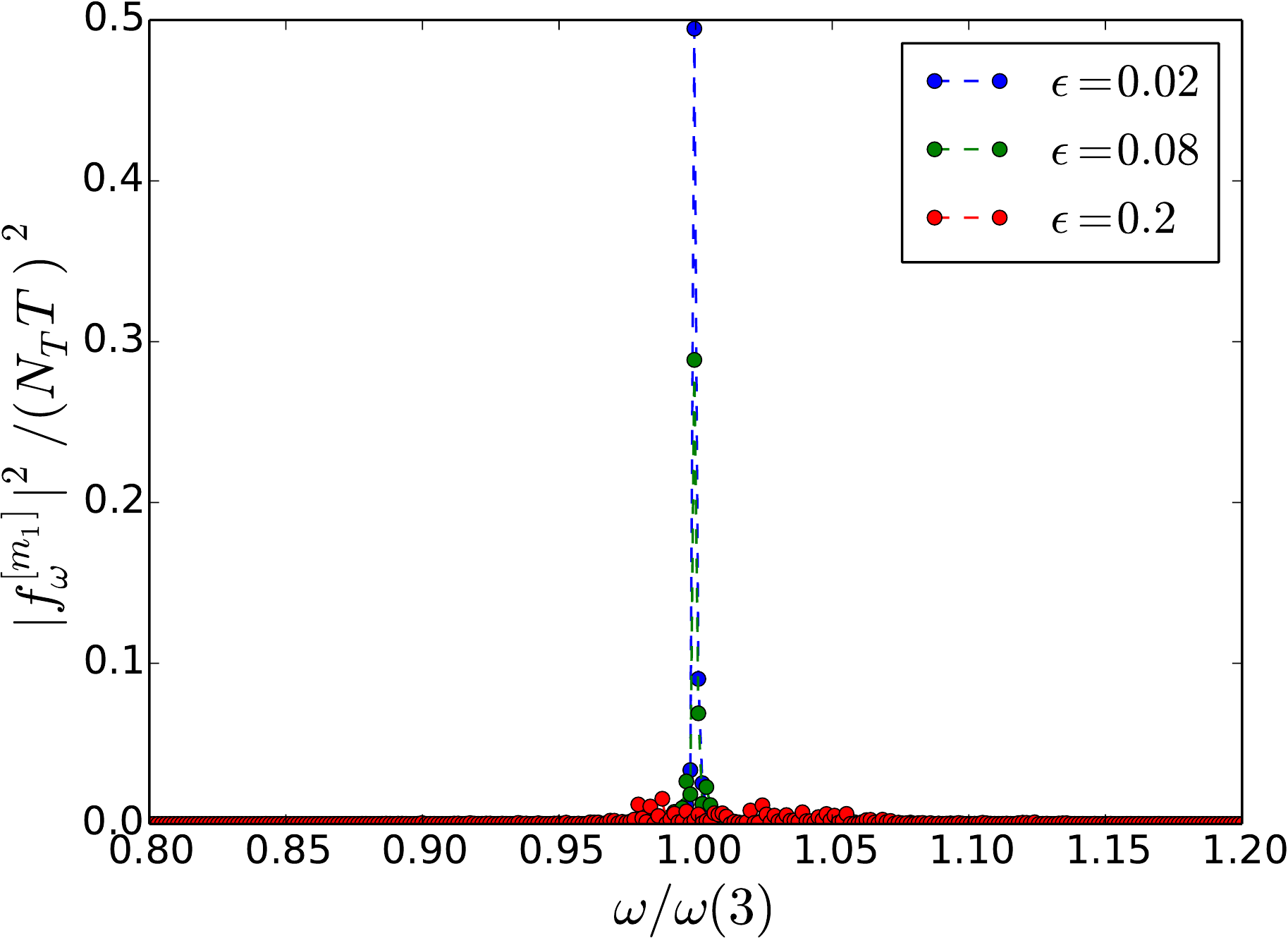}\\
\hspace{0cm}
\includegraphics[width=8cm]{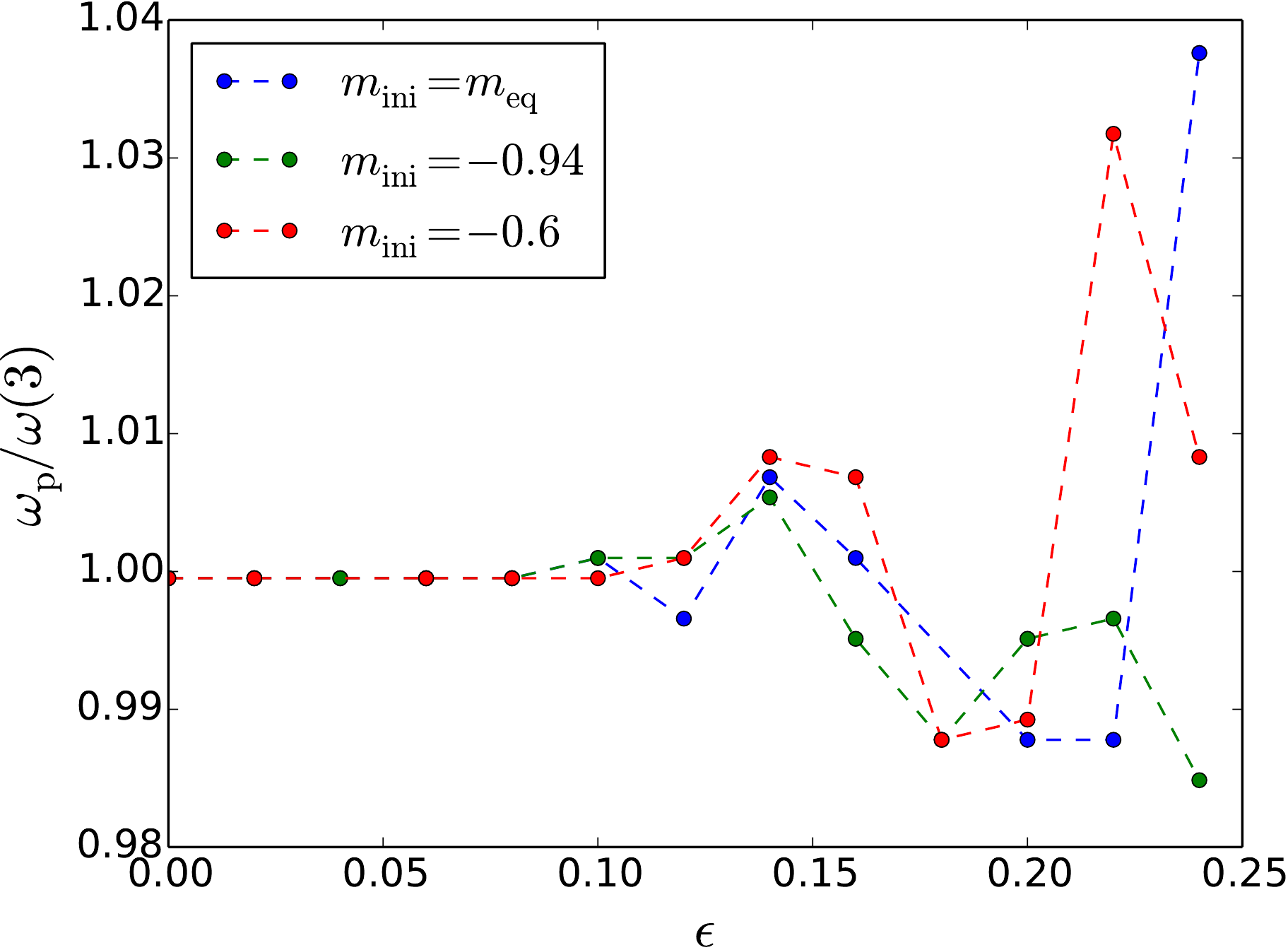}&
\includegraphics[width=8cm]{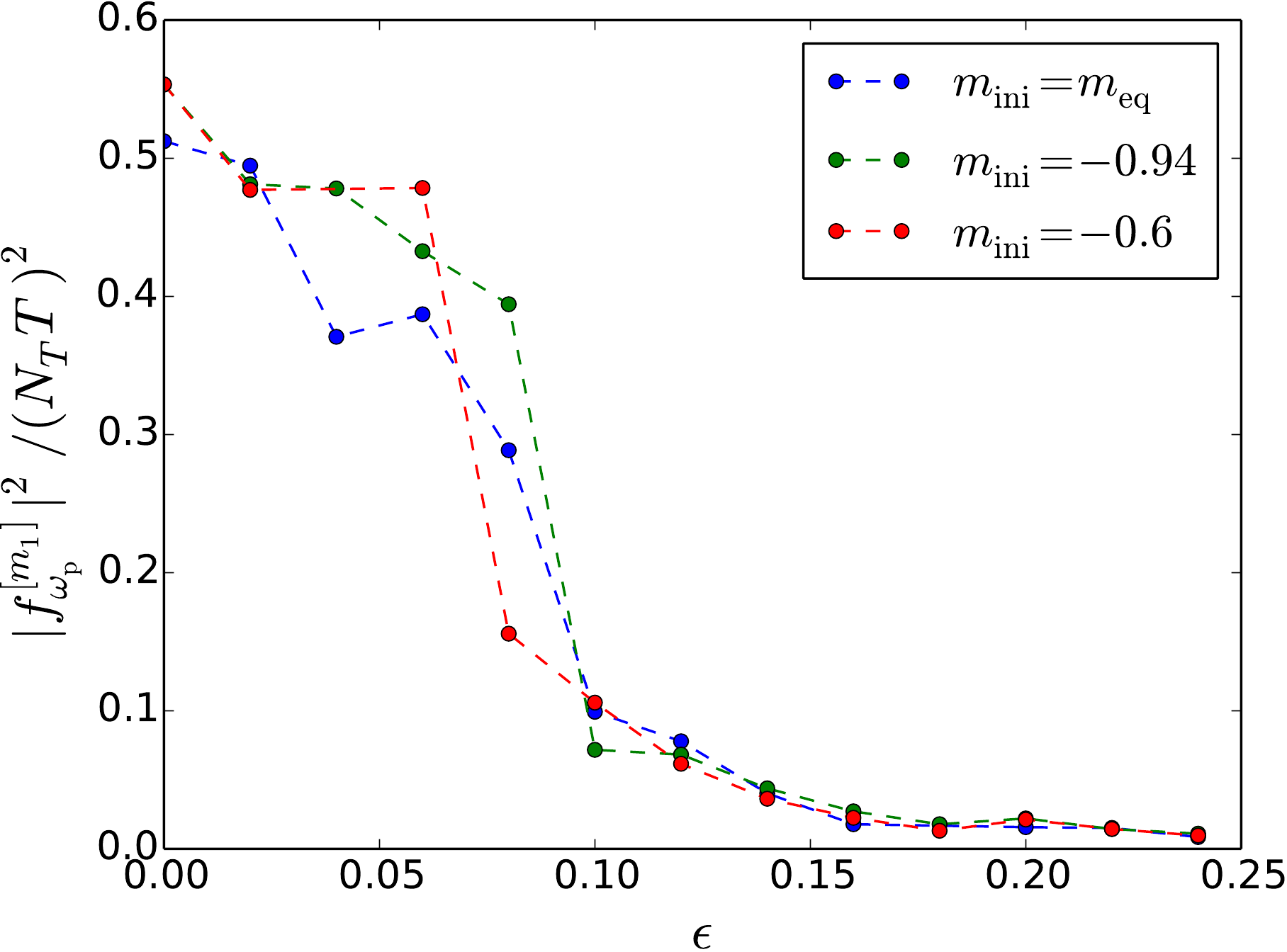}
\end{tabular}
\end{center}
        \caption{ \textbf{(Upper panels)} Evolution of $m(t)$ with the Hamiltonian Eq.~\eqref{accaccona:eqn} and the $\epsilon\neq 0$-periodic 
		kicking Eq.~\eqref{eq:kicko31} (Numerical parameters: $h=0.36$, $J=1.0$, $T=0.1$, $m_1(0)=m_2(0)=m_{\rm eq}$ in the left panel 
		and $m_1(0)=m_2(0)=-0.6$ in the right one). \textbf{(Central panels)} Corresponding Fourier transforms for $N_T=2048$ periods: for small $\epsilon$ 
		we see a marked peak at the period-tripling frequency $\omega(3)=2\pi/(3T)$. 
		\textbf{(Lower panels)} Position $\omega_p$ (left) and height 
		$|f_{\omega_p}^{[m_1]}|^2$ (right) of the main peak of the Fourier power spectrum versus $\epsilon$ for different values of $m_1(0)=m_2(0)=m_{\rm ini}$. 
		We see persistent period-tripling oscillations (main peak at $\omega_p=\omega(3)=2\pi/3$) for $\epsilon$ small enough ($\epsilon\leq 0.08$). 
		} 
\label{plot_osc_eps:fig}
\end{figure*}

The dynamics is  robust also against perturbations in the kicking: if we apply Eq.~\eqref{eq:kicko3e} with $n=3$ we see a full interval 
of $\epsilon$ where the time crystal persists. We can see this fact by studying the Fourier transform of $m_1(t)$: we find a marked peak at the 
period-tripling frequency  for a full interval of $\epsilon$ around zero.
The symmetry breaking oscillations of $m_1(t)$ in time domain are shown in the 
upper panels of Fig.~\ref{plot_osc_eps:fig}. In the central panels the corresponding 
Fourier transforms: when $\epsilon$ is small enough there is a 
marked peak at the period-tripling frequency. In the  lower panels  it is shown how the frequency $\omega_p$ and the 
height $|f^{[m_1]}_{\omega_p}|^2$ of the peak in the Fourier transform depend  on $\epsilon$. For all the considered initial conditions, the 
peak frequency deviates from $\omega(3)$ (and then the time crystal disappears) when
$\epsilon>0.08$.

\subsubsection{$n=4$, $\eta=0$ } 
 
\begin{figure}
\includegraphics[width=8cm]{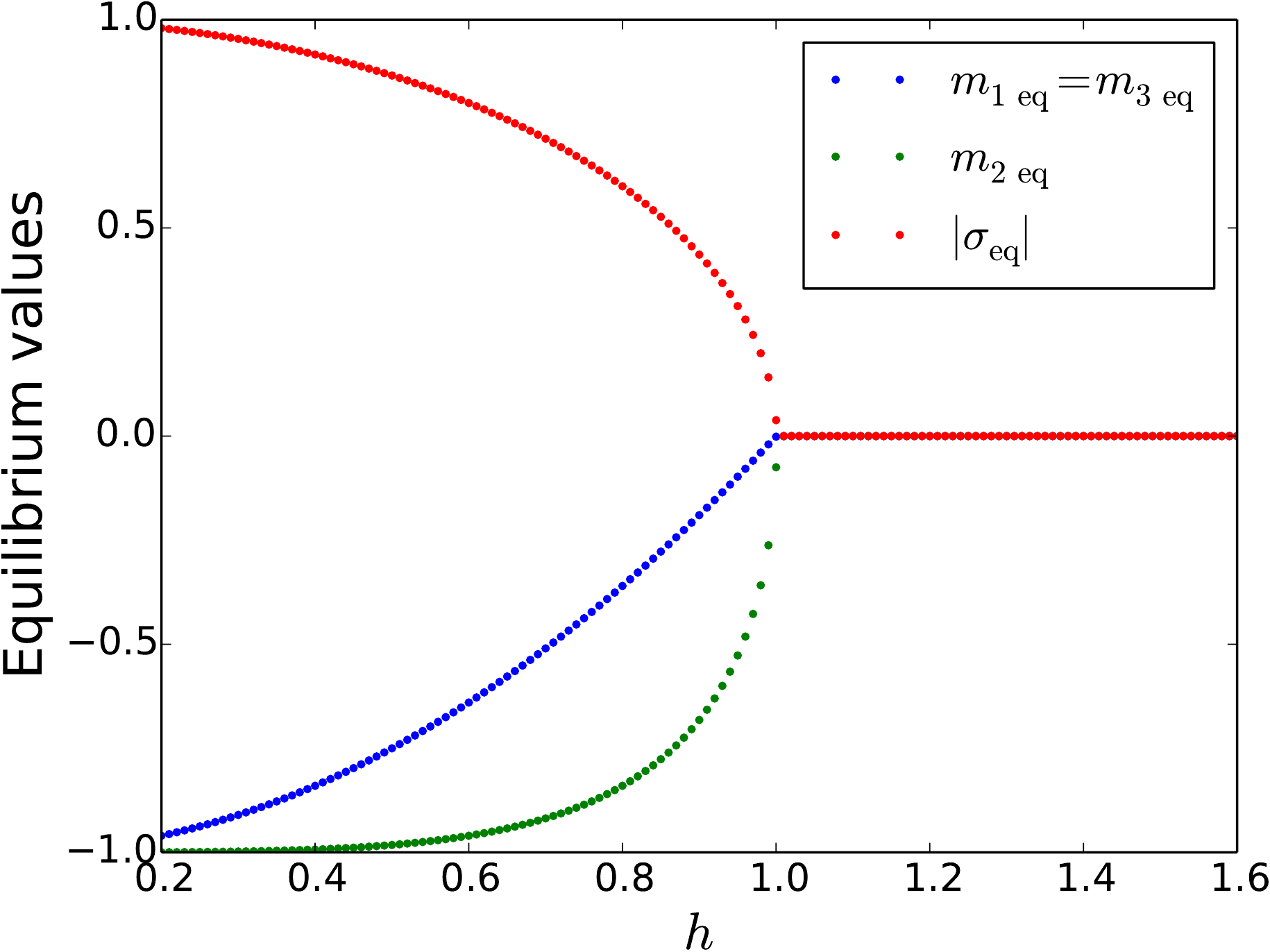}
		\caption{The values of $m_{j\,{\rm eq}}$ and $|\sigma_{\rm eq}|$ at the minimum point of the Hamiltonian Eq.~\eqref{ham4:eqn} 
		versus $h$. For $h<1$ they are non-vanishing, marking a $\mathbb{Z}_4$ symmetry-breaking phase (Numerical parameters: $J=1$).} \label{minimalia4:fig}
\end{figure}

\begin{figure}
\includegraphics[width=8cm]{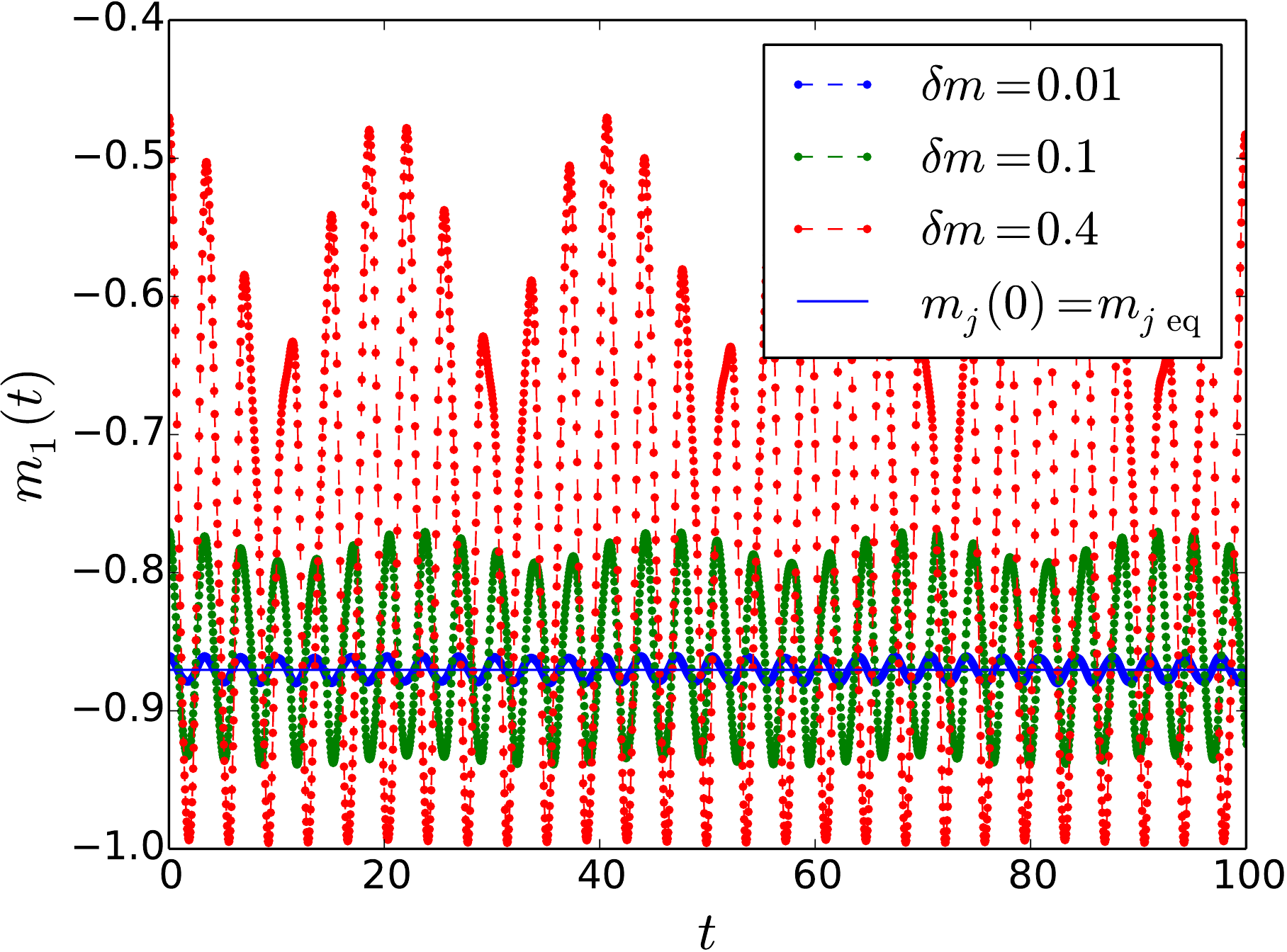}
		\caption{Evolution of $m_1(t)$ with the Hamiltonian Eq.~\eqref{ham4:eqn} and no kicking. Initial conditions $\theta_1(0)=\theta_2(0)=
		\theta_3(0)=0$, $m_{2}(0)=m_{2\,{\rm eq}}$, $m_1(0)=m_3(0)=m_{1\,{\rm eq}}+\delta m$. $J=1$, $h=0.36$.} \label{p4_evol:fig}
\end{figure}

\begin{figure*}
\includegraphics[width=8cm]{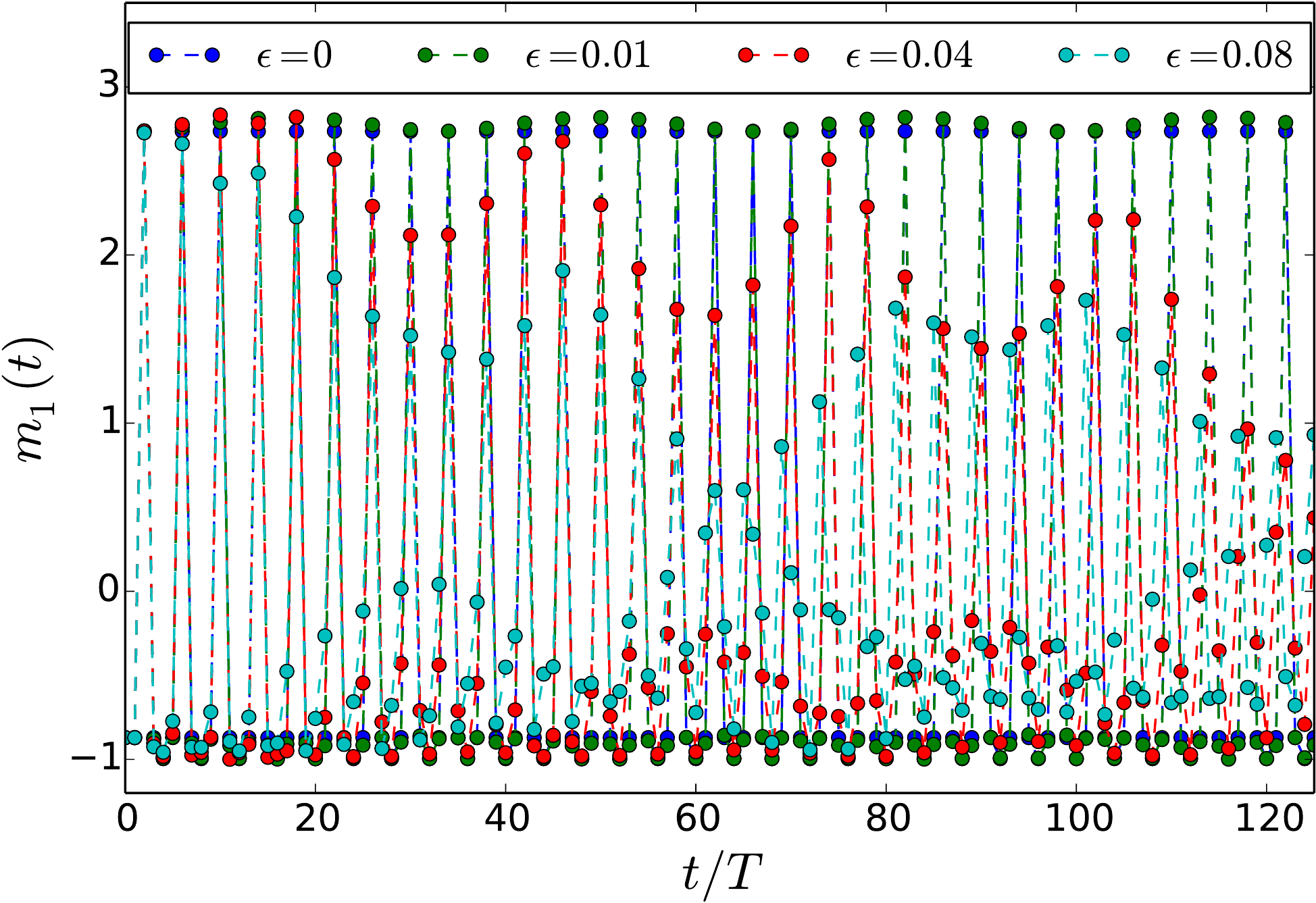}\qquad
\includegraphics[width=8cm]{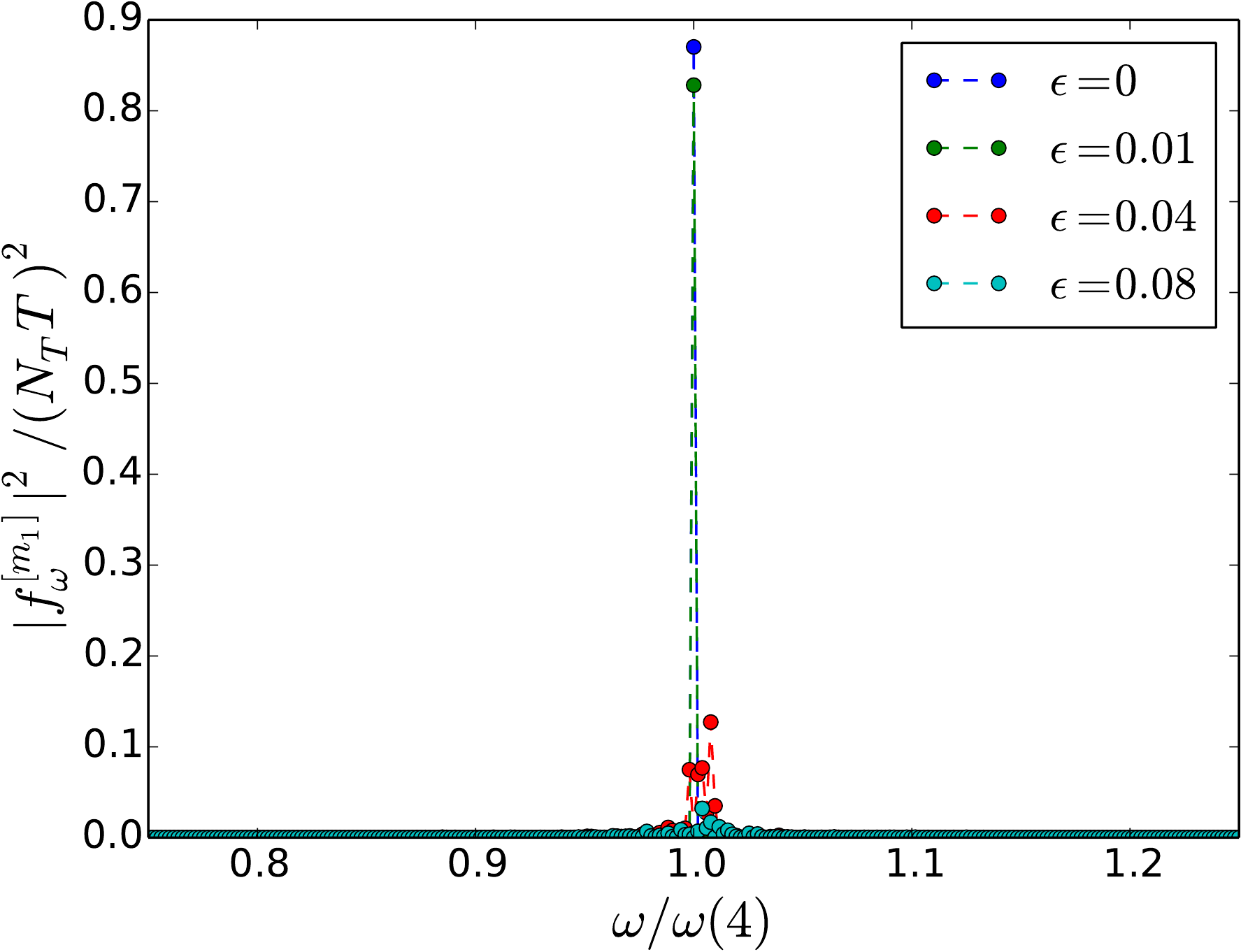}\qquad
\includegraphics[width=8cm]{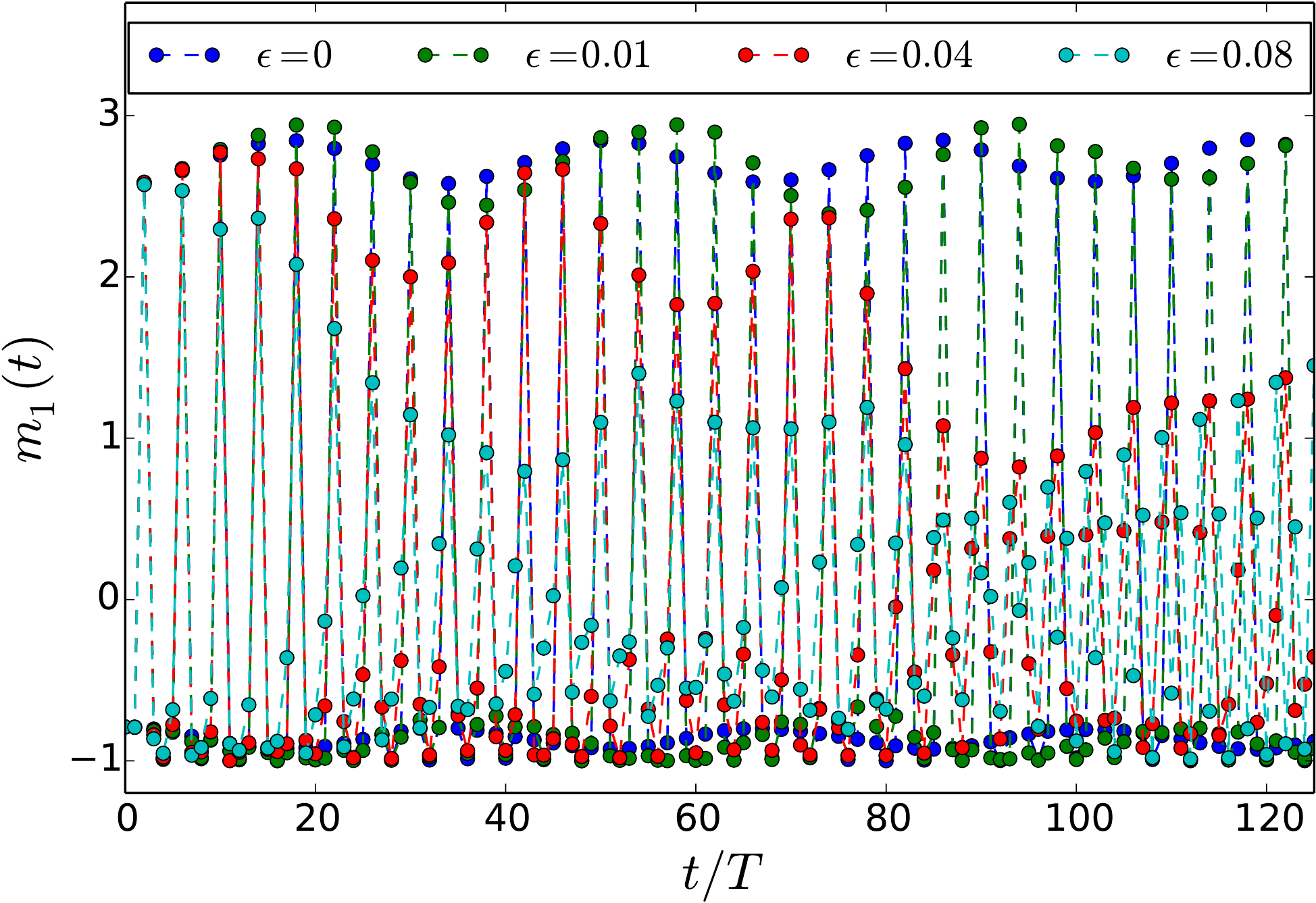}\qquad
\includegraphics[width=8cm]{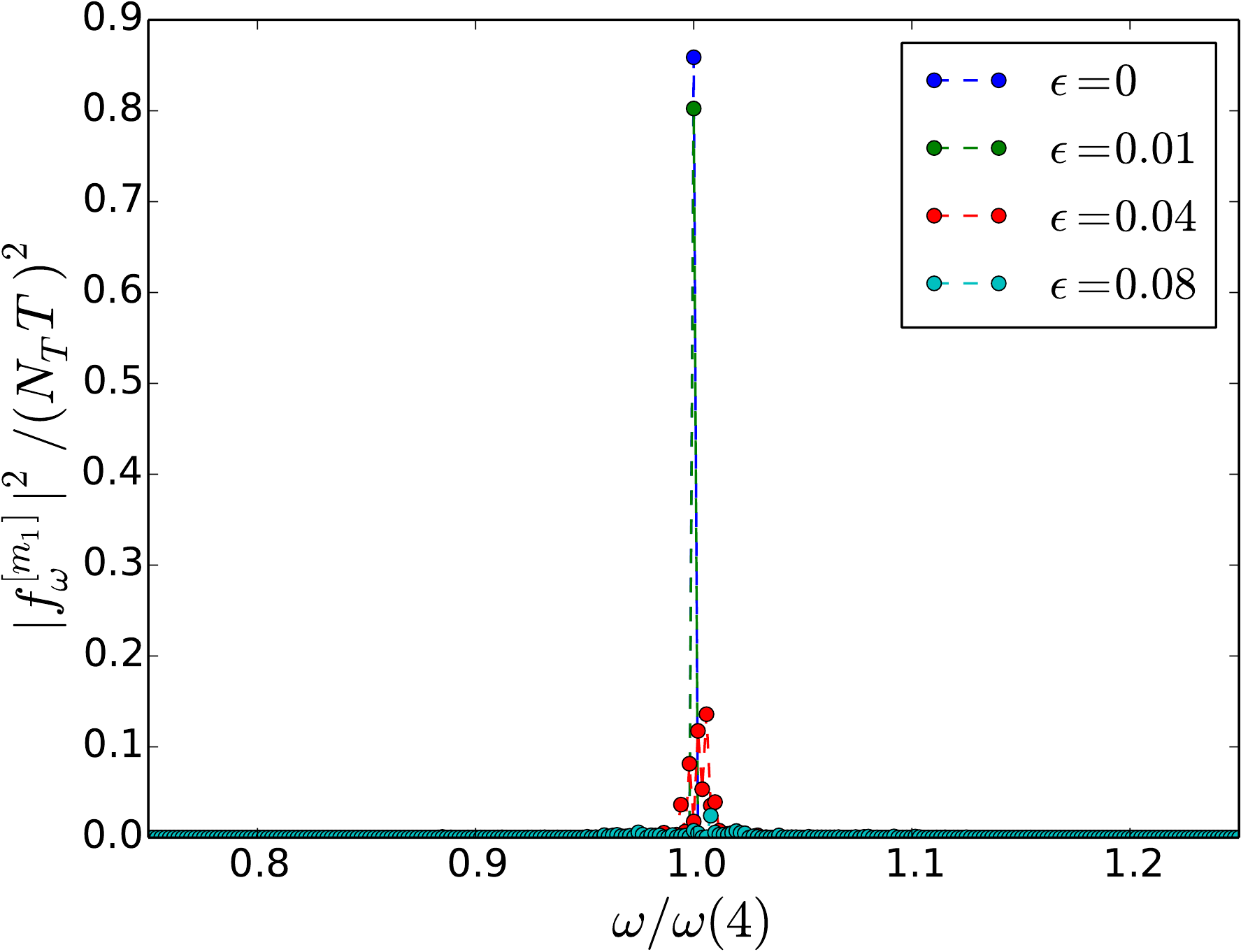}\qquad
		\caption{ Dynamics of $m_1(t)$ with the Hamiltonian Eq.~\eqref{ham4:eqn} and a perturbed kicking [Eq.~\eqref{eq:kicko3e}]. Time domain (left panels) 
		and frequency domain (right panels). For $\epsilon=0.01$ we see the period 4-tupling oscillations (appearing as a peak at $\omega(4)=\pi/(2T)$ 
		which disappear for larger $\epsilon$. Initial 
		conditions: $\theta_1(0)=\theta_2(0)=\theta_3(0)=0$, $m_{2}(0)=m_{2\,{\rm eq}}$, $m_1(0)=m_3(0)=
		m_{1\,{\rm eq}}+\delta m$; we consider two different initial conditions, $\delta m=0$ in the upper panels and $\delta m=0.08$ in the lower ones. 
		Numerical parameters: $J=1$, $h=0.36$, $T=0.1$.
		} 
\label{surveps:fig}
\end{figure*}

The approach is analogous to the case $n=3$. Using  that $p_1+p_2+p_3+p_4=1$ we can write  the effective Hamiltonian in the form
\begin{align} 
\label{ham4:eqn}
  		{\cal H}_{4,\eta=0}^{(LR)}&=-\frac{J}{4}\left[(m_1-m_3)^2+(2m_2+m_1+m_3)^2\right] \nonumber\\
   		&- 2h\sqrt{(1+m_1)(1+m_2)}\cos(\theta_1-\theta_2) \nonumber\\
   		&- 2h\sqrt{(1+m_2)(1+m_3)}\cos(\theta_2-\theta_3) \nonumber\\
   		& - 2h\sqrt{(1+m_3)(1-m_1-m_2-m_3)}\cos(\theta_3) \nonumber\\
   		&- 2h\sqrt{(1-m_1-m_2-m_3)(1+m_1)}\cos(\theta_1)\,,
\end{align}
where $\theta_j$ are canonical coordinates, $m_j$ are canonical momenta and obey the standard canonical  commutation relations. Using 
Eq.~\eqref{STERRO:eqn} we can write the order parameter $\sigma$  in terms of $m_j$ in the form
\begin{equation} 
\label{sigma4:eqn}
  		\sigma=\frac{1}{4}\left[(m_1-m_3)+i(2m_2+m_1+m_3)\right]\,.
\end{equation}
We can find the minimum of the Hamiltonian Eq.~\eqref{ham4:eqn} fixing $\theta_j=0$ and then using a steepest descent algorithm. For $h<1$ 
there is a $\mathbb{Z}_4$ broken symmetry phase where the $m_{j\,{\rm eq}}$ and the $\sigma_{\rm eq}$ are non vanishing (see Fig.~\ref{minimalia4:fig}). 
There is a full interval of energies where the trajectories break the $\mathbb{Z}_4$ symmetry; we can see this fact in Fig.~\ref{p4_evol:fig} where we simulate 
the dynamics of ${\cal H}_{4,\eta=0}^{(LR)}$ without any kicking. We choose initial conditions different from the equilibrium ones and we observe that $m_1(t)$ 
oscillates around a non-vanishing average. These are the perfect conditions for the manifestation of a period $4$-tupling  time crystal. Indeed, if we apply 
to this system the kicking Eq.~\eqref{eq:kicko3e}
 with $n=4$, we see period 4-tupling oscillations which are stable 
 if we consider
 initial conditions different from the 
lowest energy ones, $m_{2}(0)=m_{2\,{\rm eq}}$, $m_1(0)=m_3(0)=m_{1\,{\rm eq}}+\delta m$ (see Fig.~\ref{surveps:fig}; here we 
show only $m_1(t)$ for clarity, the situation is the same for  all the $m_j(t)$ and for $\sigma(t)$).
%
%

\subsubsection{$n=4$, $\eta \neq 0$ - Transition between two different time-crystal phases} 
\label{Sec4:sec}

We finally analyse the behaviour as a function of $\eta $. The order parameter $\sigma$ for the breaking of the 
$\mathbb{Z}_4$ symmetry is the one in Eq.~\eqref{sigma4:eqn}, while the $\mathbb{Z}_2$-order parameter
is expressed by the quantity
\begin{equation}
  \sigma_{[2]}\equiv \lim_{L\to\infty}\mean{\hat\sigma^2} =(1/2)(m_1+m_3) \,
\end{equation}
(see Eq.~\eqref{STERRO:eqn}). The effective Hamiltonian has the form
\begin{align} 
\label{ham42:eqn}
  		{\cal H}_{4,\eta}^{(LR)}&= (1-\eta) {\cal H}_{4,0}^{(LR)} + \eta \left[ - {J'}(m_1+m_3)^2 \right. \nonumber\\
     		&-2{h'}\sqrt{(1+m_1)(1+m_3)}\cos(\theta_1-\theta_3)\nonumber\\
     		&\left. -2{h'}\sqrt{(1+m_2)(1-m_1-m_2-m_3)}\cos\theta_2 \right] 
\end{align}
where ${\cal H}_{4,0}^{(LR)}$ is the effective Hamiltonian shown in Eq.\eqref{ham4:eqn}.  

\begin{figure}
\begin{center}
\begin{tabular}{c}
\hspace{0cm}\includegraphics[width=8cm]{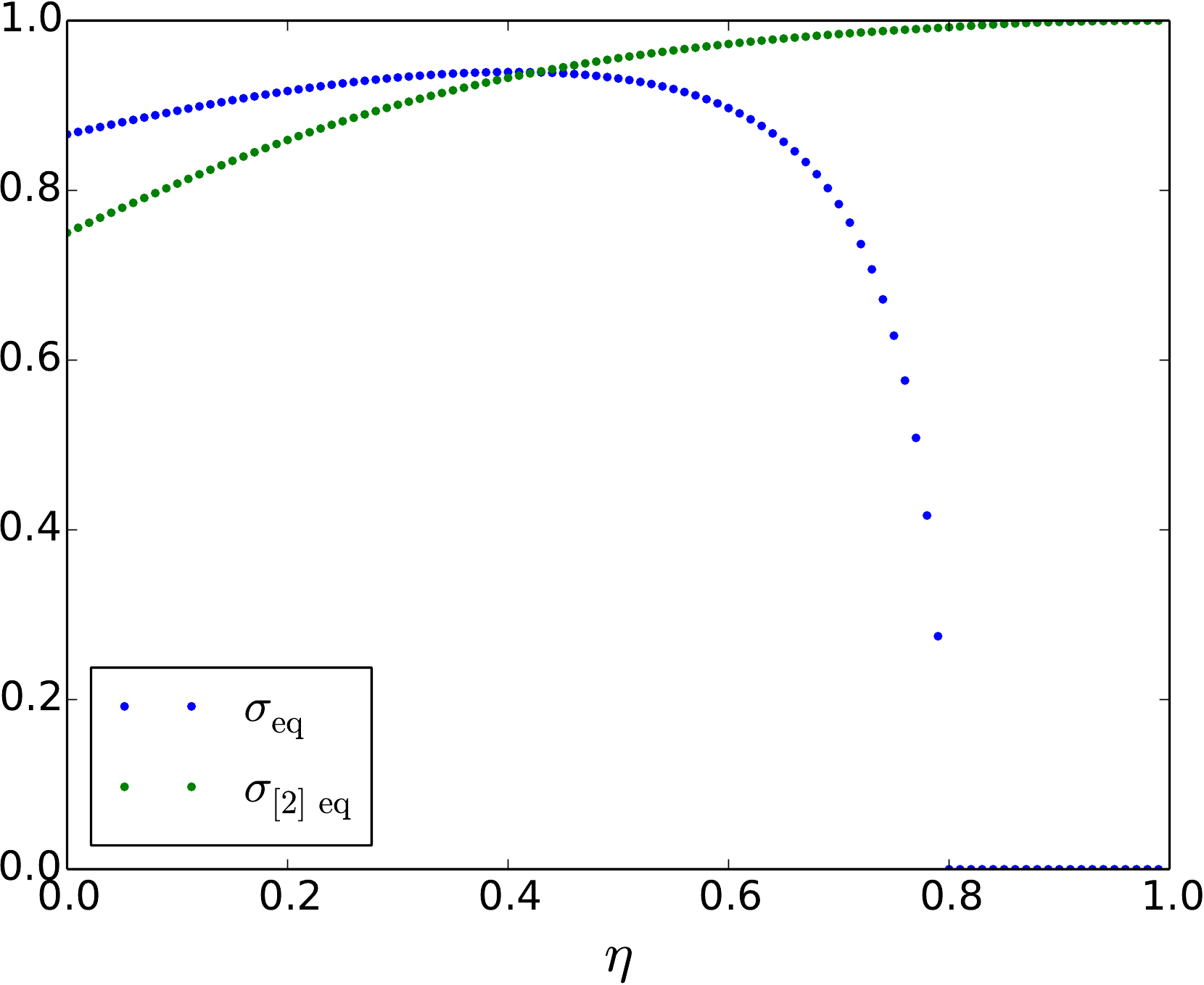}\\
\end{tabular}
\end{center}
  		\caption{Order parameters versus $\eta$ in the minimum-energy state of the Hamiltonian Eq.~\eqref{ham42:eqn} (Numerical 
		parameters $h=h'=0.5$, $J=J'=1.0$).}\label{siga4l:fig}
\end{figure}

As previously, we start from considering the properties of the minimum-energy point of the static part of this model 
(we find this point through a steepest descent algorithm). The results are reported in Fig.~\eqref{siga4l:fig}:
$\sigma_{[2]\,{\rm eq}}$ is always non vanishing, while $\sigma_{\rm eq}$ is nonvanishing only if $\eta$ is smaller than an $\eta_c$ which for this 
choice of parameters equals $0.8$. This means that the model breaks the $\mathbb{Z}_4$ symmetry for $\eta<\eta_c$ while it breaks only the 
$\mathbb{Z}_2$ symmetry otherwise.

\begin{figure}
\begin{center}
\begin{tabular}{c}
\hspace{0cm}
\includegraphics[width=8cm]{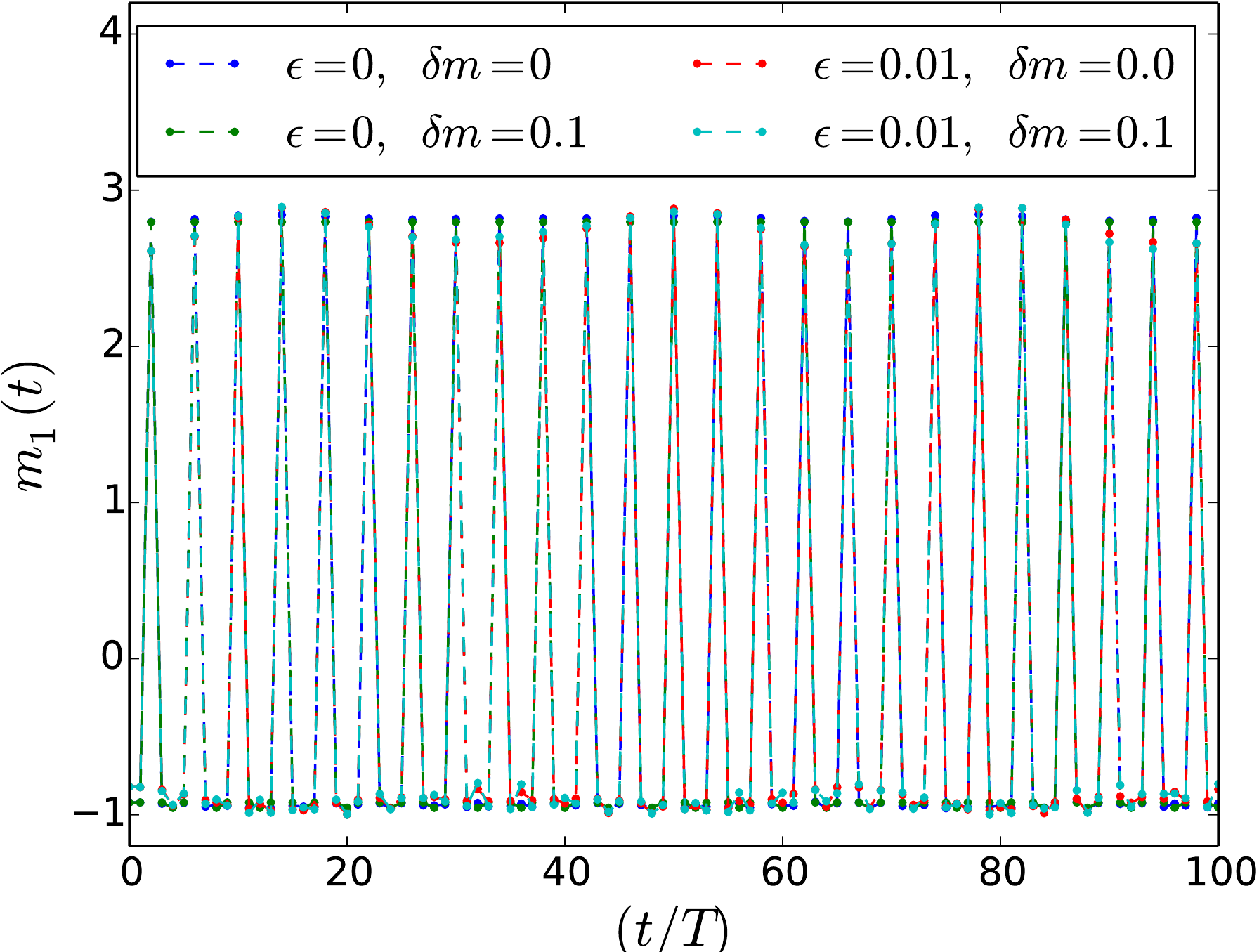}\\
\includegraphics[width=8cm]{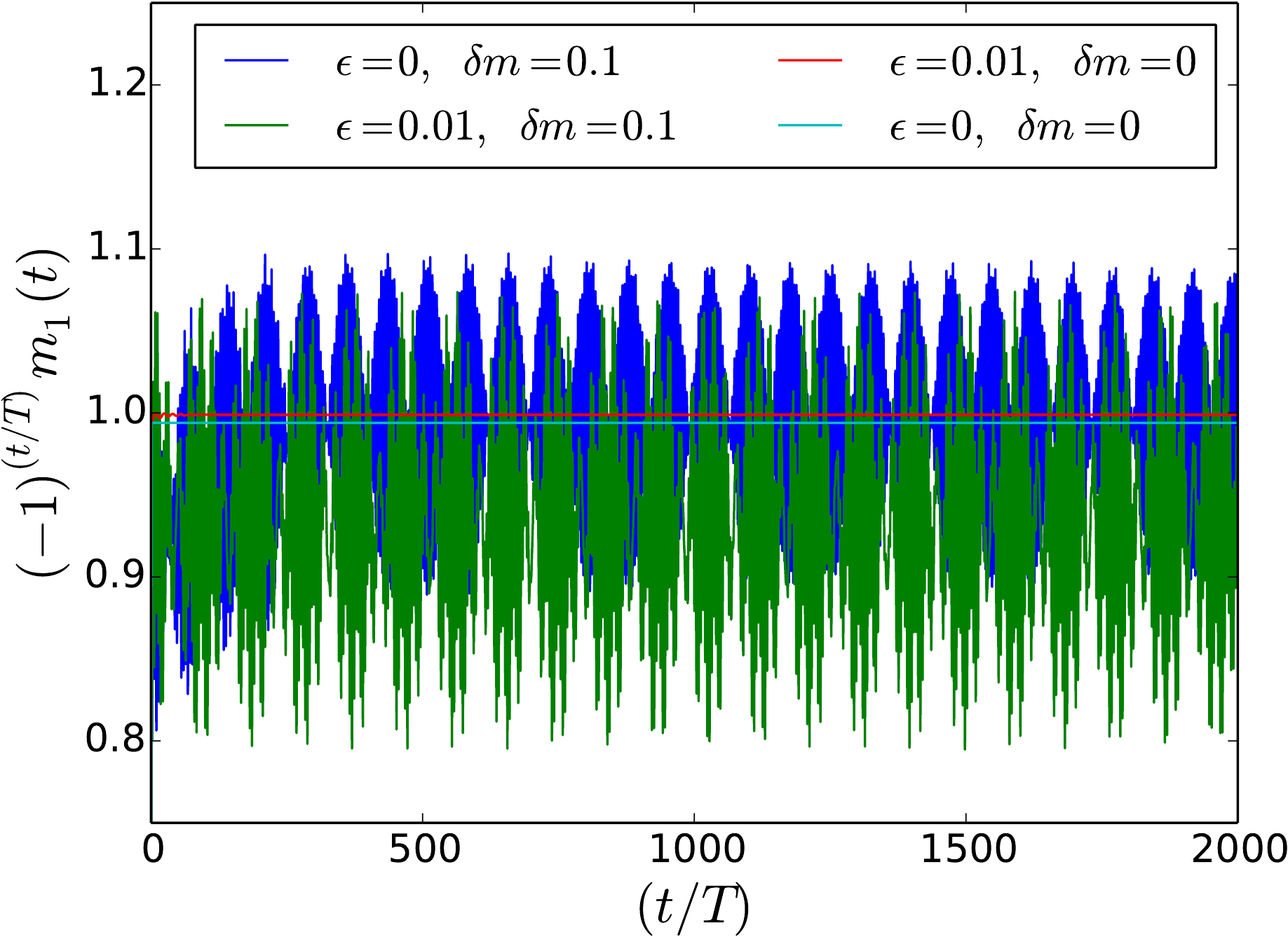}\\
\includegraphics[width=8cm]{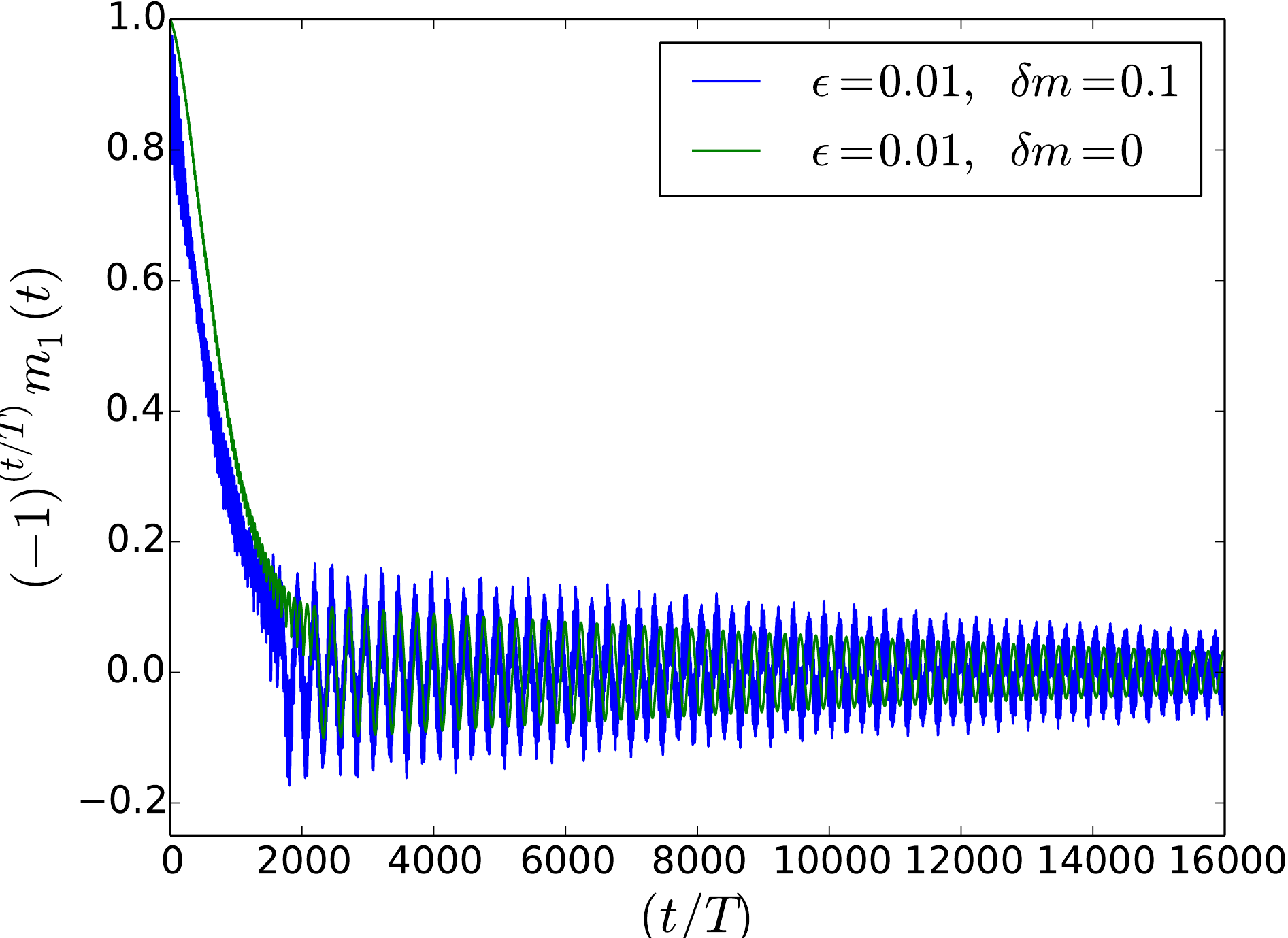}
\end{tabular}   
\end{center}
 		\caption{Evolution of $m_1(t)$ with the Hamiltonian 
 		Eq.~\eqref{ham42:eqn} and the kicking Eq.~\eqref{eq:kicko3e}. Both 
		period 4-tupling for $\eta=0.36$ (upper panel) and period doubling are present for $\eta=0.82$ (central panel; the factor $(-1)^k$ makes the period-doubling 
		oscillations to appear as an almost constant object). No time crystal whatsoever for $\eta=0.96$ and $\epsilon=0.01$ (lower panel). (Numerical 
		parameters $h=h'=0.5$, $J=J'=1.0$).}
\label{evoluzias:fig}
\end{figure}

The dynamics of $m_1(t)$ (the other $m_j$ behave exactly in the same way) in the presence of the kicking is shown in Fig.~\ref{evoluzias:fig}, where 
we consider different initial conditions, $m_{2}(0)=m_{2\,{\rm eq}}$, $m_1(0)=m_3(0)=m_{1\,{\rm eq}}+\delta m$. There 
are values of $\eta$ for which there is period 4-tupling (upper panel), and others for which there is period doubling (central panel). Taking a perturbed 
kicking with $\epsilon\neq 0$, there are value of $\eta$ where there is no time crystal (bottom panel). 

By looking at the properties of the Fourier transform, we see a value $\eta_c$ of $\eta$ where there is a direct transition from period 4-tupling to  period 
doubling. For $\epsilon=0$ this point coincides with the value of $\eta$ where ${\sigma}_{\rm eq}$ at equilibrium disappears (see Fig.~\ref{siga4l:fig}). 
For $\epsilon\neq 0$ another value $\eta_{c\,1}>\eta_c$ appears such that, for $\eta>\eta_{c\, 1}$, there is no time crystal behaviour. 
Three phases appear,  a period 4-tupling one, a period doubling one and a normal one. 

\begin{figure*}
\begin{center}
\begin{tabular}{cc}
\hspace{-2cm}
\includegraphics[width=8cm]{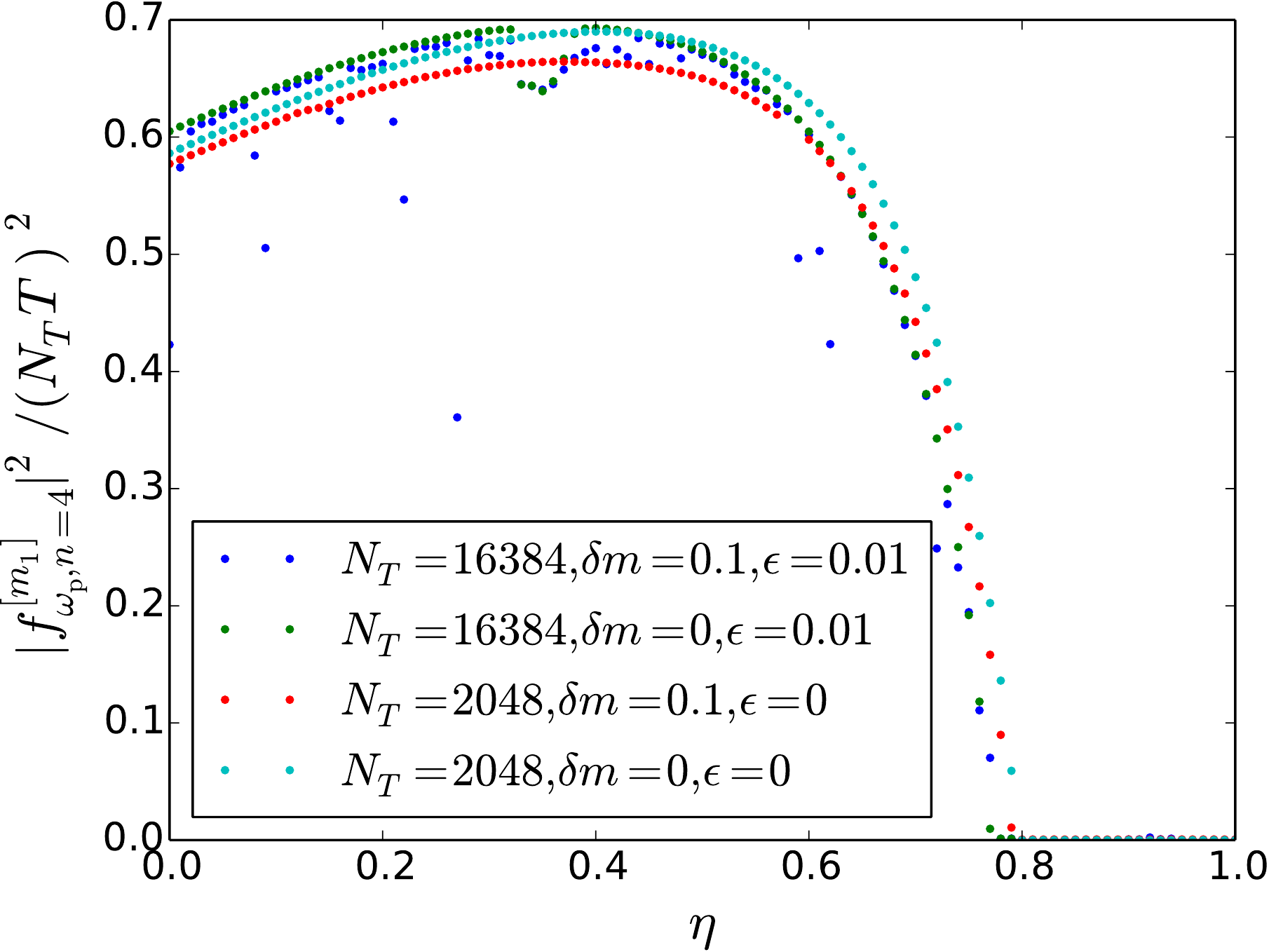}&
\includegraphics[width=8cm]{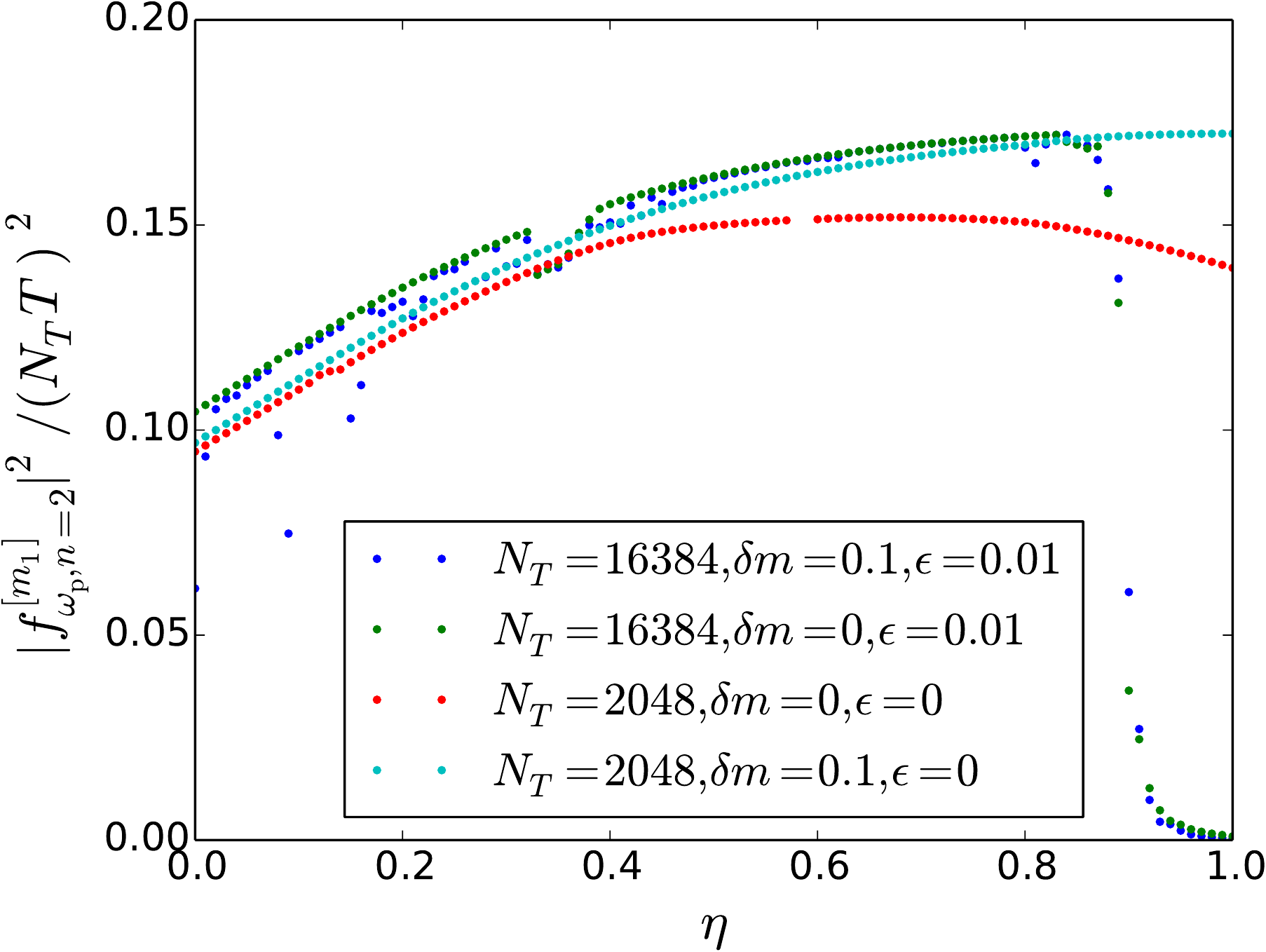}\\
\hspace{-2cm}
\includegraphics[width=8cm]{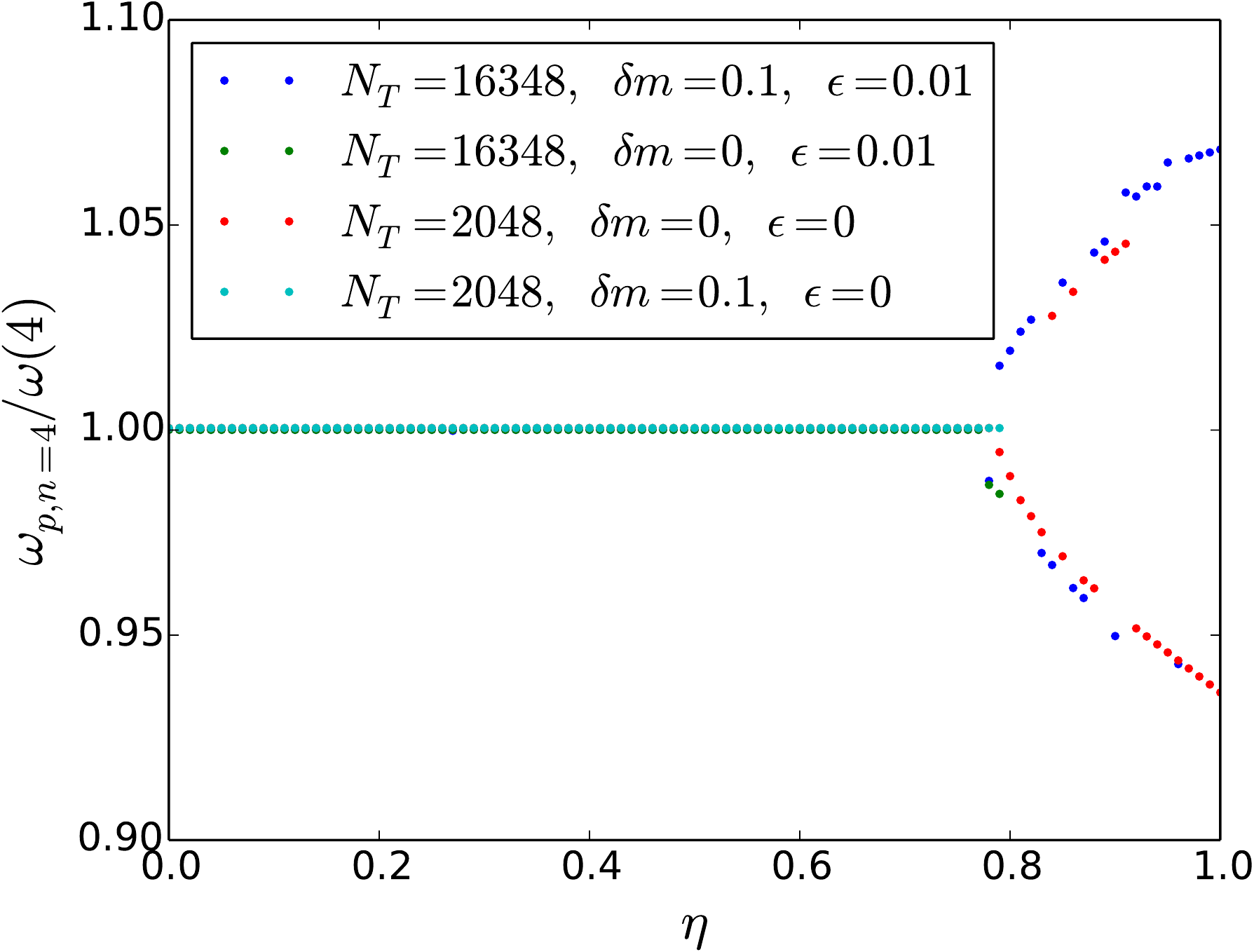}&
\includegraphics[width=8cm]{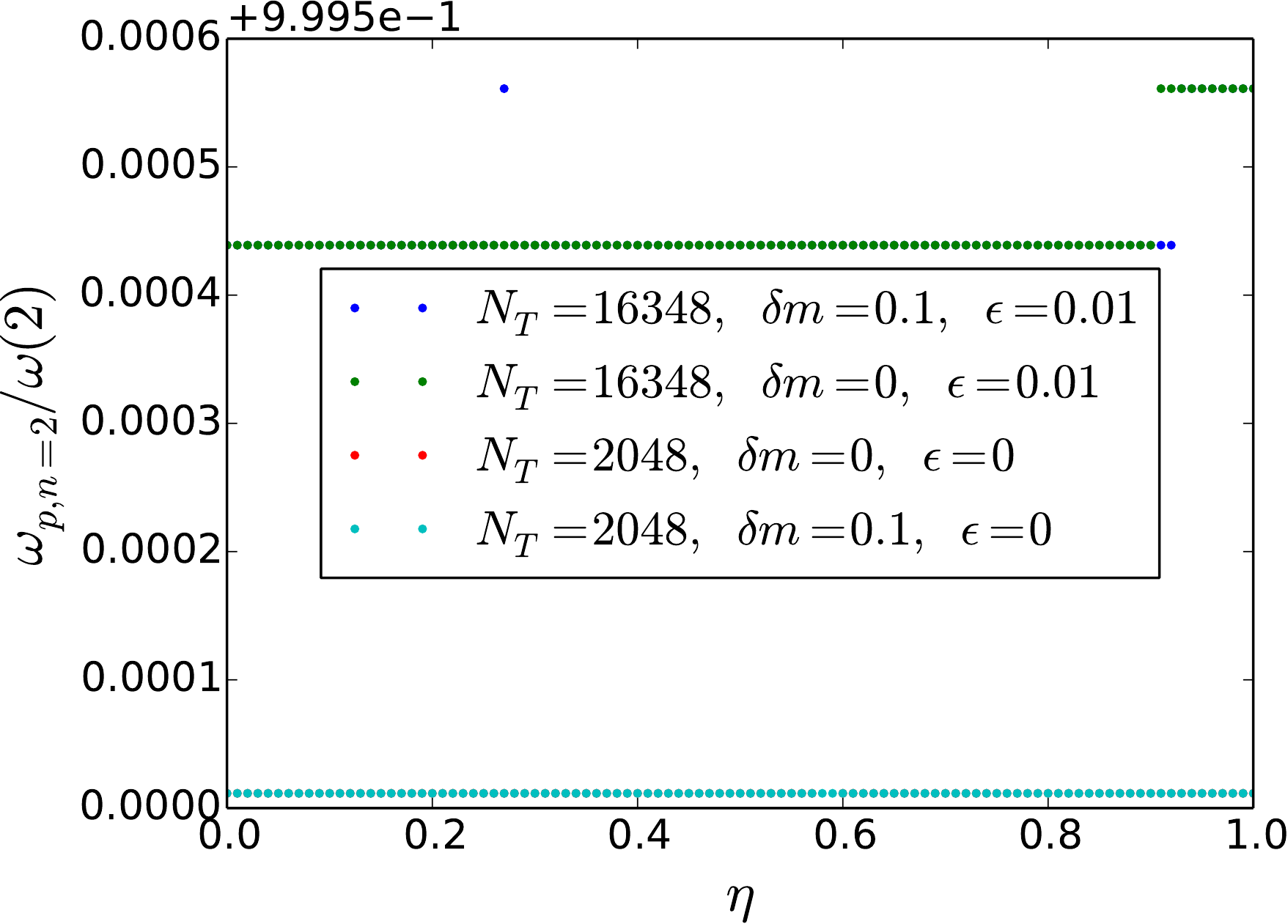}
\end{tabular}
\end{center}
 		\caption{ \textbf{(Upper left panel)} Peak in the Fourier transform at the period 4-tupling frequency ($|f_{\omega_p\,n=4}^{[m_1]}|$) versus $\eta$; 
		we notice that it disappears at a value $\eta_c \sim 0.8$ where the period 4-tupling time-crystal phase ends up. 
		\textbf{(Upper right panel)} Peak in the Fourier 
		transform at the period doubling frequency ($|f_{\omega_p\,n=2}^{[m_1]}|$) versus $\eta$; we notice that it disappears at a value $\eta_{c\,1} > \eta_c$ 
		when $\epsilon\neq 0$. For $\eta_c<\eta<\eta_{c\,1}$ there is a period-doubling time crystal phase, for $\eta>\eta_{c\,1}$ there is no time crystal. 
		\textbf{(Bottom left panel)} In the period 4-tupling phase the peak at the period 4-tupling frequency sticks at $\omega(4) = \pi/(2T)$. 
		\textbf{(Bottom right panel)} Peak at the period doubling frequency $\omega(2) = \pi/T$ (Numerical parameters: $h=h'=0.5$, $J=J'=1.0$).
		}
\label{pikkazzi:fig}
\end{figure*}

The first transition point at $\eta_c$ is marked by the disappearing of the peak in the Fourier transform at the period 4-tupling frequency 
($|f_{\omega_p,\,n=4}^{[m_1]}|^2$) (see the upper left panel of Fig.~\ref{pikkazzi:fig}). The peak at the period doubling frequency ($|f_{1\,\omega_p,\,n=2}^{[m_1]}|^2$) 
persists until $\eta_{c\, 1}$ (upper right panel of Fig.~\ref{pikkazzi:fig}). The first peak, in all the period 4-tupling phase, is locked at the period 4-tupling 
frequency $\omega(4)$ (lower left panel of Fig.~\ref{pikkazzi:fig}), while the second exists in both the time-crystal phases and is at frequency 
$\omega(2)$ (lower right panel of Fig.~\ref{pikkazzi:fig}). (This peak is not exactly at $\omega(2)$ due the finite number of period over which we 
analyse the dynamics; we have checked that it tends to the correct value if we perform the Fourier transform over a number of periods $N_T$ larger).
In the phase without time crystal, the position of the peak around the period doubling frequency slightly moves.  It is not however relevant for the dynamics, since 
its height is vanishingly small (see upper right panel of Fig.~\ref{pikkazzi:fig}). It is not surprising that in case of period 4-tupling there is a peak also at the 
period doubling frequency, being $\omega(2)$ one of the harmonics of $\omega(4)$. The remarkable thing is that the peak at $\omega(4)$ will disappear. 

This picture is stable if we slightly perturb the kicking with $\epsilon\neq 0$ and if we take an initial state different from the symmetry breaking ground 
state ($\delta m\neq 0$). As we said before, when $\epsilon\neq 0$ a trivial phase appears for $\eta>\eta_{c\,1}$. We emphasise that in this analysis 
the initial conditions we consider depend on $\eta$, because these initial conditions correspond to the minimum-energy point for that value of $\eta$ or 
some point around that minimum.

\section{Conclusions} 
\label{concola:sec}

We have studied a class of period-$n$-tupling discrete time crystals based on interacting models of $n$-clock variables. We have considered two different 
limits: a disordered short-range model and a clean infinite-range clock 
model.  

In the case of disordered short-range models the stability 
of the time crystal 
is provided by many-body localisation, which prevents the system from heating up 
to infinite temperature and makes possible the persistence of long-range order
 in the dynamics.
 We have analysed the features of these models combining analytical results 
 and perturbative arguments, and showed that the model 
 supports a time crystal when there are no degeneracies in
  its Floquet spectrum.
 In this case the main characterizing properties of a time crystal are robust to
  perturbations, namely, (i) the presence of Floquet states with 
long-range correlations, (ii) Floquet quasi-energies organized in $n$-tuplets,
 which are shifted from each other by the period-$n$-tupling frequency, 
 and (iii) an order parameter clock operator oscillating 
 with the period-$n$-tupling frequency.
 We have found that these properties are robust up to corrections exponentially 
 small in the system size.
This implies that they become exact in the thermodynamic limit were the time-translation symmetry breaking occurs. We have corroborated 
our theory with a numerical analysis 
for the case $n=3$, which shows a period-tripling time crystal,
 and for $n=4$, where we constructed a model showing period $4$-tupling in 
 one regime and period doubling in another one.

In the infinite-range case we have found that the interaction Hamiltonian has a phase where an extensive number of eigenstates breaks the 
$\mathbb{Z}_n$ symmetry in the thermodynamic limit and this was the basis for the stability of the 
period-$n$-tupling time crystal in such models. Due to its 
symmetry, generated by the invariance under permutation of its subsystems,
 the infinite-range model can be studied for larger system sizes, 
 allowing us to perform a precise finite size scaling analysis. 
 In fact, using its symmetries we have shown that the model could be mapped 
 over a bosonic model with $n$ sites whose occupation depends on 
 the system size. Within this picture we have 
 numerically studied the cases $n=3$ and $4$, showing in both cases 
 the existence of a time-translation symmetry breaking phase \textit{only} in the 
 thermodynamic limit, as appropriate for a time crystal.
 
  In the thermodynamic limit, we have also shown that the infinite-range model is 
  described by a classical effective Hamiltonian, where we have 
  studied its dynamics in more detail.
  We have showed exactly the existence of the time crystal for $n=3$ and $n=4$.
   Moreover, similarly to the short-range case, we have also constructed 
   a model whose static part could show a transition between period $n$-tupling and period $n/2$-tupling. We studied its properties in detail for the case with $n=4$.
 After showing the existence of the two time crystal phases by means
  of a finite-size scaling analysis, we used the 
  effective classical model in the thermodynamic limit to properly
  study their transition. 
    We have then verified that the model gives 
    rise to a \textit{direct} transition between 
    the time-crystal phase with period $n$-tupling to the one 
    with period $n/2$-tupling.
To the best of our knowledge, this represents the first   example in the literature of a direct
   transition between different time-crystal phases.

\acknowledgements{This work was supported in part by European Union through QUIC project (under Grant Agreement 641122)
and by the ERC under Grant No. 758329 (AGEnTh).}
 \appendix
  \section{Proof of Eqs.~\ref{avgH1} and \ref{avgH2}}\label{app1}
  \subsection{Case 1: $p$ and $n$ are coprime}
  Let us define for clarity $\hat H=\hat H_n^{(SR)}$
  By inserting a certain number of identities we can rewrite $\hat U^n$ as follows:
  \begin{align}
   \hat U^n=e^{-iT\hat H}\hat X^p e^{-iT\hat H} \hat X^{-p}\hat X^{2p}e^{-iT\hat H}\hat X^{-2p}\hat X^{3p}\dots\nonumber\\
   \dots \hat X^{np}e^{-iT\hat H} \hat X^{-np} \hat X^{np}=\nonumber\\
   =e^{-iT\hat H} e^{-iT\hat X^p\hat H\hat X^{-p}}e^{-iT\hat X^{2p}\hat H \hat X^{-2p}}\dots e^{-iT\hat X^{np}\hat H \hat X^{-np}}
  \end{align}

where we also used $\hat X^{np}=1$. Since all the exponentiated operators commute, we can write $\hat U^n=e^{-inT\bar{H}}$ with
\[\bar{H}=\frac{1}{n}\sum_{j=0}^{n-1} \hat X^{jp}\hat H\hat X^{-jp}\]
We note that $\hat H_n^{(SR)}$ contains the interaction terms, which are invariant under the transformation induced by $X^{-jp}$, and a longitudinal field containing operators $\hat \sigma_i^m$ (with $1<m<n-1$), which satisfy
\[\sum_{j=0}^{n-1} \hat X^{jp}\hat \sigma^m\hat X^{-jp}=\left(\sum_{j=0}^{n-1}\omega^{-jpm} \right)\hat \sigma^m.\]
If $n$ and $p$ are coprime, the sum in parentheses contains all the $n$-th roots of 1, so it vanishes. We obtain
\[ \bar{H}=\sum_i J_i \sum_{m=1}^{n-1} \alpha_m\, (\hat \sigma_i^\dagger \hat \sigma_{i+1})^m.\]

\subsection{Case 2: $p$ and $n$ have $gcd(p,n)=s>1$}
Similarly to the previous case we can use the fact that $\hat X^{qp}=1$ (with $q=n/s$) to rewrite $\hat U^q$ as
\begin{align}
\hat U^q&=e^{-iT\hat H} e^{-iT\hat X^p\hat H\hat X^{-p}}e^{-iT\hat X^{2p}\hat H\hat X^{-2p}}\dots\nonumber\\
&\dots e^{-iTX^{qp}\hat H \hat X^{-qp}}
\end{align}
We obtain that $\hat U^q=e^{-iqT\bar{H}}$ with
\[\bar{H}=\frac{1}{q}\sum_{j=0}^{q-1} \hat X^{jp}\hat H_n^{(SR)}\hat X^{-jp}.\]
As before, the interaction terms are not affected by the action of $X^{-jp}$, but the longitudinal field is. We see that
\[\sum_{j=0}^{q-1} \hat X^{jp}\hat \sigma^m\hat X^{-jp}=\left(\sum_{j=0}^{q-1}\omega^{-jpm} \right)\hat \sigma^m.\]
The sum in parentheses is equal to $q$ when $mp = 0 \Mod n$ (i.e. when $m$ is a multiple of $q$), it vanishes otherwise. Hence we get
    \[\bar{H} = \sum_i J_i \sum_{m=1}^{n-1} \alpha_m \,(\hat \sigma_i^\dagger \hat \sigma_{i+1})^m
     + \sum_i h_{z,i} \sum_{m=1}^{n/q-1} \gamma_{m q}\, \hat \sigma_i^{mq}.\]
  \section{Consequences of the quasi-adiabatic continuation}\label{app2}
    \subsection{Long range order}
  In this section we will generalize some results proven in \onlinecite{Keyserlingk2016} for the Ising model to the case of the clock model. In addition, we will use these generalized results to prove some important properties concerning time crystal order (persistence of oscillations, spectral properties), which were hinted to but not explicitly proven in \onlinecite{Keyserlingk2016}.

  The assumption that there exists a family of local unitaries $\hat V_\lambda$ (depending continuously on the perturbation strength $\lambda$), that connects perturbed and unperturbed eigenstates has many important consequences. First, as we now prove, it implies the stability of the long range order. Consider the perturbed eigenstates $\ket{{\psi}_\lambda(\{s_i\}, p)}=V_\lambda \ket{{\psi}_0(\{s_i\}, p)}$. 
  We define the dressed operators
 
\[\tilde{\sigma}_{i,\lambda}= \hat V_\lambda\sigma_i \hat V^\dagger_\lambda, \hspace{1cm}\tilde{\tau}_{i,\lambda}
=\hat V_\lambda\tau_i \hat V^\dagger_\lambda\]
It follows that

\begin{align}\label{lwalls}
\tilde{\sigma}^\dagger_{i,\lambda}\tilde{\sigma}_{i+1, \lambda}\ket{{\psi}_\lambda(\{s_i\}, p)}&=\hat V_\lambda\sigma^\dagger_i\sigma_{i+1}\ket{\psi_0(\{s_i\}, p)} \nonumber \\ 
&=s_i^*s_{i+1}\ket{{\psi}_\lambda(\{s_i\}, p)} 
\end{align}

The unitary $\hat V_\lambda$ is equivalent to the time evolution operator of a local Hamiltonian, as a consequence of the Lieb Robinson bound the dressed operators $\tilde{\sigma}_{i,\lambda}$ are exponentially localized. Therefore, Eq. \ref{lwalls} shows the existence of long range order.

\subsection{Persistent oscillations}
We proved that the eigenstates of $\hat U_{\lambda}$ are also eigenstates of $\tilde{\sigma}^\dagger_{i,\lambda}\tilde{\sigma}_{i+1,\lambda}$, hence
\begin{equation}\label{commlbits}
[\hat U_{\lambda},\tilde{\sigma}^\dagger_{i,\lambda}\tilde{\sigma}_{i+1,\lambda}]=0. 
\end{equation}
  
  Using the same argument as in~\onlinecite{Keyserlingk2016}, we now prove that
\begin{equation}\label{lbits}
 \hat U_{\lambda}^\dagger \tilde{\sigma}_{i,\lambda} \hat U_{\lambda} \simeq \omega^p \tilde{\sigma}_{i,\lambda}
\end{equation}
where Eq. (\ref{lbits}) is valid up to a correction that is exponentially small in the system size.

Let us consider the operator $\tilde{\sigma}^\dagger_{i,\lambda}\tilde{\sigma}_{j,\lambda}$ with $j>i$. This can be written as 
a product of ``l-wall'' operators between neighboring sites
\[\tilde{\sigma}^\dagger_{i,\lambda}\tilde{\sigma}_{j,\lambda}=(\tilde{\sigma}^\dagger_{i,\lambda}\tilde{\sigma}_{i+1,\lambda})(\tilde{\sigma}^\dagger_{i+1,\lambda}\tilde{\sigma}_{i+2,\lambda})\dots (\tilde{\sigma}^\dagger_{j-1,\lambda}\tilde{\sigma}_{j,\lambda})\]
Since each l-wall operator commutes with $\hat U_{\lambda}$, we have
\[[\hat U_{\lambda},\tilde{\sigma}^\dagger_{i,\lambda}\tilde{\sigma}_{j,\lambda}]=0.\]
We can rewrite this equation as
\begin{equation}
\hat U_{\lambda}^\dagger \tilde{\sigma}^\dagger_{i,\lambda}\tilde{\sigma}_{j,\lambda} \hat U_{\lambda} 
=(\hat U_{\lambda}^\dagger \tilde{\sigma}^\dagger_{i,\lambda} \hat U_{\lambda}) (\hat U_{\lambda}^\dagger \tilde{\sigma}_{j,\lambda} \hat U_{\lambda})=\tilde{\sigma}^\dagger_{i,\lambda}\tilde{\sigma}_{j,\lambda}
\end{equation}
We can further manipulate this last equation by taking to the left side the operators localized in $i$ and on the right side the operators localized in $j$. We obtain
\begin{equation}\label{localoper}
 \tilde{\sigma}_{i} (\hat U_{\lambda}^\dagger \tilde{\sigma}^\dagger_{i,\lambda} \hat U_{\lambda}) =\tilde{\sigma}_{j,\lambda} (\hat U_{\lambda}^\dagger \tilde{\sigma}_{j,\lambda}^\dagger \hat U_{\lambda})
\end{equation}
We already argued that $\tilde{\sigma}_{i,\lambda}$ is exponentially localized around the site $i$. The operator $\hat U_{\lambda}^\dagger \tilde{\sigma}^\dagger_{i,\lambda} \hat U_{\lambda}$ is also localized because 
it can be obtained from the localized operator $\tilde{\sigma}_{i,\lambda}$ by evolving it 
for a time $T$ with a local time-dependent Hamiltonian. Therefore, we still expect 
that $\hat U_{\lambda}^\dagger \tilde{\sigma}^\dagger_{i,\lambda} \hat U_{\lambda}$ decays exponentially with the distance from the site $i$.

From Eq. (\ref{localoper}) we deduce that the two unitary operators $\tilde{\sigma}_{i} (\hat U_{\lambda}^\dagger \tilde{\sigma}^\dagger_i \hat U_{\lambda})$ 
and $\tilde{\sigma}_{j} (\hat U_{\lambda}^\dagger \tilde{\sigma}_j^\dagger \hat U_{\lambda})$ are equal, even though they are localized possibly far apart on the chain. The distance between $i$ and $j$ can be of order $L$. In the thermodynamic limit, the only possibility is that these two operators are c-numbers.
More precisely, they are unitary so they must be phases. If the system has a finite size $L$, the exponential localization of the two operators implies that a correction of order $O(e^{-cL})$ can be present (where $c$ is a constant that depends on the localization length of the operators). It follows that
\begin{equation}\label{phasetheta}
 \hat U_{\lambda}^\dagger \tilde{\sigma}_i \hat U_{\lambda} = e^{i\theta} \tilde{\sigma}_i+O(e^{-cL}).
\end{equation}

Taking the $q$-th power of Eq.(\ref{phasetheta}) in the thermodynamic limit we have
\[\hat U_{\lambda}^\dagger \tilde{\sigma}_{i,\lambda}^n \hat U_{\lambda} = e^{in\theta} \tilde{\sigma}_{i,\lambda}^n.\]
From $\tilde{\sigma}_{i,\lambda}^n=1$, it follows that $e^{in\theta}=1$, so $e^{i\theta}$ can only assume one of the $n$ values $1,\omega,\dots, \omega^{n-1}$.
 
To determine the value of $\theta$ we consider a special case: when the perturbation 
is absent ($\lambda=0$), $\tilde{\sigma}_{i,\lambda}$ reduces to $\sigma_i$ and $\hat U_{\lambda}$ reduces 
to $\hat U_{0}$. In this case, Eq.(\ref{phasetheta}) is satisfied by $e^{i\theta}=\omega^p$:
\[\hat U_{0}^\dagger \sigma_i \hat U_{0} = \omega^p \sigma_i.\]

We assumed that $\hat V_\lambda$ depends continuously on the parameter $\lambda$. Hence, all the dressed quantities also depend continuously on $\lambda$. As a consequence, the phase $e^{i\theta}$ cannot change abruptly from $\omega^p$ to the other possible values $1$, $\omega, \dots, \omega^{n-1}$ as $\lambda$ is turned on. We must conclude that for every $\lambda$ we find $e^{i\theta}=\omega^p$. We get

\[ \hat U_{\lambda}^\dagger \tilde{\sigma}_i \hat U_{\lambda}= \omega^p \tilde{\sigma}_i+O(e^{-cL}).\]

This implies that $\tilde{\sigma}_i(mT) = \hat U_{\lambda}^{-m} \tilde{\sigma}_i \hat U_{\lambda}^m= \omega^{mp} \tilde{\sigma}_i+mO(e^{-cL})$, meaning that oscillations persist at least up to a time that is exponentially large in $L$.

We can further argue that the undressed operator $\sigma_i$ has an expansion in terms of the dressed operators of the form
\[\sigma_i=c_i\tilde{\sigma}_{i,\lambda}+\dots\]
where $c_i\simeq O(1)$ and the other terms are exponentially localized around the position $i$. It follows that
\[\sigma_i(mT) \sigma_i^\dagger(0)=|c_i|^2\tilde{\sigma}_{i.\lambda}(mT) \tilde{\sigma}_i^\dagger(0)+\dots\]
As a consequence, while $\tilde{\sigma}_{i\lambda}$ oscillates with amplitude 1, the oscillations of $\sigma_i$ will have an amplitude $|c_i|^2<1$ for not too large times. The additional oscillations given by the other terms of the sum will average to 0 when we consider different disorder realizations. Hence we expect $\overline{\braket{\sigma_i}}$ to have finite amplitude oscillations, decaying to 0 after a time $t^* \sim O(e^{cL})$.

\subsection{Spectral properties}
In the exactly solvable case we showed that Floquet eigenstates are found in multiplets with $2\pi/q$ quasi-energy splitting. We are now going to show that this also happens for the perturbed system in the thermodynamic limit as long as we are in the time crystal regime.

Eq.(\ref{lbits}) implies that $[{\hat U_{\lambda}}^q,\tilde{\sigma}_{i,\lambda}]=O(e^{-cL})$, which means 
that the $\tilde{\sigma}_{i,\lambda}$ are approximate constants of motion in the stroboscopic evolution with 
period $qT$ for finite size systems. Only in the limit $L\rightarrow \infty$ they become exact constants 
of motion. Since all the $\tilde{\sigma}_{i,\lambda}$ commute among themselves and 
(approximately) commute with ${\hat U_{\lambda}}^q$, it follows that the 
transformed states $\hat V_{\lambda}\ket{\{s_i\}}$, being eigenstates of all 
the $\tilde{\sigma}_{i,\lambda}$, are (approximate) eigenstates of ${\hat U_{\lambda}}^q$. 
The $q$ states $\hat V_\lambda\ket{\{s_i\}}$, $\hat V_\lambda\ket{\{\omega^p s_i\}}$,$\dots$,
$\hat V_\lambda\ket{\{\omega^{p(q-1)} s_i\}}$ are linear combinations of 
the $q$ Floquet eigenstates $\ket{{\psi}_\lambda(\{s_i\}, k)}$ with $k=0,1,\dots, q-1$ defined 
in section \ref{solvn}. But Floquet eigenstates are, by definition, also eigenstates 
of ${\hat U_{\lambda}}^q$: a linear combination of them can be 
an eigenstate of $\hat U_{\lambda}^q$ only if they are degenerate (with respect 
to  ${\hat U_{\lambda}}^q$). This means that, in thermodynamic limit, the 
$q$ Floquet eigenstates $\ket{{\psi}_\lambda(\{s_i\}, k)}$ must have the same eigenvalue that 
we denote $\exp{(-qi\tilde{E}^+(\{s_i\}))}$.
\[ {\hat U_{\lambda}}^q\ket{\tilde{\psi}(\{s_i\}, k)}=e^{-qi\tilde{E}^+(\{s_i\})} \ket{\tilde{\psi}(\{s_i\}, k)}\]

Therefore, they can have as eigenvalues of $U_{f,\lambda}$ one of the $q$-th roots of $\exp{(-qi\tilde{E}^+)}$: $\exp{(-i\tilde{E}^+)}$, $\omega^p\exp{(-i\tilde{E}^+)}$, $\dots$, $\omega^{p(q-1)}\exp{(-i\tilde{E}^+)}$. Hence, the possible values of the quasi-energy gaps are $0$, $2\pi/q$, $\dots$, $2\pi(q-1)/q$. Using the continuity of the unitary $V_\lambda$, we can deduce that the gaps can only change continuously: since they can only assume one of the $q$ discrete values, they cannot change at all.
This proves that the exact $2\pi/q$ splitting is preserved in the thermodynamic limit. For finite size systems, this fact is only valid up to corrections of the order $O(e^{-cL})$.

\section{Disordered $\mathbb{Z}_4$ clock model: from period 2 to period 4}\label{app:plot}
\begin{figure*}
\includegraphics[width=.35\textwidth]{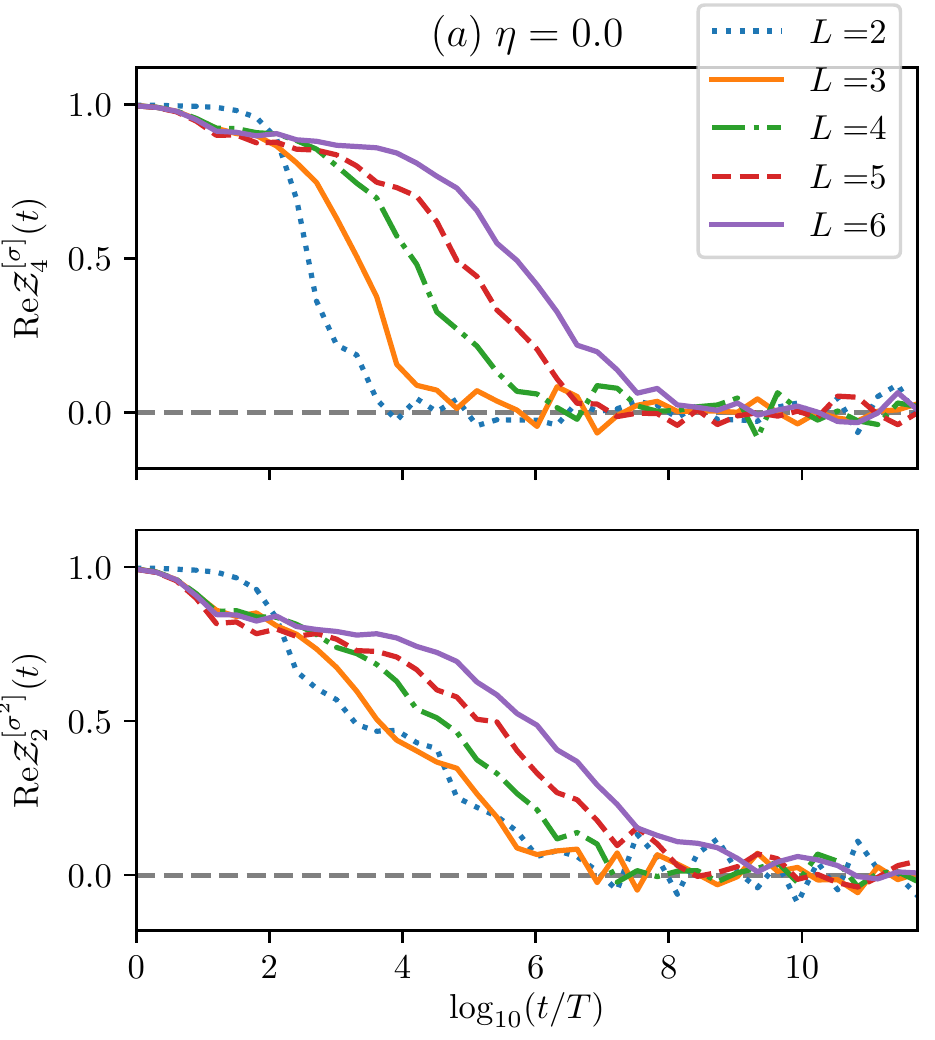}
\includegraphics[width=.35\textwidth]{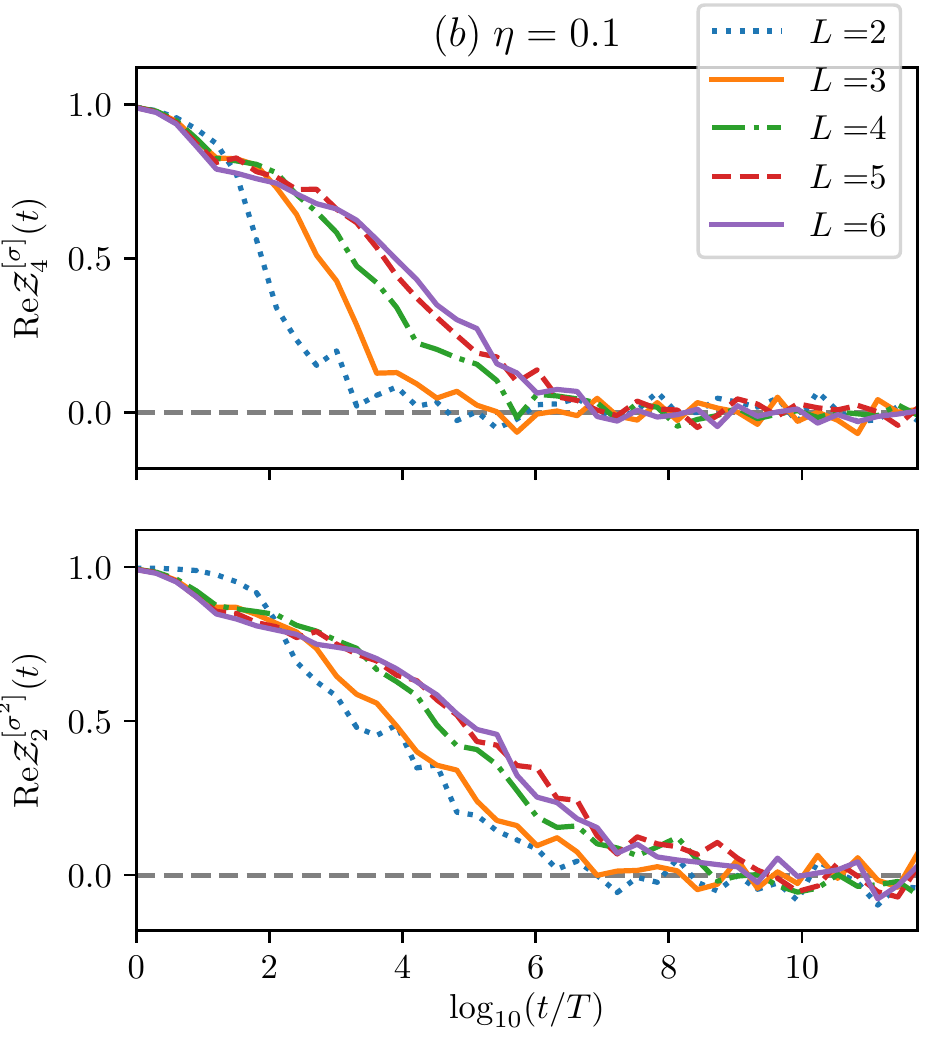}
\includegraphics[width=.35\textwidth]{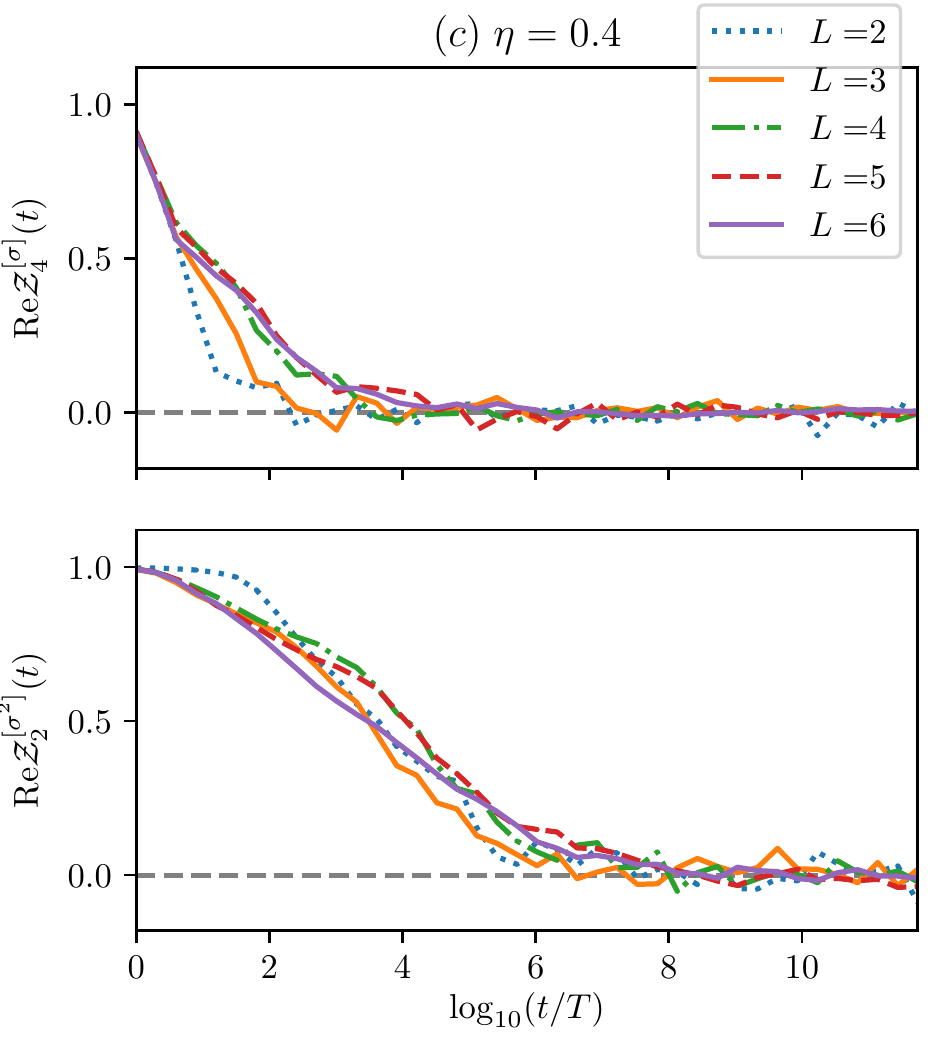}
\includegraphics[width=.35\textwidth]{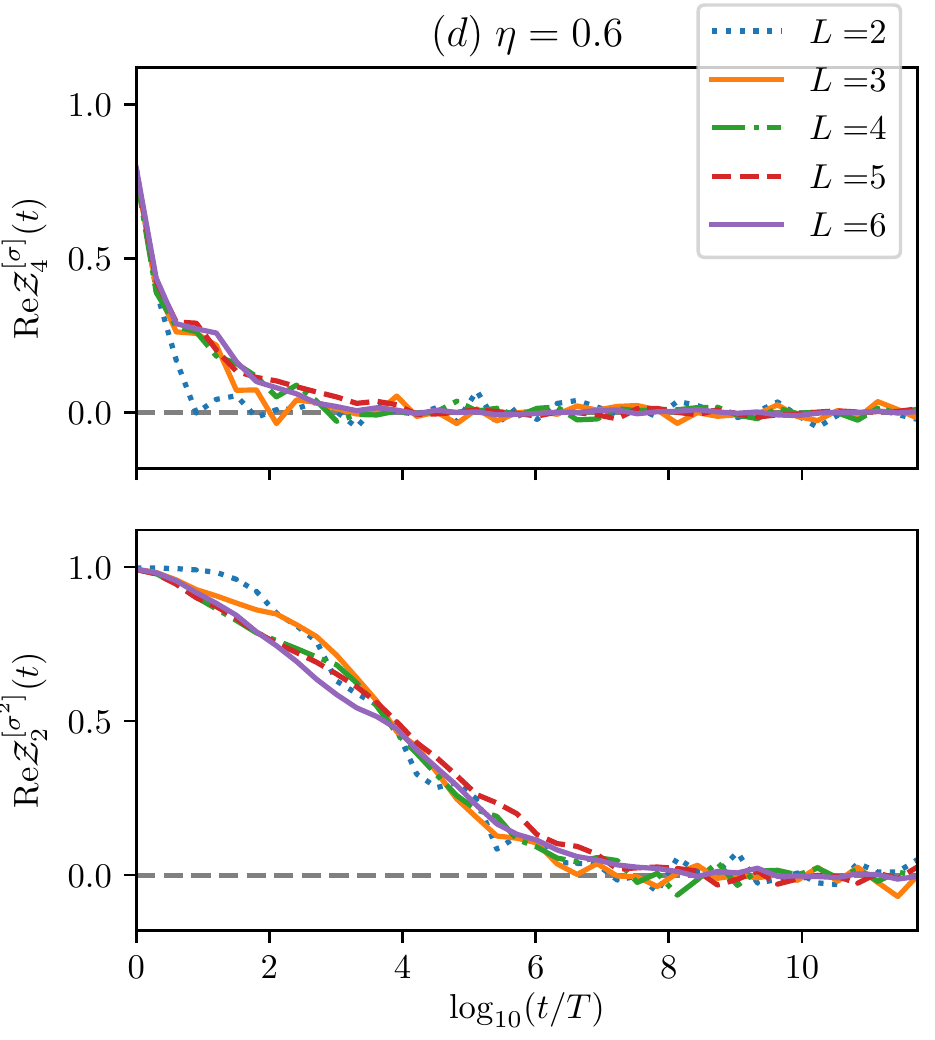}
\includegraphics[width=.35\textwidth]{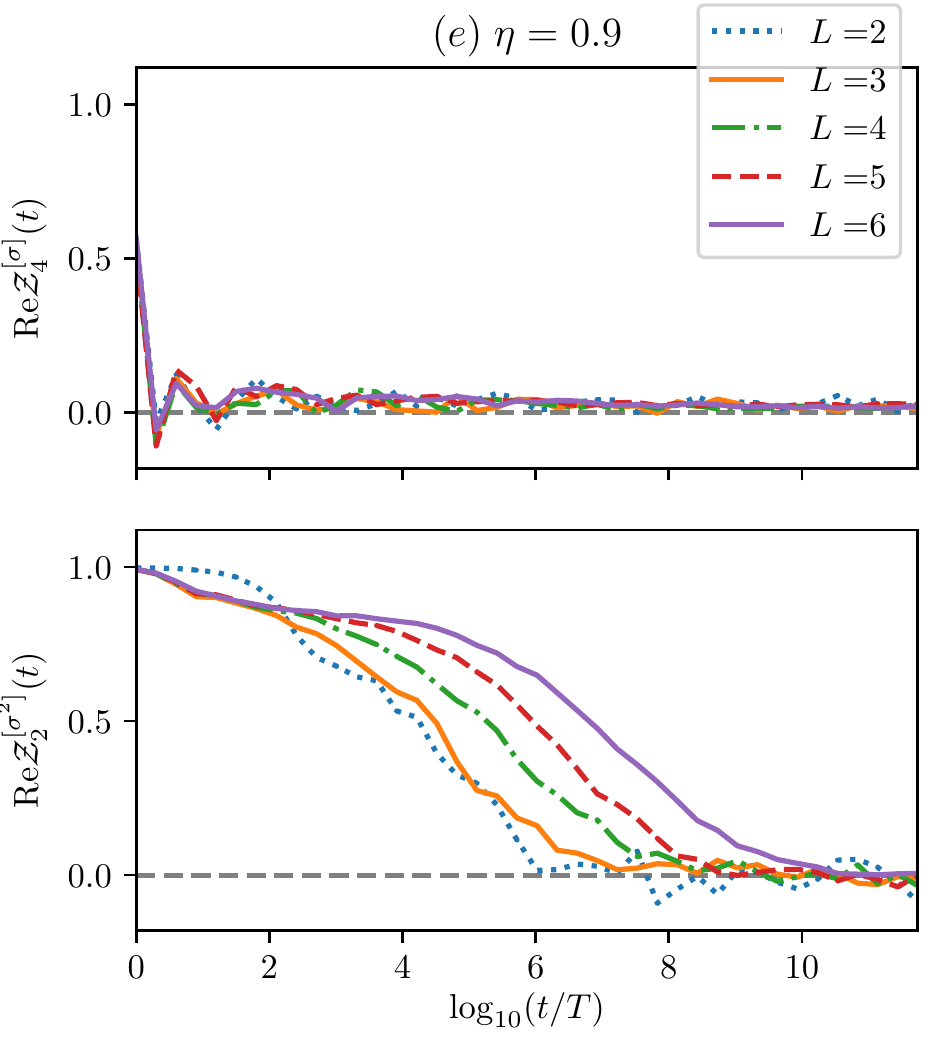}
\includegraphics[width=.35\textwidth]{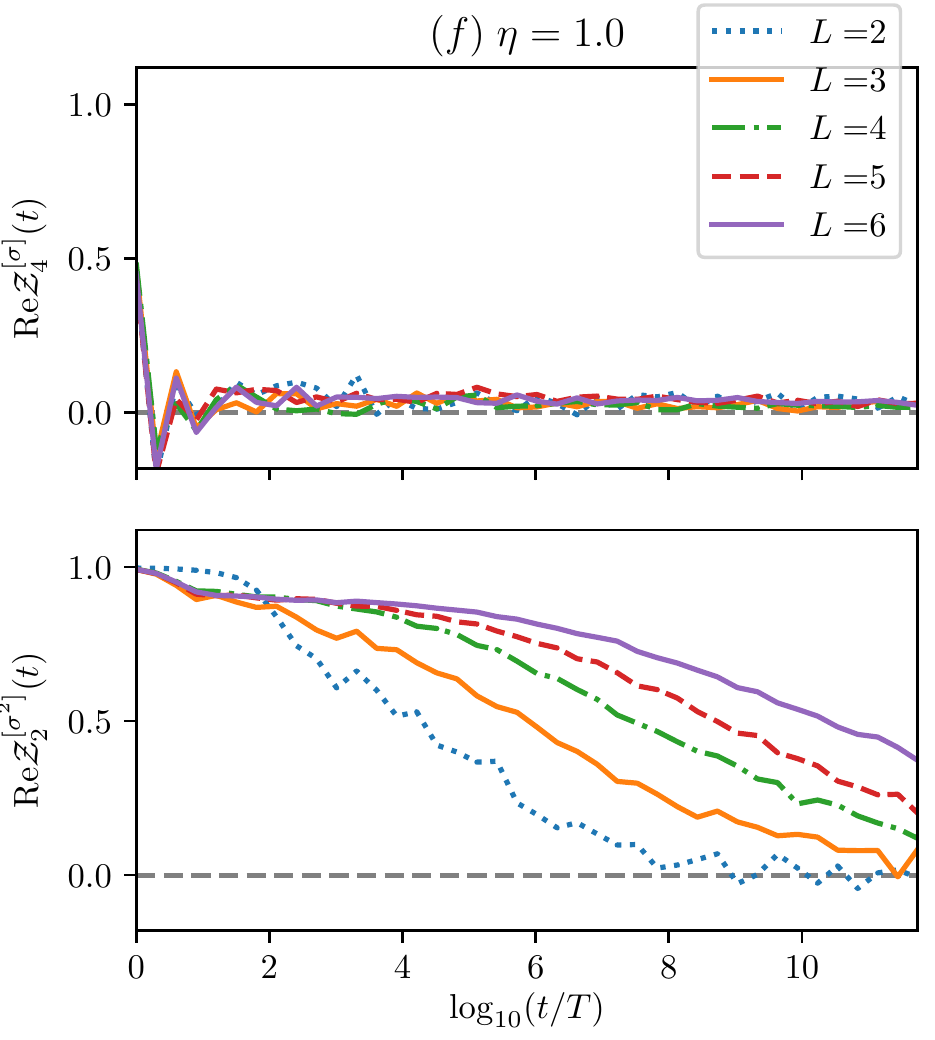}
    \caption{ Time evolution of the order parameters $Z(t)$ (period $4$ time crystal) and
     $Z_{[2]}(t)$ (period doubling time crystal), for varying $\lambda$ parameters. Results are obtained with the following choice of parameters: $J_i$ from the uniform distribution $[1/2,3/2]$, $h_{z,i}$ from $[0,1]$, $g_i$ from $[0,1]$, $\alpha_1=\frac{e^{i\pi/3}}{2}$, $\epsilon=0.1$
      }
    \label{fig.z4toz2complete}
\end{figure*}

Supplementary numerical results for section \ref{sec:disorder4} are shown in Fig.~\ref{fig.z4toz2complete}.

  \section{Mapping to a bosonic representation}
  \label{app_bosonazzi:sec}
  We start defining the symmetrization operator $\hat P$ for our system with $L$ subsystems, each one 
  composed by a clock variable of order $n$, as 
\begin{equation}
 \hat P = \sum_{\{ \hat \Pi \}} \hat \Pi_{j_1,...,j_L },
\end{equation}
 where $j_i = 1,...,L$ and $\hat \Pi_{j_1,...,j_L }$ permute the subsystems according
  to the $j_i$ indexes. As an illustrative example,
$\hat \Pi_{1,3,2} | \sigma_1 \sigma_2 \sigma_3 \rangle = | \sigma_1 \sigma_3 \sigma_2 \rangle$,
 where $\sigma_j=1,\omega,...,n-1$ represents the direction of the $j$'th clock spin.
 
We know that symmetric subspace for a Hilbert space with $L$ subsystems 
can always be represented, in second quantization, in terms of 
bosonic operators $\{\hat b_j\}$ (Eq.\eqref{eq:bosons.commutation}).
We then define a basis $\{|n_1,n_2,...,n_n\rangle \}$ for this subspace as follows:
\begin{widetext}
   \begin{eqnarray}
    |n_1 n_2 ... n_n \rangle &\equiv& 
    \frac{1}{\sqrt{N! \prod_{k=1}^n n_k!}} \hat P 
    | (1...1)_{n_1} (\omega ... \omega)_{n_2} ... (\omega^*...\,\omega^*)_{n_n} \rangle 
    \label{eq:def.symmbasis1}\\
    &=& 
    \frac{1}{\sqrt{N! \prod_{k=1}^n n_k!}} (\bd{1})^{n_1}(\bd{2})^{n_2}...(\bd{p})^{n_n} |vac\rangle \label{eq:def.symmbasis2}
   \end{eqnarray}
\end{widetext}
   where the index $n_j$ represents the number of clock operators in the $\sigma_j$ direction, 
   or alternatively, the number of bosons in the $j$'th bosonic mode,
    and $N = \sum_{j=1}^n n_j$  is the total number of bosons.
   
   Since the Hamiltonian is invariant under permutation, therefore commuting with $\hat P$,
   the study of its representation in bosonic language becomes significantly simpler:
    we must simply analyse how it acts in a single representative clock spin 
    configuration (right side of Eq.\eqref{eq:def.symmbasis1}).
  
  The bosonic representation for the operator $\hat \sigma = (1/L) \sum_j \hat \sigma_j$ is obtained by
  \begin{widetext}
  \begin{eqnarray}
   \hat \sigma |n_1 n_2 ... n_n \rangle &=&  
     \hat P \frac{1}{\sqrt{N! \prod_{k=1}^n n_k!}} \hat \sigma 
    | (1...1)_{n_1} (\omega ... \omega)_{n_2} ... (\omega^*...\,\omega^*)_{n_n} \rangle  \\
    &=& (1/L) (\sum_{j=1}^n n_j \omega^{j-1}) |n_1 n_2 ... n_n \rangle
  \end{eqnarray}
  \end{widetext}
  where in the first line we used commutativity between $\hat P$ and $\hat \sigma$.
Thus, we clearly see that 
\begin{equation}\label{eq:Sz.operator.bosons}
 \hat \sigma = (1/L) \sum_j \hat n_j \omega^{j-1},
\end{equation}
where $\hat n_j = \bd{j} \hat b_j$. The operator $\hat R = \sum_j \hat \tau_j$ follows analogously,
\begin{widetext}
\begin{eqnarray}
 \hat R |n_1 n_2 ... n_p \rangle &=& \hat P \frac{1 }{\sqrt{N! \prod_{k=1}^n n_k!}} \hat R
 | (1...1)_{n_1} (\omega...\omega)_{n_2} ... (\omega^*...\omega^*)_{n_n} \rangle \\
 &=&  \frac{\hat P }{\sqrt{N! \prod_{k=1}^n n_k!}}
 \left( n_1 | (1...1)_{n_1-1} (\omega...\omega)_{n_2+1} ... (\omega^*...\omega^*)_{n_n} \rangle  \right. \\
 & & \left. + n_2 | (1...1)_{n_1} (\omega...\omega)_{n_2-1} (\omega^2...\omega^2)_{n_3+1} ... (\omega^*...\omega^*)_{n_p} \rangle + ... \right) \\
 &=& \sum_j \sqrt{n_j} \sqrt{n_{j+1} + 1} |\,...\,(\omega^j...\,\,\omega^j)_{(n_j-1)} 
 (\omega^{j+1}...\,\,\omega^{j+1})_{(n_{j+1}+1)}\, ...\,\rangle
 \end{eqnarray}
 \end{widetext}
 Thus, 
 \begin{equation}
  \hat R = \sum_j \hat b_{j} \bd{j+1}
 \end{equation} 
 
 Exactly the same reasoning follows for the operators
 $\hat \sigma_{[2]} = (1/L)\sum_j \hat \sigma_j^2 $ and $\hat R_{[2]} = \sum_j \hat \tau_j^2$.
 We see in this case that
 \begin{eqnarray}
\hat{\sigma_{[2]}} &=& \frac{1}{L} \sum_{j=1}^n \hat n_j \omega^{2(j-1)}), \\
\hat R_{[2]} &=& \sum_{j=1}^n \hat b_{j} \bd{j+2} 
\end{eqnarray}

 The unperturbed kicking operator $\hat X_{\epsilon = 0} = \prod_j \hat \tau_j$ acts as
 \begin{equation}
 \hat X_{\epsilon = 0} |n_1 n_2 ... n_n \rangle = |n_p n_1 ... n_{n-1} \rangle
 \end{equation}
 and is thus described as a global translation of a single mode ($j \rightarrow j+1$) in the bosonic system.
 
 Global Hamiltonian terms which are invariant under permutation, such as the kicking
  operator with perturbations, can also be easily described in bosonic language.
 Consider a general unitary operator $\hat V^{\rm{[sp]}}$ acting in all 
 of the $L$ clock operators, as follows,
\begin{equation}
 \hat O_{\rm{global}} = \hat V^{\rm{[sp]}^{\otimes L} }
\end{equation}
This operator is translated to a single particle bosonic transformation in the
 bosonic language,
 \begin{equation}
  \hat b_j' = \sum_\ell \hat  V^{\rm{[sp]}}_{\ell,j} \hat b_\ell
 \end{equation}

\section{Spontaneous symmetry breaking in the infinite-range case}\label{subsec.SSB.LR}

\begin{figure*}
\centering
\begin{overpic}[width=0.325\linewidth]{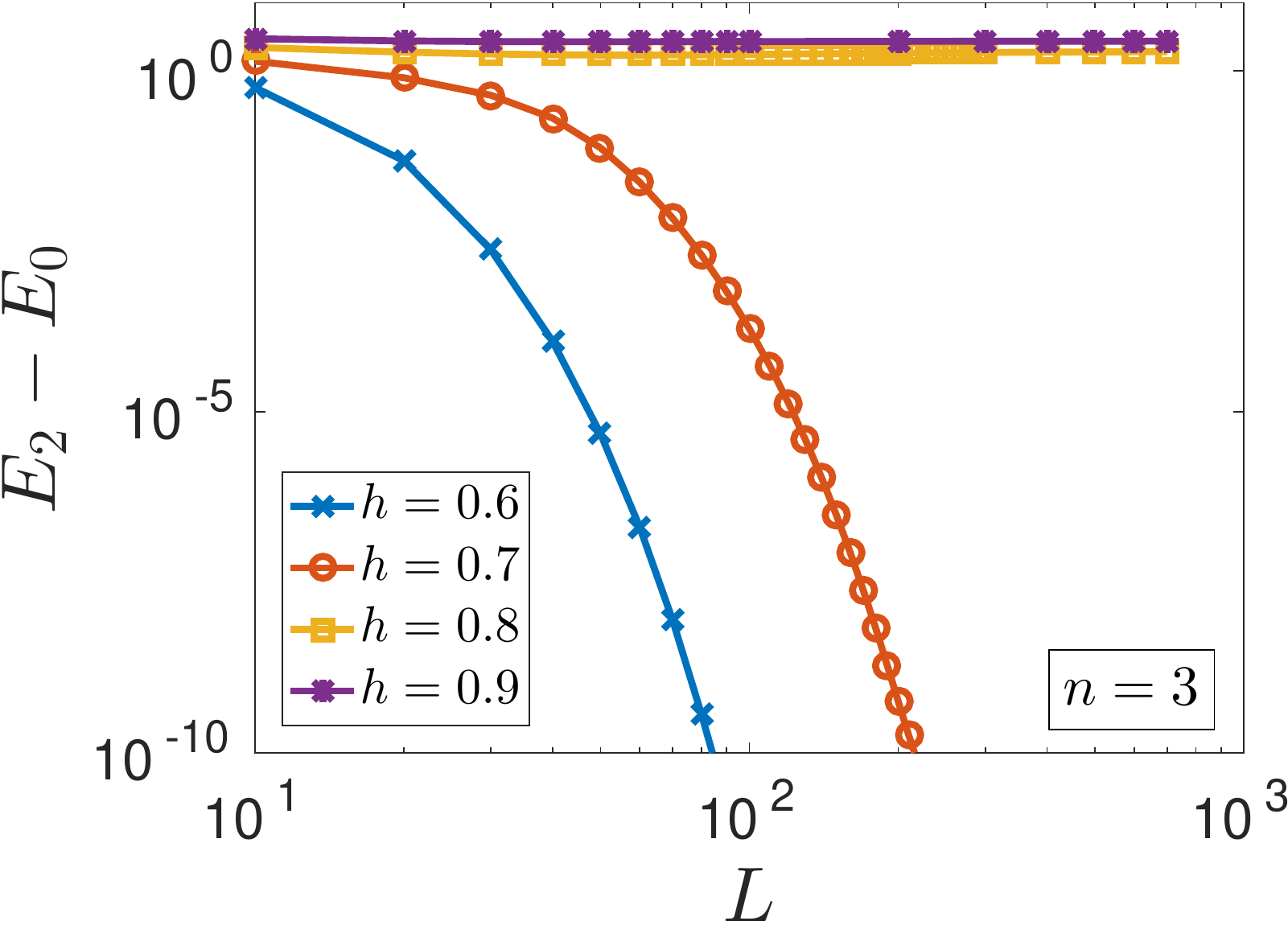}\put(-1,71){(a)}\end{overpic}
\begin{overpic}[width=0.325\linewidth]{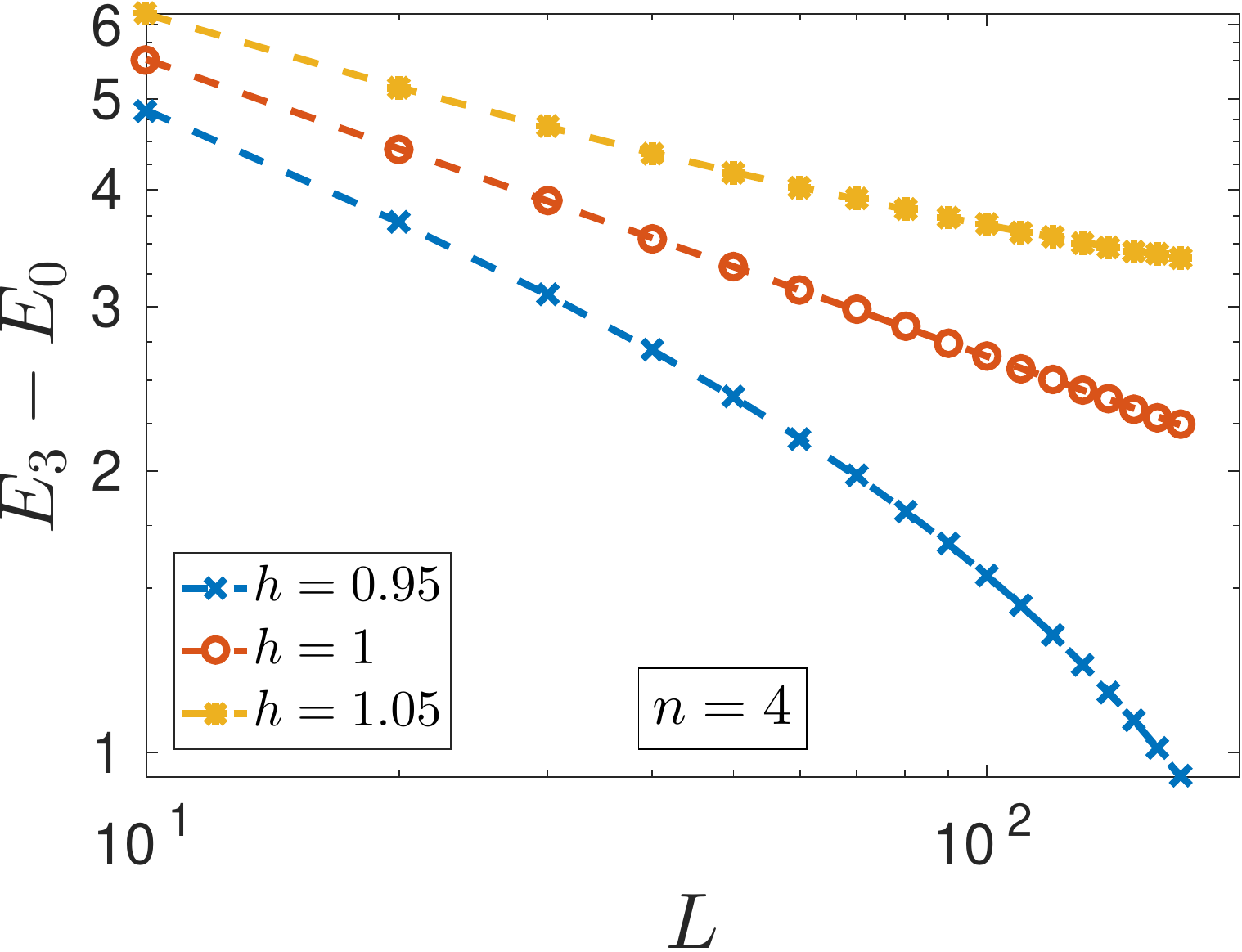}\put(-1,71){(b)}\end{overpic}
\begin{overpic}[width=0.325\linewidth]{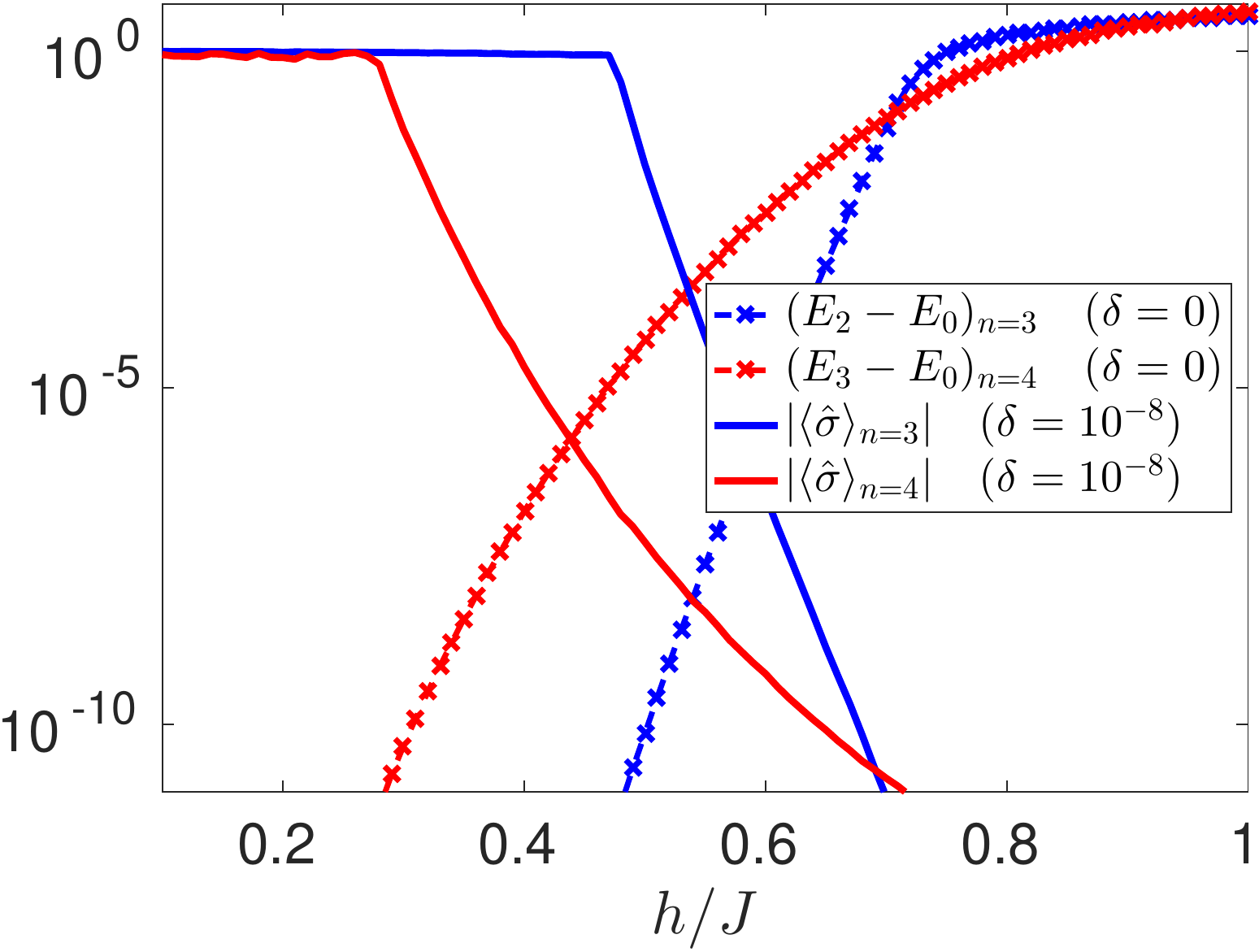}\put(-1,75){(c)}\end{overpic}
 \caption{
 (a) and (b) Scaling of the $n$-fold gap $E_{n-1} - E_0$ with the system size, for $\eta=0$  and
 $n=3$ and $4$, respectively.
  (c) Under a small perturbation $\hat W$ explicitly breaking the 
  symmetry of the Hamiltonian,
 the (non-degenerated) ground state
  acquires a macroscopic value for the order parameter $\langle \hat \sigma_i \rangle_{\rm GS} \sim 1$
   in the region where the $n$-fold gap is roughly smaller than the perturbation.
  System sizes: $L=50$ and $30$ for $n=3$ and $4$, respectively;
  $J=1$ in all the plots.
   }
 \label{fig.ssb}
 \end{figure*}
\begin{figure*}
\centering
\includegraphics[width=0.325\linewidth]{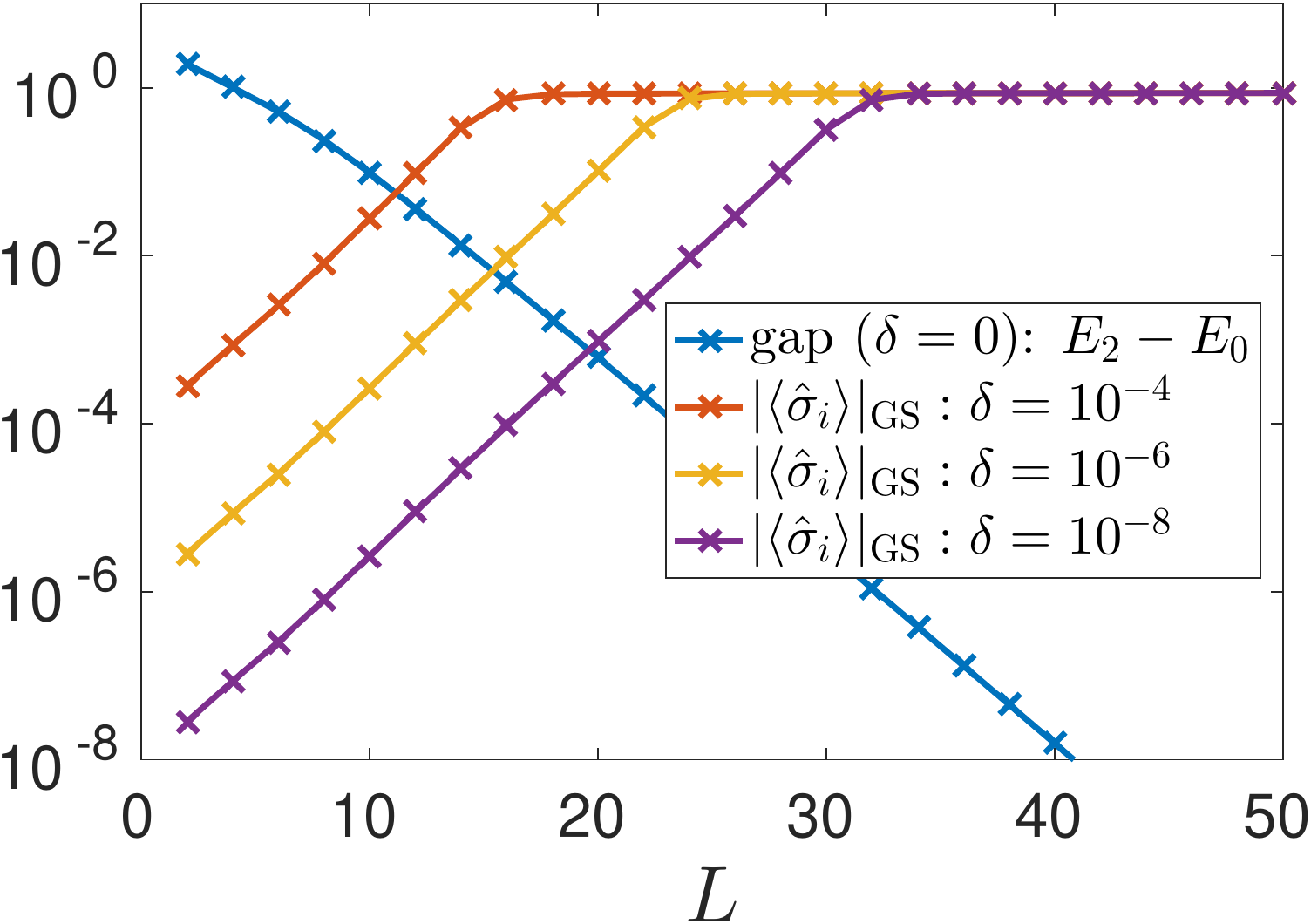}
\includegraphics[width=0.325\linewidth]{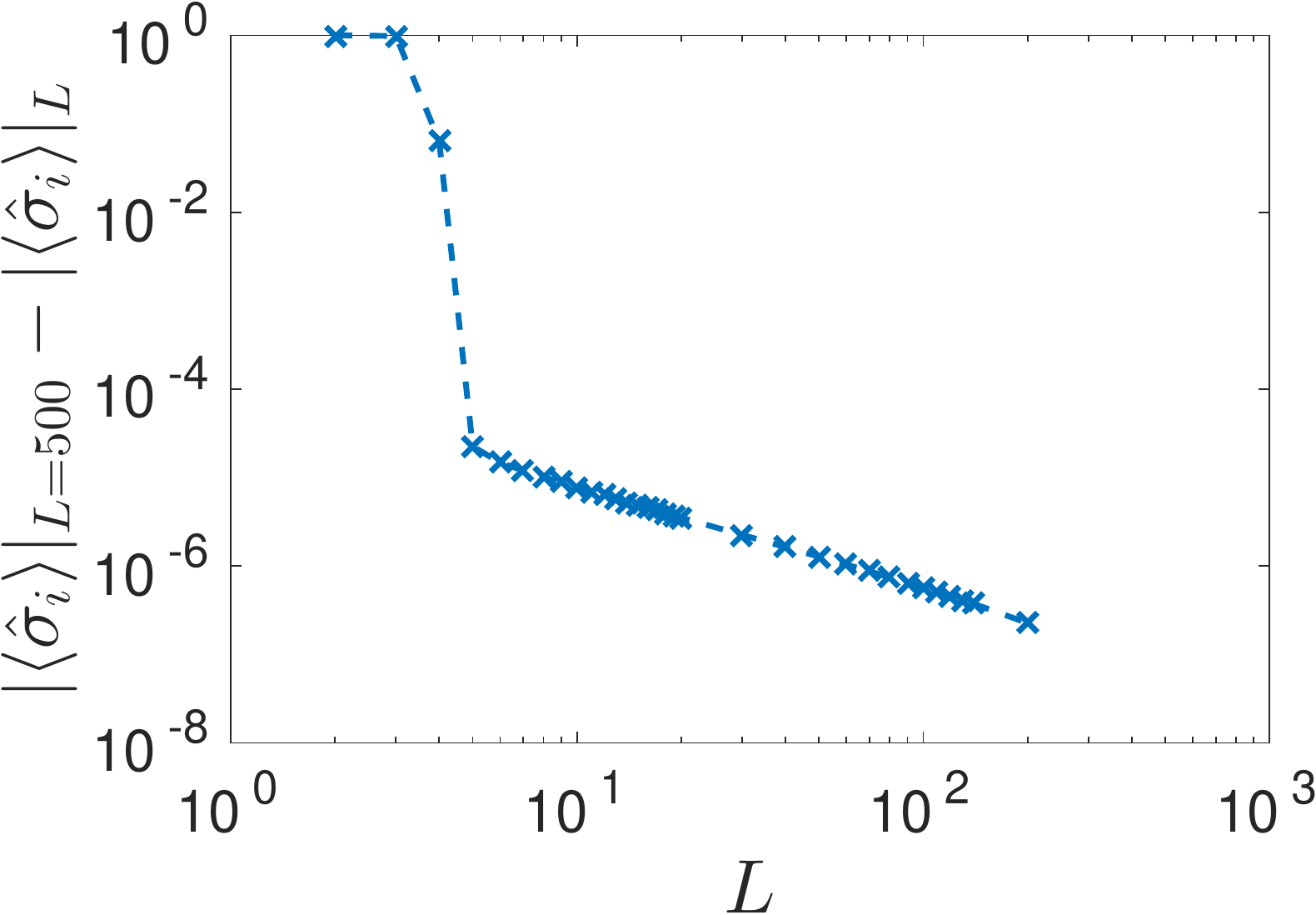}
 \caption{Finite-size effects and scaling with the size of the system for the order parameter $\langle \sigma_i \rangle$ in the ground state of the 
 Hamiltonian $\hat H_{n}^{(\rm{LR})} +  \hat W$, with 
 $\hat W = -\delta\,\sum_{i=1}^L (\hat \sigma_i + \hat \sigma_i^{\dagger})$ a small perturbation 
 breaking explicitly the symmetry of the model.
  We show here results for the case with $n=3$. In the left-panel we 
  set $h/J=0.5$ and see the exponential corrections to the order parameter when $\delta \lesssim (E_n-E_0)$,
  while in the opposed case it scales to a finite value. In the right-panel 
  we set $h/J = 0.01$ where we clearly see the polynomial scaling of the order 
  parameter to a finite value in the thermodynamic limit.
   }
 \label{fig.ssb.fss.scaling}
 \end{figure*}
\begin{figure*}
\centering
%
\includegraphics[width=0.32\textwidth]{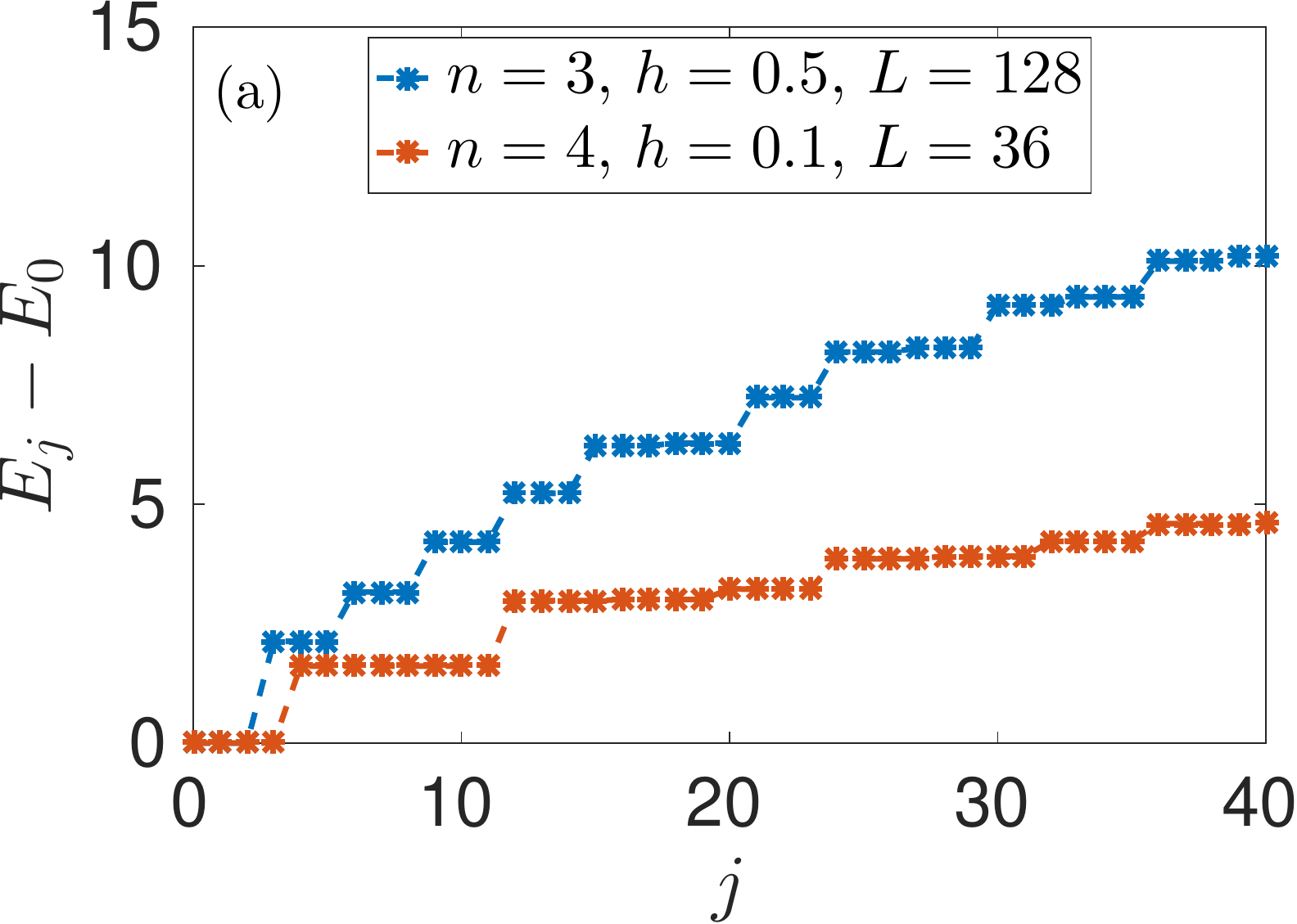}
\includegraphics[width=0.32\textwidth]{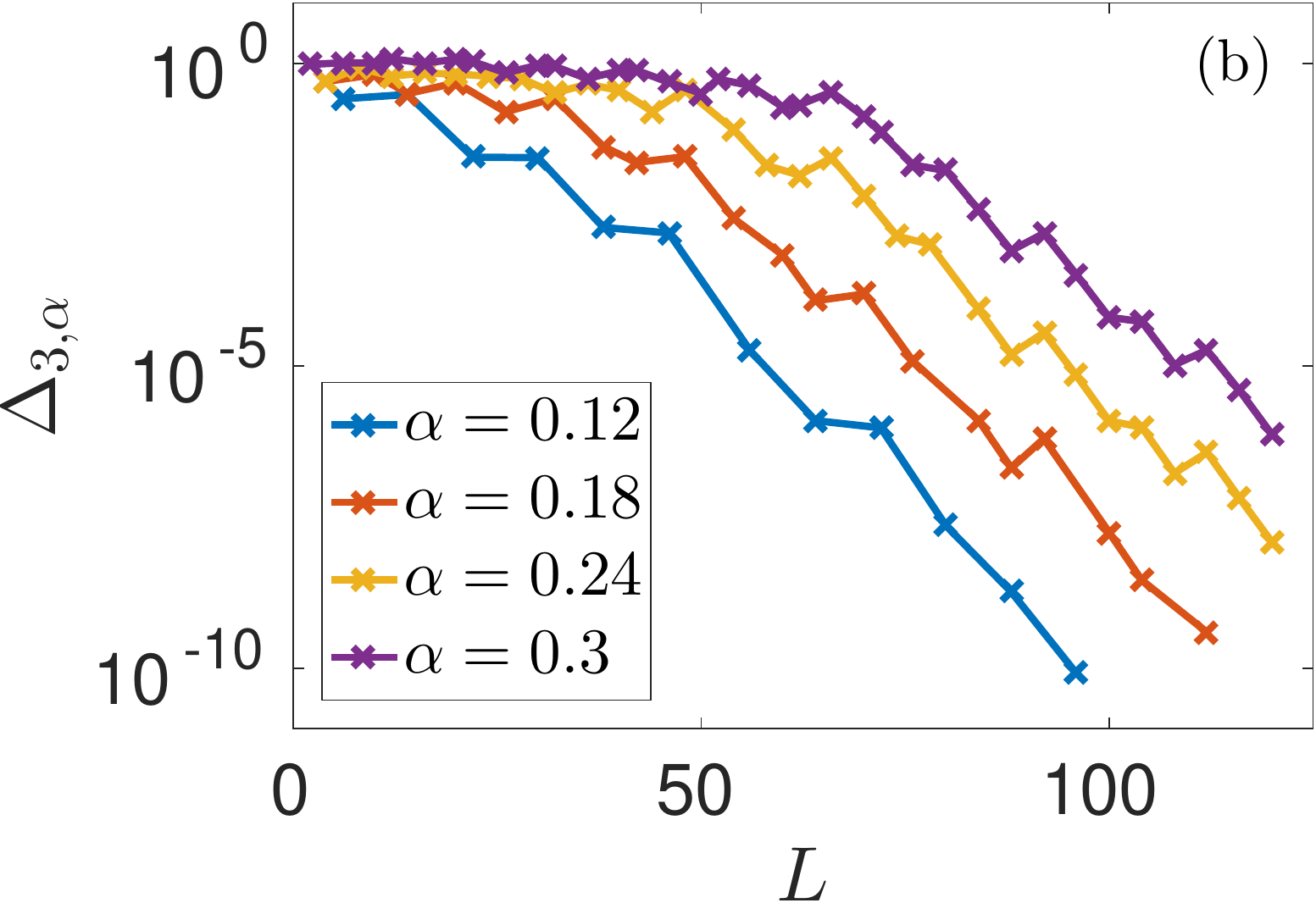}
\includegraphics[width=0.32\textwidth]{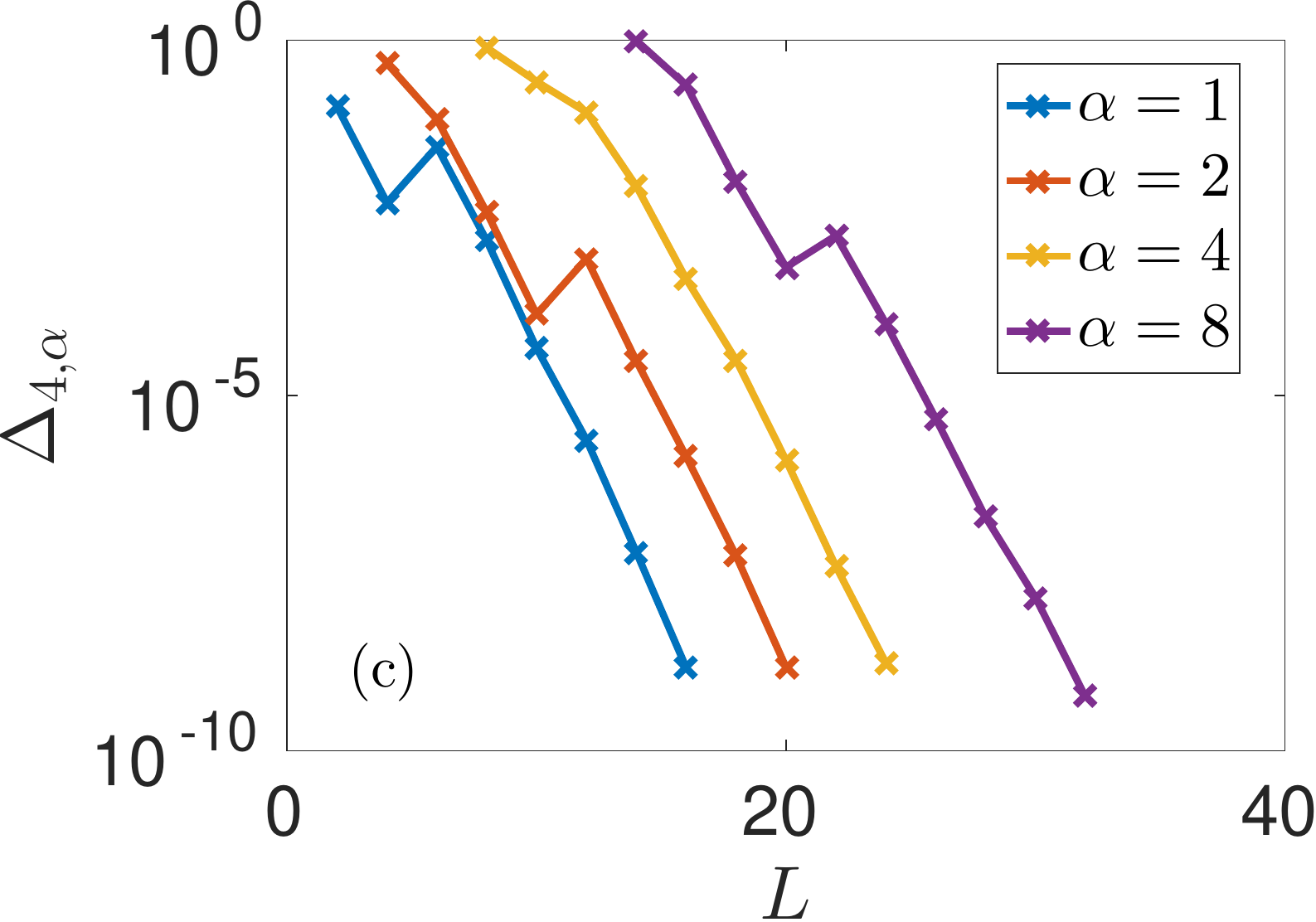}
\includegraphics[width=0.32\textwidth]{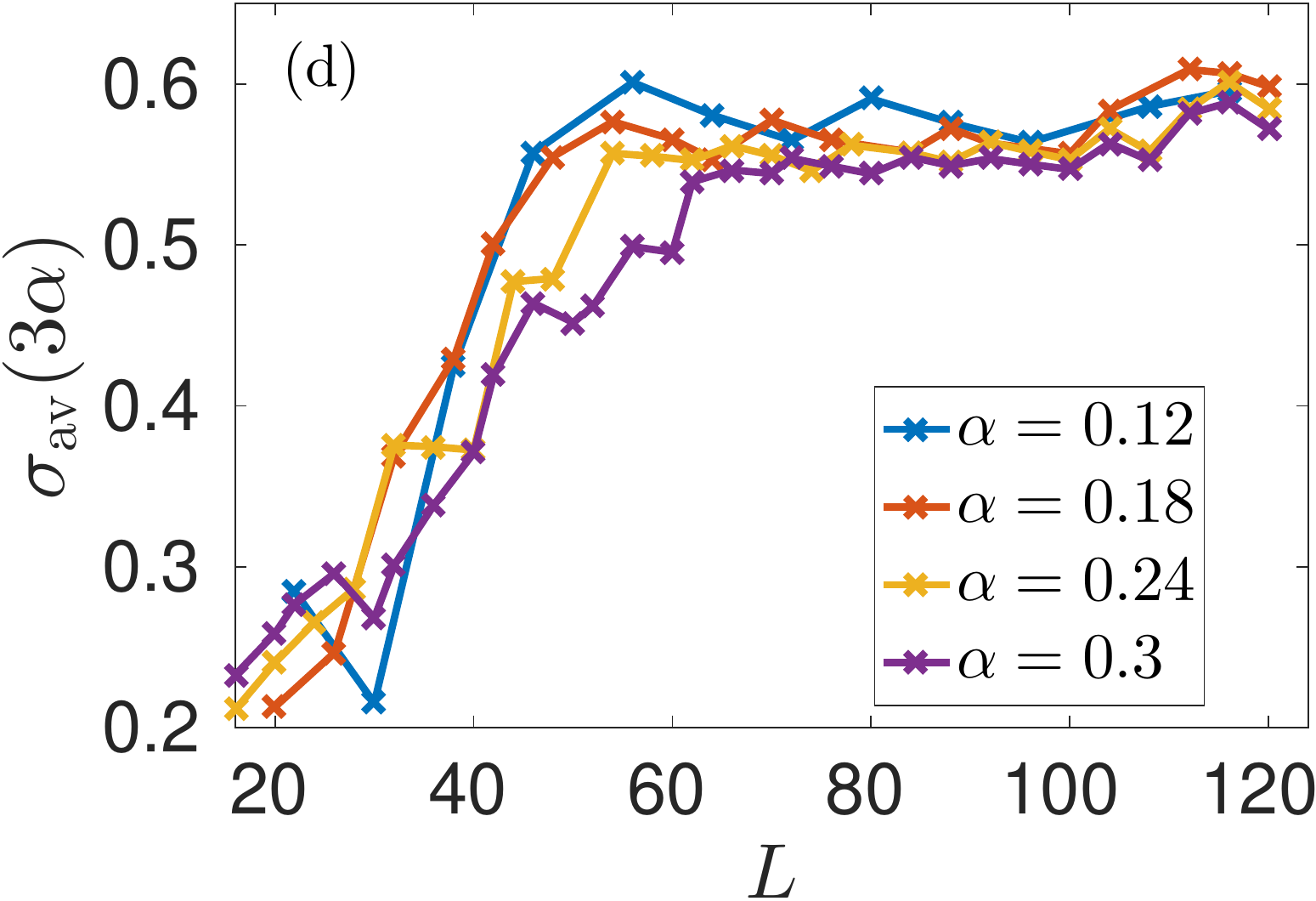}
\includegraphics[width=0.32\textwidth]{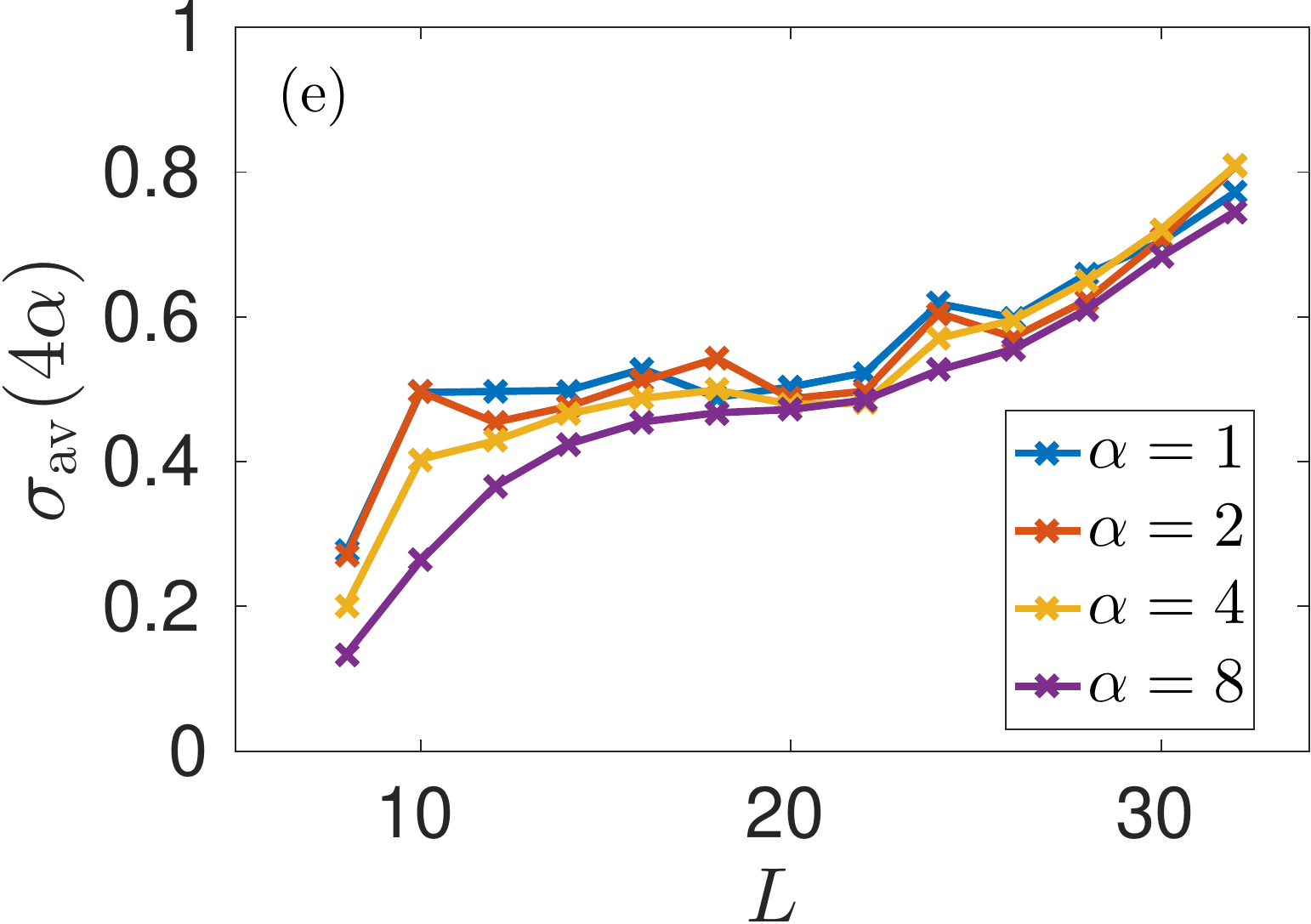}
 \caption{
  (a) Low-energy spectrum (shifted by the ground state energy) of the Hamiltonian $\hat H_{n,\eta=0}^{(LR)}$
  for $n=3$ and $n=4$: the spectrum is organized in $n$-tuplets.
  In panels (b) and (c) we show the scaling of the $n$-tuplets
  energy splittings
  ($\Delta_{q=n,\alpha}$ as defined in Eq.~\eqref{eq:ext.ngap}) with the system size.
  In panel (b) we show the case $n=3$ with $h=0.5$, and
    in (c) $n=4$ with $h=0.1$. 
    In the panels (d) and (e)
    we show the scaling of the order 
    parameter averaged over such $n$-tuplets,
     \textit{i.e.}, the averaged order parameter over the $n \alpha$ lowest eigenstates
     ($\sigma_{\rm av}(n \alpha)$ as defined in Eq.~\eqref{eq:ext.orderparam}).
     In panel (d) we show the case $n=3$ with $h=0.5$, and in
   (e) $n=4$ with $h=0.1$.
  $J=1$ in all the plots. 
  }
 \label{fig.ssb.extensive}
 \end{figure*}
%

We focus here on the infinite-range version of the Hamiltonian Eq.~\eqref{Hamiltonian-H} which we denote as $\hat H_{n,\eta}^{(LR)}$ in Sec.~\ref{time_infty:sec}.
As we have remarked in Sec.~\ref{time_infty:sec}, the presence of a period $q$-tupling time-crystal phase  is intimately related to the existence of an extensive amount of 
states that spontaneously break a $\mathbb{Z}_q$ symmetry and this will be the subject of this appendix. In the bosonic representation, this 
maps to the breaking of the translation symmetry of the model Hamiltonian.   

\subsubsection{ Cases $n=3$ and $n=4$ with $\eta = 0$} 

In this case, the $\mathbb{Z}_n$ symmetry is clearly broken when $h=0$ and, for not too large fields $h$, we should expect that this symmetry breaking persists.
The symmetry breaking manifests in the thermodynamic limit as a $n$-fold degeneracy in the ground-state subspace.
All the states of the system below a threshold energy, extensive in the size $L$  (broken symmetry edge $Le^*$), break the symmetry and the corresponding eigenenergies organize in $n$-tuplets. The order parameter characterising the symmetry breaking (in ground and excited states) is $\hat{\sigma}_i$.

We start considering the properties of the ground state. In Fig.~\eqref{fig.ssb}-(a,b,c) we analyse the properties of the ground state for $n=3$ and $4$ for finite sizes. 
In order to probe the existence of the $\mathbb{Z}_n$ symmetry breaking ground states we study the  
$n$-fold gap $E_{n-1} - E_0$ of the Hamiltonian, where $\{E_\mu\;(\mu=0,\,1,\,2,\ldots)\}$ are the eigenvalues of the  Hamiltonian in increasing order $E_\mu\leq E_{\mu+1}$, 
with $E_0$ the ground state energy.  In Fig.~\eqref{fig.ssb}-(a,b,c) we show the $n$-fold gap for different values of the system size and the coupling.
For $n=3$ and $h \lesssim 0.7 $ the $n$-fold gap closes  exponentially fast with the system size,
  while for larger $h \gtrsim 0.8$ the system is $n$-fold gapped [Fig.~\eqref{fig.ssb}-(a)]. A similar behavior occurs for $n=4$ [Fig.~\eqref{fig.ssb}-(b)],  where for 
  $h<1$ the $n$-fold gap closes exponentially with the system size, while
   for $h=1$ the closing is polynomial $(E_3-E_0) \sim L^{-1/3}$, and for larger $h>1$
  the system is $n$-fold gapped.
   
 In order to show that this $n$-fold degeneracy is  actually related to a spontaneous symmetry breaking of the interaction Hamiltonian, we add a 
  vanishingly small perturbation 
%
  $\hat W = -\delta\,\sum_{i=1}^L (\hat \sigma_i + \hat \sigma_i^{\dagger})$
%
to the Hamiltonian Eq.~\eqref{bosonic_ham:eqn} (we use $\delta = 10^{-8}$),  breaking explicitly its symmetry. In Fig.~(\ref{fig.ssb})-(c)   
the (non-degenerate) ground state  acquires then a macroscopic value for the order parameter $\langle \hat \sigma_i \rangle_{\rm GS} \sim 1$
in the region where the $n$-fold gap is roughly smaller than the perturbation, showing the existence of the symmetry breaking.
It is interesting to understand how the order parameter
  signaling the symmetry breaking phase depends on the perturbation  $\delta$ 
  and the size of the system. For a small 
  perturbation $\delta \ll (E_n-E_0)$, \textit{i.e.}, small compared to the gap of the system, we expect from first order
 in perturbation theory corrections which scale with the inverse of the gap. Thus,
   in a symmetry broken phase, these corrections due to finite-size effects
    should scale exponentially with the system size, 
    see Fig.~(\ref{fig.ssb.fss.scaling})-(left-panel). For larger perturbations $\delta \gtrapprox (E_n - E_0)$ this picture is not valid anymore, and we find 
    [Fig.~(\ref{fig.ssb.fss.scaling})-(right-panel)] that the order 
  parameter scales polynomially with the system size to a finite value in the thermodynamic limit.
   
Now we move to the excited states. In order to see if the Hamiltonian supports an extensive fraction of $\mathbb{Z}_n$ spontaneously symmetry breaking (SSB) states, we study if the spectrum is organised 
in $n$-tuplets  [see Fig.~\ref{fig.ssb.extensive}-(a)].
In general, in order to quantify the existence of an extensive amount of $q$-tuplets, we define the quantity $\Delta_{q,\alpha}$ 
 \begin{equation}\label{eq:ext.ngap}
  	\Delta_{q,\alpha} = \sum_{\mu=1}^{\mu_\alpha } (E_{q\mu-1} - E_{q\mu-q})\,,
 \end{equation}
where $\mu_\alpha=\alpha L$ ($\alpha$ a finite positive number). In Figs.\eqref{fig.ssb.extensive}-(b,c) we fix the coupling $h$ so that the ground states show spontaneous symmetry breaking, and we study the dependence of $\Delta_{q=n,\alpha}$ on the system size: we observe that 
there is 
an extensive fraction of the spectrum ($\alpha > 0$) which is organised in $n$-tuplets, where $\Delta_{n,\alpha}$ decays exponentially fast with the system size.  
 
\begin{figure*}
	\centering
	\begin{overpic}[width=0.4\linewidth]{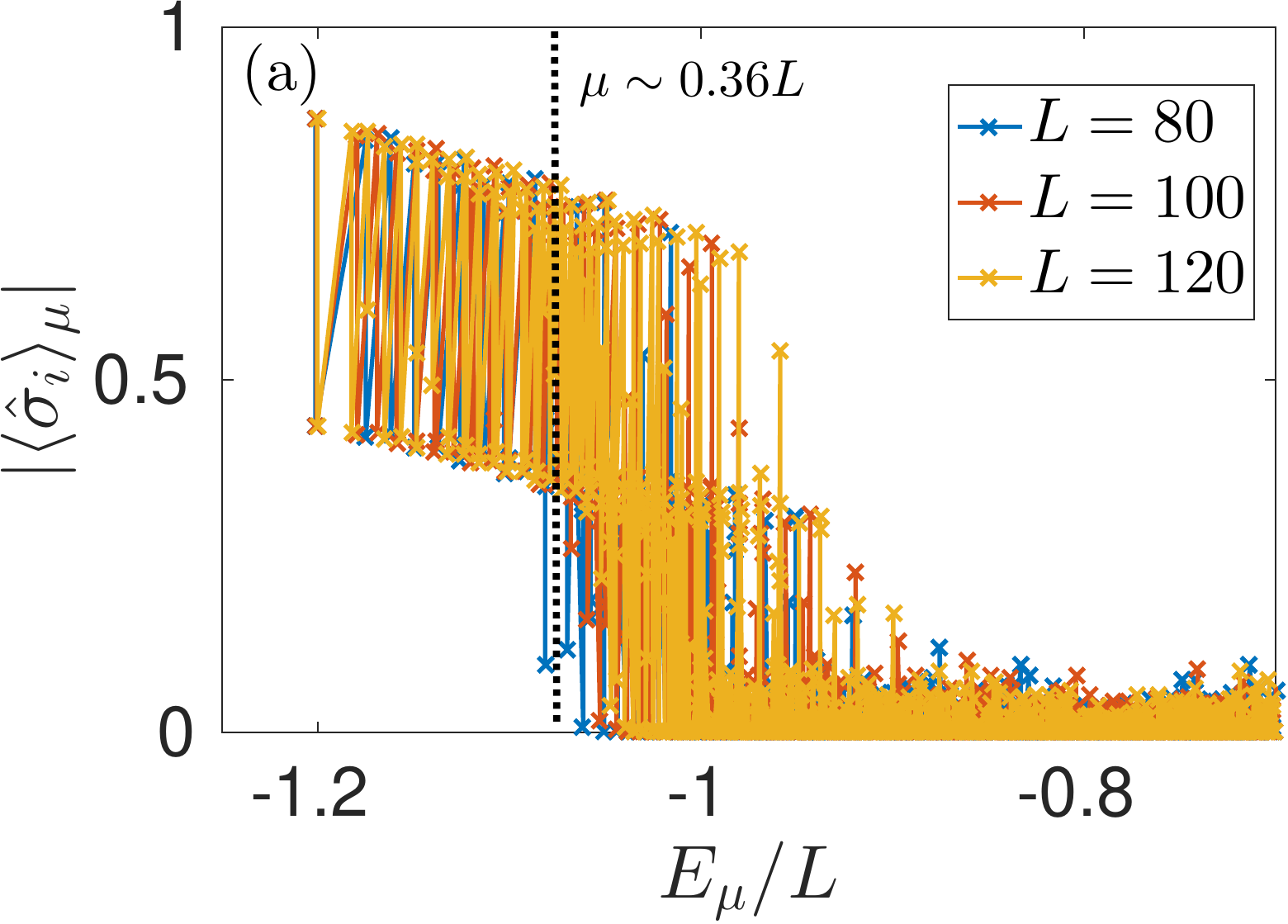}\put(-1,75){}\end{overpic}
	\begin{overpic}[width=0.4\linewidth]{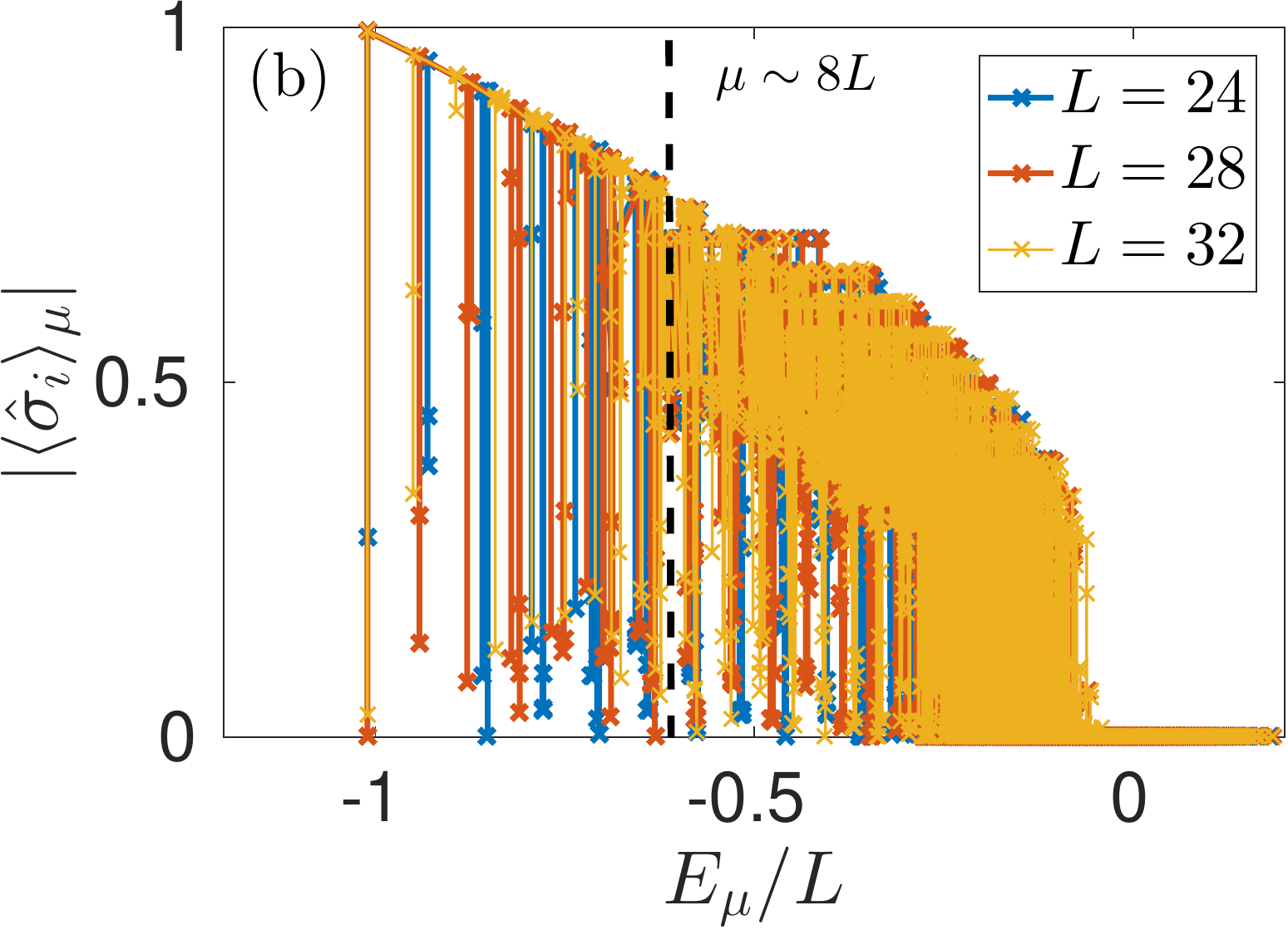}\put(-1,75){}\end{overpic}
	\caption{$|\mean{\hat{\sigma}_i}_\mu|$ versus $E_\mu/L$ for (a) $n=3$ with
	 $h = 0.5$ and (b) $n=4$ with $h=0.1$ in a symmetry-breaking phase. $J=1$ in both cases.
	  We see an extensive number of symmetry-breaking states below the broken-symmetry edge.
	}
		\label{broken-edge}
\end{figure*}

In order to show that these $q$-tuply (with $q=n$) degenerate subspaces are actually related to symmetry-breaking states, we apply the vanishingly small perturbation $\hat{W}$ defined above and compute the order parameter averaged over all the states up to $\mu_\alpha$
%
 \begin{equation}\label{eq:ext.orderparam}
  {\sigma}_{\rm av}(\alpha) = \frac{1}{\mu_\alpha} \sum_{\mu=1}^{\mu_\alpha} |\mean{\hat{\sigma}_i}_\mu|\,.
 \end{equation}
%
 In Figs.\eqref{fig.ssb.extensive}-(d,e) we notice that in the case where the 
 extensive gap $\Delta_{q=n,\alpha}$ decays exponentially fast with system 
 size [Figs.~\ref{fig.ssb.extensive}-(b,c)],
  the $n$-tuple eigenstates are indeed related to a SSB, showing a
  finite value for ${\sigma}_{\rm av}(\alpha)$. The last case corresponds to the existence of a size-independent broken-symmetry edge; we can actually see it by
plotting $|\mean{\hat{\sigma}_i}_\mu|$ versus $E_\mu/L$ [Fig.~\ref{broken-edge}].
  
  \subsubsection{Different symmetry-breaking phases for $n=4$, $\eta \neq 0$}

\begin{figure*}
\centering
\includegraphics[width=0.4\textwidth]{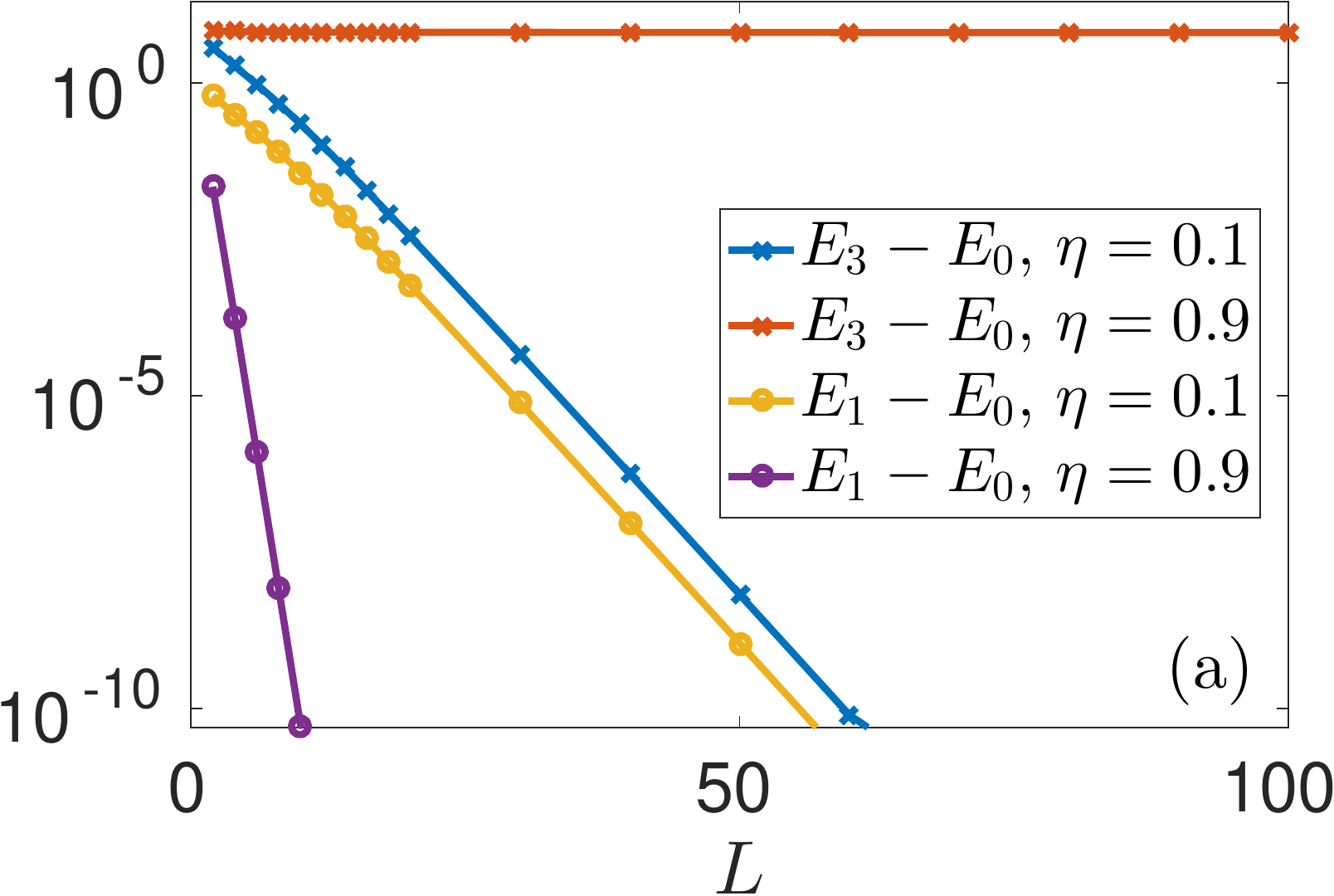}
\includegraphics[width=0.4\textwidth]{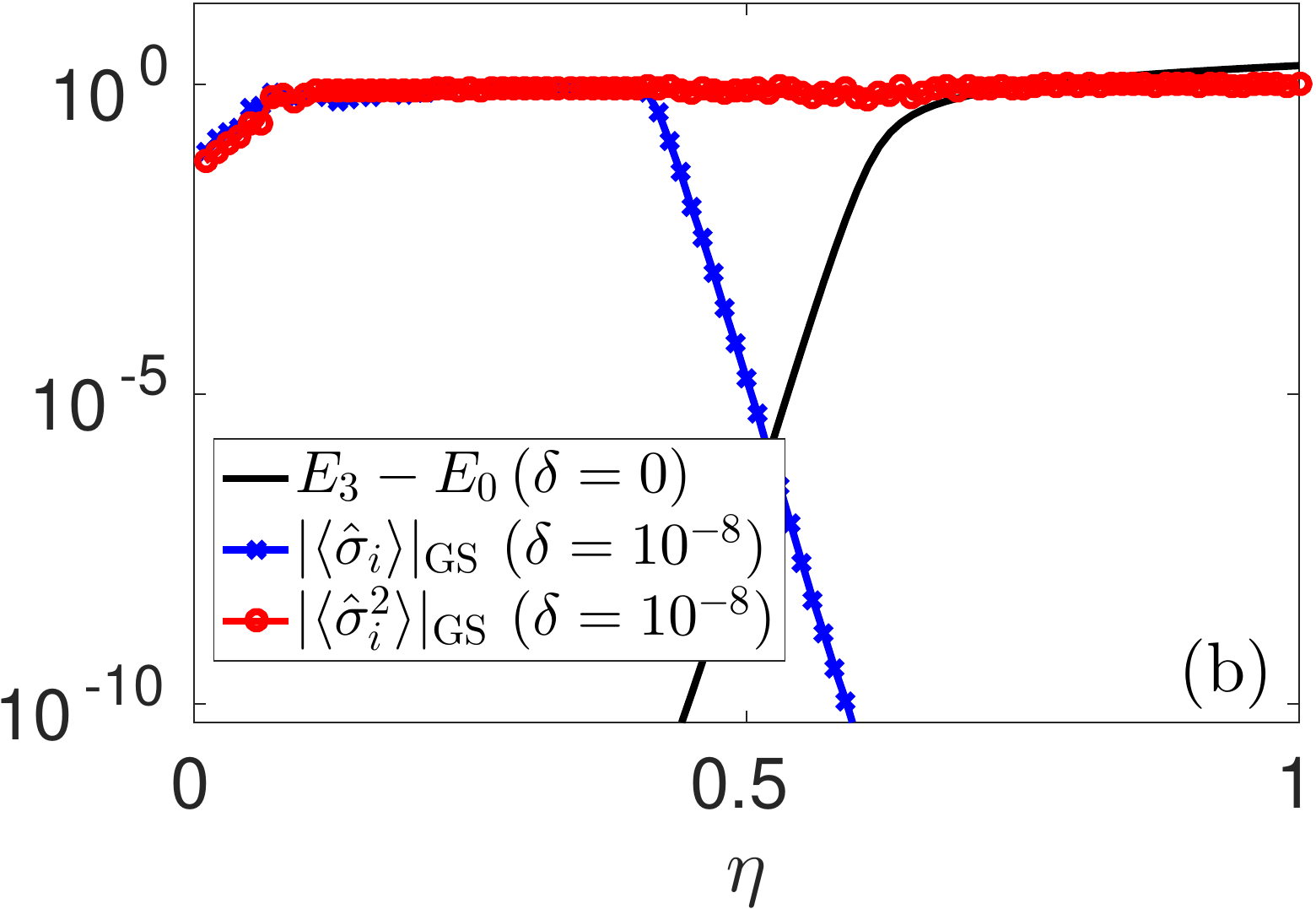}\\
\includegraphics[width=0.4\textwidth]{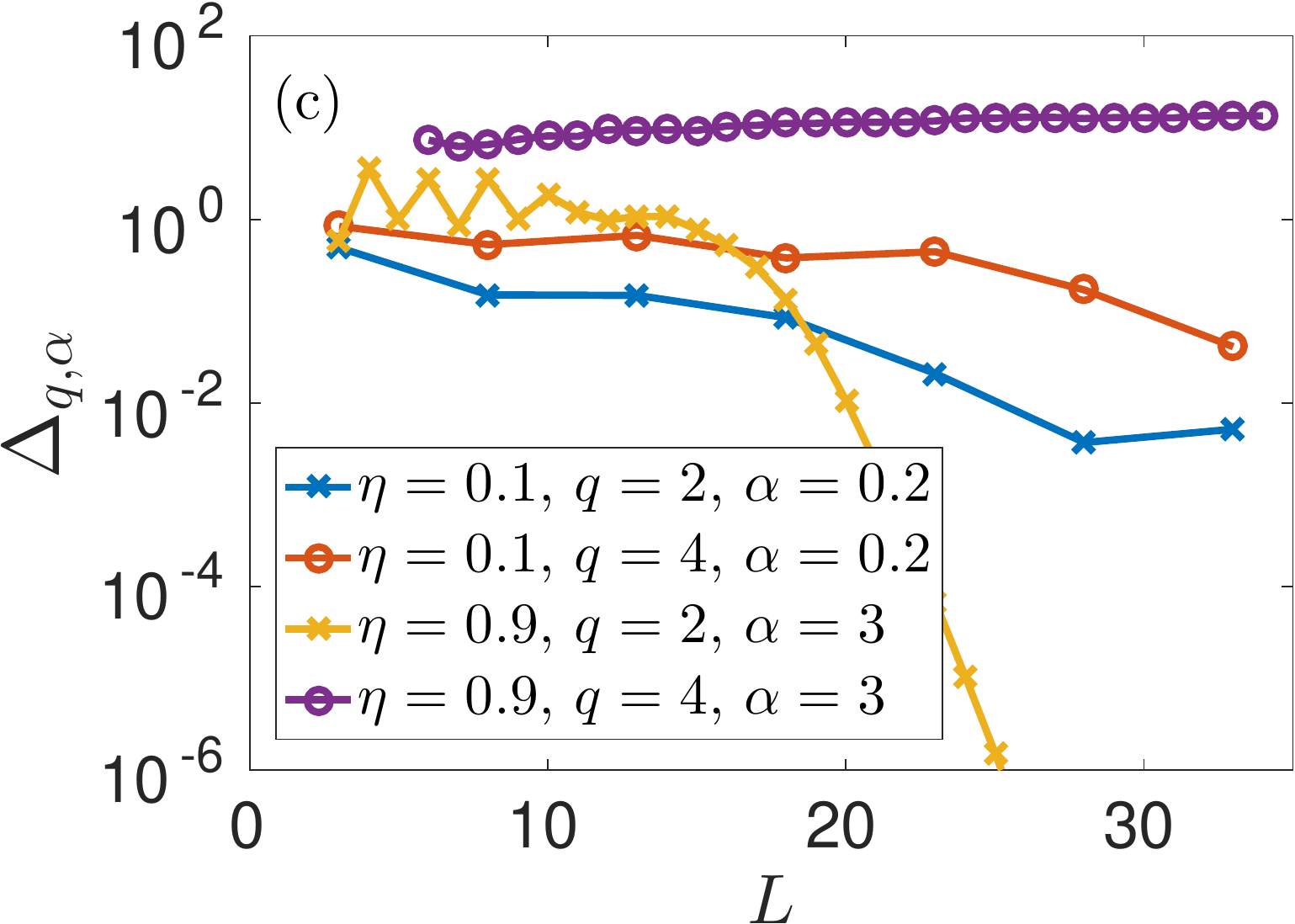}
\includegraphics[width=0.4\textwidth]{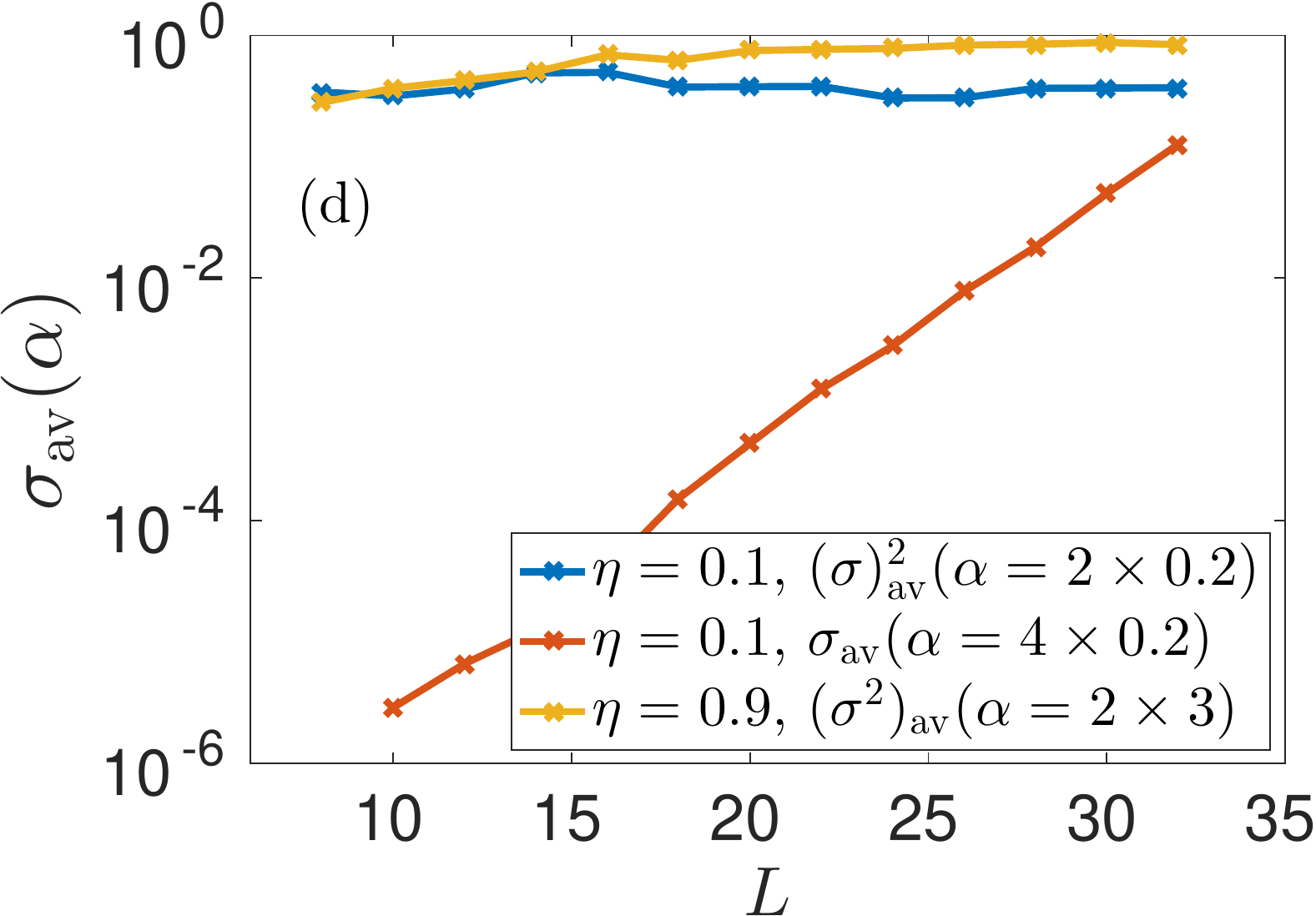}
 \caption{ 
 Different symmetry breakings in $\hat H_{4,\eta}^{(LR)}$.
 (a) Scaling 
 of the $4$-fold gap $E_3 - E_0$ and 
 the $2$-fold gap $E_1-E_0$ with the size of the system, for parameters $\eta = 0.1$ and $0.9$.
  (b) Under a small perturbation $\hat W_{[2]}$ 
   breaking explicitly the 
  symmetry of the Hamiltonian,
 the (non-degenerated) ground state
  acquires a macroscopic value for the order parameters $\langle \hat \sigma_i \rangle_{\rm GS} \sim 1$
  or $\langle \hat \sigma_i^2 \rangle_{\rm GS} \sim 1$, 
   in the region where its corresponding $q$-fold gap is roughly smaller than the perturbation strength $\delta\sim 10^{-8}$.
   Here the system size is $L=30$ and the $2$-fold gap is negligible ($<10^{-10}$),
    being omitted from the figure.
  (c) Scaling of
  $\Delta_{4,\alpha}$ and $\Delta_{2,\alpha}$ [Eq.~\eqref{eq:ext.ngap}] with the system size.
  (d)  Scaling of $\sigma_{\rm av}(\alpha)$ and 
  $(\sigma^2)_{\rm av}(\alpha)$ (Eq.~\eqref{eq:ext.orderparam})
   with the system size. $\sigma_{\rm av}(\alpha)$ is null in the case
   of $\eta = 0.9$, and we thus omit it from the figure. 
  }
 \label{fig.ssb_zptozp2}
 \end{figure*}

In this section we focus on the case $n=4$. The Hamiltonian (Eq.~\eqref{bosonic_ham:eqn}) for $\eta =0$ breaks the $\mathbb{Z}_n$ symmetry (as we have demonstrated above), while for $\eta =1$ it breaks a lower $\mathbb{Z}_{n/2}$ symmetry (being the bosonic representation of the well-known Lipkin-Meshkov-Glick model~\cite{symbreak}). The natural order parameters for these two phases are $\hat \sigma_i$ and $\hat \sigma_i^2$,  respectively. It is interesting to understand if there is a sharp transition between these two phases at a finite value of $\eta$. In order to make this analysis, we fix $J=J'=1$ and $h=h'=1/2$ and study the existence of an extensive number of SSB states
for different values of $\eta$. We find the persistence of the $\mathbb{Z}_n$ SSB phase
for $\eta$ close to zero ($\eta = 0.1$), and the persistence of the $\mathbb{Z}_{n/2}$ SSB phase
for $\eta $ close to one ($\eta = 0.9$). We find clues for a transition between the two phases at $\eta\sim 0.5$.

Let us start focusing on the case $\eta = 0.1$. Concerning the ground state properties, we can see in Fig.~\ref{fig.ssb_zptozp2}-(a) that both the 
$4$-fold and $2$-fold gap decay to zero exponentially fast with the system size, marking the existence of a $4$-fold degeneracy in the ground state. This
corresponds to a breaking of the $\mathbb{Z}_4$ symmetry of an extensive part of the spectrum: we can see this fact in Fig.~\ref{fig.ssb_zptozp2}-(c)
where both the extensive doubling gap $\Delta_{2,\alpha}$ and the 4-tupling gap $\Delta_{4,\alpha}$ decay to zero exponentially fast with $L$. In agreement
with this we find that, adding a vanishingly small perturbation
%
    $\hat W_{[2]} = -\delta\,\sum_{i=1}^L(\hat \sigma_i + \hat \sigma_i^2/2 + {\rm H.~c.})$
%
there is an extensive amount of states with a macroscopic expectation of both $\hat \sigma_i$ and $\hat \sigma_i^2$. This means that both ${\sigma}_{\rm av}(\alpha)$ and $({\sigma}^2)_{\rm av}(\alpha)$ scale to a finite value for $L\to\infty$ -- the definition for $({\sigma}^2)_{\rm av}(\alpha)$ is the same as in Eq.~\eqref{eq:ext.orderparam}. We can see this fact in Fig.~\ref{fig.ssb_zptozp2}-(d)
where for $\eta=0.1$, both ${\sigma}_{\rm av}(\alpha)$ and $({\sigma}^2)_{\rm av}(\alpha)$ tend to a finite value when $L\to\infty$. We conclude that for $\eta=0.1$ the system breaks the $\mathbb{Z}_4$ and also the $\mathbb{Z}_2$ symmetry (which is a subgroup of $\mathbb{Z}_4$).
Different is the case $\eta = 0.9$. Here there is only the breaking of the $\mathbb{Z}_{2}$ symmetry. We can see this in the ground-state properties (only $E_1-E_0$ scales to zero when $L\to\infty$  -- see Fig.~\ref{fig.ssb_zptozp2}-(a)) and in the properties of the excited states (only $\Delta_{2,\alpha}$ scales to 0 for $L\to\infty$ -- panel (c) -- and only $({\sigma}^2)_{\rm av}(\alpha)$ tends to a finite value -- panel (d)). 
The value of ${\sigma}_{\rm av}(\alpha)$ is always null in this case.
There is a transition between these two phases when $\eta$ is changed and we can see this fact in Fig.~\ref{fig.ssb_zptozp2}-(b). The symmetry-breaking ground state acquires a macroscopic 
value for the order parameter $\langle \hat \sigma_i \rangle_{\rm GS}$
in the region $\eta\gtrsim 0.5$. Here there is $\mathbb{Z}_4$ symmetry breaking and the $4$-fold gap is roughly smaller than the perturbation $\hat W_{[2]}$. On the opposite, 
$\langle \hat \sigma^2\rangle_{\rm GS}  \sim 1$
for all $\eta $, both for $\mathbb{Z}_4$ and $\mathbb{Z}_4$ symmetry breaking. The model indeed provides 
    a transition between different symmetry-breaking phases. 

\begin{thebibliography}{99}

\bibitem{goldenfeld}
	N. Goldenfeld, {\it Lectures on Phase Transitions and the Renormalization Group} (Addison Wesley, New York, 1992).
\bibitem{Wilczek2012}
	F.~Wilczek, Physical Review Letters {\bf 109}, 160401 (2012).
\bibitem{Shapere2012}
	A.~Shapere and F.~Wilczek, Physical Review Letters {\bf 109}, 160402 (2012).
\bibitem{Wilczek2013}
	F.~Wilczek,  Physical Review Letters {\bf 111}, 250402 ((2013)).
\bibitem{first-papers}
	T. Li, Z.-X. Gong, Z.-Q. Yin, H. T. Quan, X. Yin, P. Zhang, L.-M. Duan, and X. Zhang, 
	Physical Review Letters {\bf 109}, 163001 (2012); P. Bruno, {\it ibid} {\bf 110}, 118901 (2013);  
	P. Nozieres, EPL {\bf 103}, 57008 (2013); G. E. Volovik, JETP Letters 98, 491 (2013); 
	K. Sacha, Physical Review A {\bf 91}, 033617 (2015).
\bibitem{Watanabe2015}
	H.~Watanabe and M.~Oshikawa,  Physical Review Letters {\bf 114}, 251603 (2015).
\bibitem{Else2016a}
	D.~V. Else, B.~Bauer, and C.~Nayak, Physical Review Letters {\bf 117}, 090402 (2016).
\bibitem{Zhang2016}
	J.~Zhang, P.W. Hess, A. Kyprianidis, P. Becker, A. Lee, J. Smith, G. Pagano, I.-D. Potirniche, A.C. Potter, A. Vishwanath, N.Y. Yao, 
	and C. Monroe, Nature {\bf 543}, 217 (2017).
\bibitem{Choi2016}
	S.~Choi,  J. Choi, R. Landig, G. Kucsko, H. Zhou, J. Isoya, F. Jelezko, S. Onoda, H. Sumiya, V. Khemani, C. von Keyserlingk,
	N.Y. Yao, E. Demler, and M.D. Lukin1, Nature {\bf 543}, 221 (2017).
\bibitem{Khemani_2016}
	V.~Khemani, A.~Lazarides, R.~Moessner, and S.~L. Sondhi, Physical Review Letters {\bf 116}, 250401 (2016).
\bibitem{Russomanno2017}
	A.~Russomanno, F.~Iemini, M.~Dalmonte, and R.~Fazio, Physical Review  B {\bf 95}, 214307 (2017).
\bibitem{Khemani2016}
	V. Khemani, C. W. von Keyserlingk, and S. L. Sondhi, Physical Review B {\bf 96}, 115127 (2017).
\bibitem{Yao2017}
	N.~Y. Yao, A.~C. Potter, I.-D. Potirniche, and A.~Vishwanath, Physical Review Letters {\bf 118}, 030401 (2017).
\bibitem{Ho2017}
	W.~W. Ho, S.~Choi, M.~D. Lukin, and D.~A. Abanin, Physical Review Letters {\bf 119}, 010602 (2017).
\bibitem{Lazarides2017}
	A.~Lazarides, R.~Moessner, Physical Review B {\bf 95}, 195135 (2017).
\bibitem{Huang2017}
	B.~Huang, Y.-H. Wu, and W.~V. Liu, Physical Review Letters {\bf 120}, 110603 (2018).
\bibitem{Berdanier2018}
	W.~Berdanier, M.~Kolodrubetz, S.~A. Parameswaran, and R.~Vasseur, ArXiv:1803.00019.
\bibitem{Else2017}
	D.~V. Else, B.~Bauer, and C.~Nayak, Physical Review X {\bf 7}, 011026 (2017).
\bibitem{Syrwid2017}
	A.~Syrwid, J.~Zakrzewski, and K.~Sacha, Physical Review Letters {\bf 119}, 250602 (2017).
\bibitem{Fernando2017}
	F. Iemini, A. Russomanno, J. Keeling, M. Schir\'o, M. Dalmonte, and R. Fazio, Physical Review Letters {\bf 121}, 035301 (2018).
\bibitem{Gong2018}	
	Z. Gong, R. Hamazaki, and M. Ueda,  Physical Review Letters {\bf 120}, 040404 (2018).
\bibitem{Smale2018}	
	S. Smale, et al, arXiv:1806.11044. 
\bibitem{Shammah2018}
	N. Shammah, et al, arXiv:1805.05129.  
\bibitem{PhysRevLett.120.180603}
	J.~Rovny, R.~L. Blum, and S.~E. Barrett, Physical Review Letters {\bf 120}, 180603 (2018).
\bibitem{PhysRevB.97.184301}
	J.~Rovny, R.~L. Blum, and S.~E. Barrett, Physical Review B {\bf 97}, 184301 (2018).
\bibitem{Pal2018}	
	S. Pal, N. Nishad, T.~S. Mahesh, and G.~J. Sreejith, Physical Review Letters 120,180602 (2018).
\bibitem{Huse2013}
	D.~A. Huse, R.~Nandkishore, V.~Oganesyan, A.~Pal, S.~L. Sondhi, Physical Review B {\bf 88}, 014206 (2013).
\bibitem{Ponte2014}
	P.~Ponte, Z.~Papic, F.~Huveneers, and D.~A. Abanin, Physical Review Letters {\bf 114}, 140401 (2015).
\bibitem{D_Alessio_2014}
	L.~D'Alessio and M.~Rigol, Physical Review X {\bf 4}, 041048 (2014).
\bibitem{von_Keyserlingk_2016}
	C.~W. von Keyserlingk and S.~L. Sondhi, Physical Review B {\bf 93}, 245146 (2016).
\bibitem{Keyserlingk2016}
	C.~W. von Keyserlingk, V.~Khemani, and S.~L. Sondhi, Physical Review B {\bf 94}, 085112 (2016).
\bibitem{Sreejith_2016}
	G.~J. Sreejith, A.~Lazarides, and R.~Moessner, Physical Review B {\bf 94}, 045127 (2016).
\bibitem{Lukin_Nat}
 	H.~Bernien, S. Schwartz, A. Keesling, H. Levine, A. Omran, H. Pichler, S. Choi, A. S. Zibrov, M. Endres, M. Greiner, 
	V. Vuletic, and M. D. Lukin,  Nature {\bf 551}, 579 (2017). 
\bibitem{Note4}
	This argument is similar to the one leading to synchronization with the driving in~\cite{Russomanno12}.
\bibitem{Note1}
	This fact is strictly valid in systems when there are no infinite range interactions. Otherwise those
	terms can give rise to oscillations which are nevertheless overwhelmed by the 
	period-$n$-tupling ones as far as the parameter dynamics is concerned.
\bibitem{Baxter89}
	R. J. Baxter, Physics Letters A {\bf 140}, 155 (1989); Journal of  Statistical Physics {\bf 57}, 1 (1989).


\bibitem{Lipkin}
H.~Lipkin, N.~Meshkov and A.~Glick, 
\newblock Nucl. Phys. {\bf 62}, 188 (1965).
\bibitem{Note3}
These complex coefficients are chosen in such a way to make the Hamiltonian Hermitian.
\bibitem{Pollmann14}
J.~A. Kj{\"a}ll, J.~H. Bardarson, and F.~Pollmann,
\newblock Physical Review Letters {\bf 113}, 107204 (2014).
\bibitem{Fendley2012}
	P. Fendley,  Journal of  Statistical Mechanics {\bf 2012}, P11020 (2012).

\bibitem{Jermyn2014}
A.~S. Jermyn, R.~S.~K. Mong, J.~Alicea, and P.~Fendley,
\newblock 1407.6376v3.

\bibitem{Hastings2010}
M.~B. Hastings,
\newblock 1001.5280v2.

\bibitem{Liebello}
E. H. Lieb and D. W. Robinson,
\newblock Communications in Mathematical Physics {\bf 28}, 3, 251–257 (1972).

\bibitem{De_roeck_2015}
W.~D. Roeck and M.~Schütz,
\newblock Journal of Mathematical Physics {\bf 56}, 061901 (2015).


\bibitem{Zhuang2015}
Y. Zhuang, H. J. Changlani, N. M. Tubman, T. L. Hughes,
\newblock Physical Review B {\bf 92}, 035154 (2015).

\bibitem{Samajdar2018}
R. Samajdar, S. Choi, H. Pichler, M. Lukin, S. Sachdev,
\newblock arXiv:1806.01867

\bibitem{Note3}
After the disorder-average, the off-diagonal terms in Eq.~\eqref{diagon_time:eqn}
applied to $\hat{O}=\hat{\sigma}$ cancel out in long times, due to destructive interference.

\bibitem{Note4}
 From a technical point of view, in the numerics we consider the ground state of $\hat H_{n,\eta=0}^{(LR)} + \delta\sum_{i=1}^L(\hat\sigma_i+\hat\sigma_i^\dagger)$, with $\delta = 10^{-8}$.

\bibitem{Oganesyan2006}
V.~Oganesyan and D.~A. Huse,
\newblock Physical Review B {\bf 75}, 155111.

 \bibitem{Note5}
 
 This can be shown as follows:
 \[{\hat U_{\eta}}^2=\hat X e^{-i\hat H_{\eta}^{(SR)}}\hat X e^{-i\hat H_{\eta}^{(SR)}}=\hat X^2 e^{-i\hat X^\dagger \hat H_{\eta}^{(SR)}\hat X} e^{-i\hat H_{\eta}^{(SR)}}\]
 and since $\hat X^\dagger \hat H_{\eta}^{(SR)}\hat X$ commutes with $\hat H_{\eta}^{(SR)}$, we get
 \begin{align} 
&\hat U_{\eta}^2= \left(\prod_i \tau_i^2 \right)\exp\left[-2iT\left(\sum_i J_i \left[\hat \sigma_i^2 \hat \sigma_{i+1}^2+\right.\right.\right. \nonumber\\
&\left.\left.\left.(1-\eta)(\alpha_1 \hat \sigma_i^\dagger \hat \sigma_{i+1}+{\rm H.\,c.})\right]+\eta \sum_i g_i \hat \tau_i^2\right)\right]\nonumber
 \end{align}

\bibitem{Note6}
From a technical point of view, in the numerics we consider the ground state of $\hat H_{n,\eta}^{(LR)} + \delta\sum_{i=1}^L\left(\hat{\sigma}_i^2+{\rm H.~c.}\right)$, with $\delta = 10^{-8}$.

\bibitem{Note7}
We notice that $h_{n,\eta}^{(LR)}$ coincides with $\mean{\hat{H}_{n,\eta}^{(LR)}}/L$ in the correspondence limit $L\to\infty$.

\bibitem{Russomanno12}
A.~Russomanno, A.~Silva and G.~E. Santoro,
\newblock Physical Review Letters {\bf 109}, 257201 (2012).

\bibitem{symbreak}
For the symmetry breaking in this model see for instance~\cite{Lipkin} or
\newblock G.~Mazza and M.~Fabrizio,
\newblock Physical Review B {\bf 86}, 184303, (2012).







  
  




\bibitem{PhysRevE.97.020202}
R.~R.~W. Wang, B.~Xing, G.~G. Carlo, and D.~Poletti,
\newblock Physical Review E {\bf 97}, 020202 (2018).


\bibitem{smerza}
A.~Smerzi, S.~Fantoni, S.~Giovanazzi and S.~R. Shenoy,
\newblock Physical Review Letters, {\bf 79}, 4950 (1997).

\end{thebibliography}
%

\end{document}